\documentclass[twocolumn, twocolappendix, times]{aastex631}

\usepackage{amsmath}
\usepackage{xspace}
\usepackage{txfonts}

\newcommand{\tco}{$^{13}$CO\xspace}
\newcommand{\ceo}{C$^{18}$O\xspace}

\shorttitle{eDisk IV. L1489 IRS first-look}
\shortauthors{Yamato et al.}
\graphicspath{{./}{figures/}}
\begin{document}

\title{Early Planet Formation in Embedded Disks (eDisk). IV. \par The Ringed and Warped Structure of the Disk around the Class~I Protostar L1489~IRS}

\author[0000-0003-4099-6941]{Yoshihide Yamato}
\affiliation{Department of Astronomy, Graduate School of Science, The University of Tokyo, 7-3-1 Hongo, Bunkyo-ku, Tokyo 113-0033, Japan}

%%%%%%%%%%%%%%%%%%%%%%%%%%%%%%%%%%%%%%%%%%%%%%
%%%%% AUTHOR LIST:
%%%%% Add your name + ORCID and affiliation here if you want to be part of and have contributed to the paper.
%%%%% The exact order of the author list will be determined later.

% \author{ORDER OF THE CO-AUTHORS IS TBD} 

% supervisor
\author[0000-0003-3283-6884]{Yuri Aikawa}
\affiliation{Department of Astronomy, Graduate School of Science, The University of Tokyo, 7-3-1 Hongo, Bunkyo-ku, Tokyo 113-0033, Japan}

% co-PIs
\author[0000-0003-0998-5064]{Nagayoshi Ohashi}
\affiliation{Academia Sinica Institute of Astronomy \& Astrophysics, 11F of Astronomy-Mathematics Building, AS/NTU, No.\ 1, Sec.\ 4, Roosevelt Rd, Taipei 10617, Taiwan, R.O.C.}

\author[0000-0002-6195-0152]{John J. Tobin}
\affiliation{National Radio Astronomy Observatory, 520 Edgemont Rd., Charlottesville, VA 22903, USA}

\author[0000-0001-9133-8047]{Jes K. J{\o}rgensen}
\affiliation{Niels Bohr Institute, University of Copenhagen, \O ster Voldgade 5-7, 1350, Copenhagen K, Denmark}

% those who contributed significantly
\author[0000-0003-0845-128X]{Shigehisa Takakuwa}
\affiliation{Department of Physics and Astronomy, Graduate School of Science and Engineering, Kagoshima University, 1-21-35 Korimoto, Kagoshima, Kagoshima 890-0065, Japan}
\affiliation{Academia Sinica Institute of Astronomy \& Astrophysics, 11F of Astronomy-Mathematics Building, AS/NTU, No.\ 1, Sec.\ 4, Roosevelt Rd, Taipei 10617, Taiwan, R.O.C.}

\author[0000-0002-8238-7709]{Yusuke Aso}
\affiliation{Korea Astronomy and Space Science Institute, 776 Daedeok-daero, Yuseong-gu, Daejeon 34055, Republic of Korea}

\author[0000-0003-4361-5577]{Jinshi Sai (Insa Choi)}
\affiliation{Academia Sinica Institute of Astronomy \& Astrophysics, 11F of Astronomy-Mathematics Building, AS/NTU, No.\ 1, Sec.\ 4, Roosevelt Rd, Taipei 10617, Taiwan, R.O.C.}

% alphabetical order below

\author[0000-0002-8591-472X]{Christian Flores}
\affiliation{Academia Sinica Institute of Astronomy \& Astrophysics, 11F of Astronomy-Mathematics Building, AS/NTU, No.\ 1, Sec.\ 4, Roosevelt Rd, Taipei 10617, Taiwan, R.O.C.}

\author[0000-0003-4518-407X]{Itziar de Gregorio-Monsalvo}
\affiliation{European Southern Observatory, Alonso de Cordova 3107, Casilla 19, Vitacura, Santiago, Chile}

\author[0000-0002-4317-767X]{Shingo Hirano}
\affiliation{Department of Astronomy, Graduate School of Science, The University of Tokyo, 7-3-1 Hongo, Bunkyo-ku, Tokyo 113-0033, Japan}

\author[0000-0002-9143-1433]{Ilseung Han}
\affiliation{Division of Astronomy and Space Science, University of Science and Technology, 217 Gajeong-ro, Yuseong-gu, Daejeon 34113, Republic of Korea}
\affiliation{Korea Astronomy and Space Science Institute, 776 Daedeok-daero, Yuseong-gu, Daejeon 34055, Republic of Korea}

\author[0000-0002-2902-4239]{Miyu Kido}
\affiliation{Department of Physics and Astronomy, Graduate School of Science and Engineering, Kagoshima University, 1-21-35 Korimoto, Kagoshima, Kagoshima 890-0065, Japan}

\author[0000-0003-2777-5861]{Patrick M. Koch}
\affil{Academia Sinica Institute of Astronomy \& Astrophysics, 11F of Astronomy-Mathematics Building, AS/NTU, No.\ 1, Sec.\ 4, Roosevelt Rd, Taipei 10617, Taiwan, R.O.C.}

\author[0000-0003-4022-4132]{Woojin Kwon}
\affiliation{Department of Earth Science Education, Seoul National University, 1 Gwanak-ro, Gwanak-gu, Seoul 08826, Republic of Korea}
\affiliation{SNU Astronomy Research Center, Seoul National University, 1 Gwanak-ro, Gwanak-gu, Seoul 08826, Republic of Korea}

\author[0000-0001-5522-486X]{Shih-Ping Lai}
\affiliation{Institute of Astronomy, National Tsing Hua University, No. 101, Section 2, Kuang-Fu Road, Hsinchu 30013, Taiwan}
\affiliation{Center for Informatics and Computation in Astronomy, National Tsing Hua University, No. 101, Section 2, Kuang-Fu Road, Hsinchu 30013, Taiwan}
\affiliation{Department of Physics, National Tsing Hua University, No. 101, Section 2, Kuang-Fu Road, Hsinchu 30013, Taiwan}
\affiliation{Academia Sinica Institute of Astronomy \& Astrophysics, 11F of Astronomy-Mathematics Building, AS/NTU, No.\ 1, Sec.\ 4, Roosevelt Rd, Taipei 10617, Taiwan, R.O.C.}

\author[0000-0002-3179-6334]{Chang Won Lee}
\affiliation{Division of Astronomy and Space Science, University of Science and Technology, 217 Gajeong-ro, Yuseong-gu, Daejeon 34113, Republic of Korea}
\affiliation{Korea Astronomy and Space Science Institute, 776 Daedeok-daero, Yuseong-gu, Daejeon 34055, Republic of Korea}

\author[0000-0003-3119-2087]{Jeong-Eun Lee}
\affiliation{Department of Physics and Astronomy, Seoul National University, 1 Gwanak-ro, Gwanak-gu, Seoul 08826, Korea}

\author[0000-0002-7402-6487]{Zhi-Yun Li}
\affiliation{University of Virginia, 530 McCormick Rd., Charlottesville, Virginia 22903, USA}

\author[0000-0001-7233-4171]{Zhe-Yu Daniel Lin}
\affiliation{University of Virginia, 530 McCormick Rd., Charlottesville, Virginia 22903, USA}

\author[0000-0002-4540-6587]{Leslie W. Looney}
\affiliation{Department of Astronomy, University of Illinois, 1002 West Green St, Urbana, IL 61801, USA}

\author[0000-0002-7002-939X]{Shoji Mori}
% \affiliation{Astronomical Institute, Tohoku University}
\affiliation{Astronomical Institute, Graduate School of Science, Tohoku University, Sendai 980-8578, Japan}

\author[0000-0002-0244-6650]{Suchitra Narayanan}
\affiliation{Institute for Astronomy, University of Hawai`i at Mānoa, 2680 Woodlawn Dr., Honolulu, HI 96822, USA}

\author[0000-0002-4372-5509]{Nguyen Thi Phuong}
\affiliation{Korea Astronomy and Space Science Institute, 776 Daedeok-daero, Yuseong-gu, Daejeon 34055, Republic of Korea}
\affiliation{Department of Astrophysics, Vietnam National Space Center, Vietnam Academy of Science and Techonology, 18 Hoang Quoc Viet, Cau Giay, Hanoi, Vietnam}

\author{Kazuya Saigo}
\affiliation{Department of Physics and Astronomy, Graduate School of Science and Engineering, Kagoshima University, 1-21-35 Korimoto, Kagoshima, Kagoshima 890-0065, Japan}

\author[0000-0001-6267-2820]{Alejandro Santamaría-Miranda}
\affiliation{European Southern Observatory, Alonso de Cordova 3107, Casilla 19, Vitacura, Santiago, Chile}

\author[0000-0002-0549-544X]{Rajeeb Sharma}, 
\affiliation{Niels Bohr Institute, University of Copenhagen, \O ster Voldgade 5-7, 1350, Copenhagen K, Denmark}

\author[0000-0003-0334-1583]{Travis J. Thieme}
\affiliation{Institute of Astronomy, National Tsing Hua University, No. 101, Section 2, Kuang-Fu Road, Hsinchu 30013, Taiwan}
\affiliation{Center for Informatics and Computation in Astronomy, National Tsing Hua University, No. 101, Section 2, Kuang-Fu Road, Hsinchu 30013, Taiwan}
\affiliation{Department of Physics, National Tsing Hua University, No. 101, Section 2, Kuang-Fu Road, Hsinchu 30013, Taiwan}

\author[0000-0001-8105-8113]{Kengo Tomida}
\affiliation{Astronomical Institute, Graduate School of Science, Tohoku University, Sendai 980-8578, Japan}

\author[0000-0002-2555-9869]{Merel L.R. van 't Hoff}
\affil{Department of Astronomy, University of Michigan, 1085 S. University Ave., Ann Arbor, MI 48109-1107, USA}

\author[0000-0003-1412-893X]{Hsi-Wei Yen}
\affiliation{Academia Sinica Institute of Astronomy \& Astrophysics, 11F of Astronomy-Mathematics Building, AS/NTU, No.\ 1, Sec.\ 4, Roosevelt Rd, Taipei 10617, Taiwan, R.O.C.}

%%%%%%%%%%%%%%%%%%%%%%%%%%%%%%%%%%%%%%%%%%%%%%

%%%%%%%%%%%%%%%%%%%%%%%%%%%%%%%%%%%%%%%%%%%%%%
%people who provided comments so far:
%Tobin, Ohashi, Jorgensen, Aikawa, Takakuwa, Lin, Koch, Li, Sai, Aso, Gregorio-Monsalvo, Sharma, Theme



\begin{abstract}
Constraining the physical and chemical structure of young embedded disks is crucial to understanding the earliest stages of planet formation. As part of the Early Planet Formation in Embedded Disks Atacama Large Millimeter/submillimeter Array Large Program, we present high spatial resolution ($\sim$0\farcs1 or $\sim$15$\,$au) observations of the 1.3~mm continuum and \tco $J=$ 2--1, \ceo $J=$ 2--1, and SO $J_N=$ $6_5$--$5_4$ molecular lines toward the disk around the Class~I protostar L1489~IRS. The continuum emission shows a ring-like structure at 56\,au from the central protostar
% , confirmed by the visibility analysis, 
and a tenuous, optically thin emission extending beyond $\sim$300\,au.
% At the outer disk, two shoulder-like structures are also identified at $\sim$220\,au and $\sim$340\,au in the continuum emission. 
The \tco emission traces the warm disk surface, while the \ceo emission originates from near the disk midplane. 
% In addition, both dust continuum and CO isotopologue emission show a brightness asymmetry along the disk major axis, where the southwestern side of the disk is brighter than the other side. 
% We confirmed the Keplerian rotation of the disk with a central stellar mass of $1.5$--$1.9\,M_\odot$ using \ceo. 
% The central protostellar mass is estimated to be $1.5$--$1.9\,M_\odot$, based on the Keplerian rotation of the disk traced by the \ceo emission.
The coincidence of the radial emission peak of \ceo with the dust ring may indicate a gap-ring structure in the gaseous disk as well. The SO emission shows a highly complex distribution, including a compact, prominent component at $\lesssim$ 30\,au, which is likely to originate from thermally sublimated SO molecules. The compact SO emission also shows a velocity gradient along a slightly ($\sim15\arcdeg$) tilted direction with respect to the major axis of the dust disk, which we interpret as an inner warped disk in addition to the warp around $\sim$200\,au suggested by previous work. These warped structures may be formed by a planet or companion with an inclined orbit, or by a gradual change in the angular momentum axis during gas infall. 

\end{abstract}

%% \keywords{Classical Novae (251) --- Ultraviolet astronomy(1736) --- History of astronomy(1868) --- Interdisciplinary astronomy(804)}


\section{Introduction} \label{sec:intro}
Circumstellar disks are the birthplaces of planets. Constraining the structure of disks is thus essential for understanding the processes of planetary system formation. The physical and chemical structure of Class~II disks has been studied in great detail through dust continuum and molecular line observations with the Atacama Large Millimeter/submillimeter Array (ALMA) \citep[e.g.,][]{Andrews2018, Oberg2021}. These studies have revealed that substructures such as rings and gaps are present in both dust and gas, suggesting that planet formation is ongoing in Class~II disks. Furthermore, the presence of formed planets has recently become more apparent through the detection of velocity kinks in channel maps of bright molecular line emission \citep[e.g.,][]{Pinte2018_HD163296, Teague2019}, in addition to direct detections of circumplanetary disks within gaps \citep[e.g.,][]{Keppler2018, Benisty2021, Bae2022}.

% \citep[e.g.,][]{Isella2018, Guzman2018, Law2021_MAPSIV, Zhang2021}

% Observations with Atacama Large Millimeter/submillimeter Array (ALMA) at high spatial resolutions have revealed that substructures such as rings and gaps are present in both dust and gas in the disks around pre-main sequence stars \citep[i.e., Class~II sources; e.g.,][]{Andrews2018, Oberg2021}, which suggests that planets have already been formed in Class~II phase. Indeed, the existence of formed planets has recently been more evident by detection of velocity kinks in channel maps of bright molecular line emission \citep[e.g.,][]{Pinte2018_HD163296, Teague2019}. The physical and chemical structures of Class~II disks have been studied in greater details using high spatial resolution dust continuum and molecular line observations \citep[e.g.,][]{Isella2018, Guzman2018, Law2021_MAPSIV, Zhang2021}, providing the key information about the planet-forming environment. 

The presence of substructures and planets in Class~II disks suggests that planet formation could have started in the earlier evolutionary stages. It is thus essential to study what is happening in disks around younger Class~0/I protostars in order to obtain a complete picture of planet formation processes. 
% However, it is now emerging that the substructures of the disk has already been built in earlier evolutionary phases. 
Recent ALMA observations have revealed substructures in dust continuum emission in a handful of Class~0/I disks \citep[e.g.,][]{ALMAPartnership2015, Sheehan2018, Sheehan2020, Segura-cox2020}, suggesting that the first steps toward planet formation may be ocurring while these young disks are still embedded in their natal envelopes. Grain growth has also been suggested in embedded young disks \citep[e.g.,][]{Harsono2018}. In addition, recent surveys have shown that the mass of Class~II disks is insufficient for giant planets to be formed, while Class~0/I disks are massive enough \citep[e.g.,][]{Tychoniec2021}. 
% Furthermore, distributions of molecular gas in Class 0/I disks are still unclear due to the lacks of observations with sufficient spatial resolution to spatially separate the disk and envelope.
% Therefore, planet formation is likely to have already started in the young embedded disks. Young embedded disks provide initial conditions of planet formation, but their physical/chemical structures are poorly constrained.

Motivated by these previous studies, the Early Planet Formation in Embedded Disks (eDisk) ALMA Large Program \citep{eDisk_overview} was initiated with the main goal of understanding how early substructures form in the disks around young protostars. The core of eDisk is a high-resolution systematic survey of the dust continuum substructures down to $\sim 5$ au scales in 19 young embedded sources. In addition to the dust continuum, several molecular lines, mainly CO isotopologue lines, have been observed to probe the gas distributions and kinematics in the embedded disks as well as in the envelopes. 

One of the sources of the eDisk sample is the Class~I protostar L1489~IRS (or IRAS 04016+2610), located in the Taurus molecular cloud. While several recent studies assume $\sim$140\,pc as the distance to L1489~IRS \citep[e.g.,][]{Sai2022, Mercimek2022} based on the average distance over the Taurus region \citep[141$\pm$7\,pc;][]{Zucker2019}, \citet{Roccatagliata2020} estimate the distances to six different stellar population groups in the Taurus region using Gaia measurements. One of the groups (Taurus F in the original paper) with an estimated distance of 146\,pc includes a nearby source to L1489~IRS (with a separation of $\sim12\arcmin$). Thus, in the present work we assume that the distance to L1489~IRS is 146\,pc\footnote{We note that there is an independent estimate of the distance toward a neighboring molecular cloud L1498 of 129\,pc \citep{Zucker2020}, which is significantly different from 146\,pc adopted here.}. Taking into account this distance, as well as the updated photometry from near-infrared to millimeter wavelengths, the bolometric luminosity and bolometric temperature are 3.4~$L_\odot$ and 213~K, respectively \citep{eDisk_overview}, from which L1489~IRS is classified as a Class~I source.

L1489~IRS is embedded in a $\sim$$2000$\,au scale infalling-rotating envelope as revealed by single-dish observations \citep{Hogerheijde2000, Motte2001}. A bipolar outflow along the northwest-southeast direction is also identified by infrared and submillimeter line observations \citep{Tamura1991, Hogerheijde1998}. \citet{Brinch2007} first detected a rotation signature at $\sim$$200$--$300$\,au scale using $\sim$$1\arcsec$ resolution HCO$^+$ line observations with the Submillimeter Array (SMA), which was interpreted as a Keplerian disk. ALMA observations of $^{12}$CO and \ceo ($J=2$--$1$) lines confirmed the presence of a Keplerian disk with a radius of $\sim$$600$--$700$\,au and a central stellar mass of $1.6\,M_\odot$ \citep{Yen2014}. \citet{Yen2014} also detected a streamer-like infalling flow feeding the material to the central protostellar disk with the \ceo $J=$ 2--1 emission. Higher-resolution ($\sim0\farcs3$) \ceo line observations with ALMA revealed that the disk is warped at the boundary ($r\sim200$\,au) between the inner and outer disks \citep{Sai2020}. Dust continuum images obtained with ALMA \citep{Yen2014, Sai2020} and the Combined Array for Research in Millimeter-wave Astronomy \citep[CARMA;][]{Sheehan2017} also show a disk-like structure elongated along the northeast-southwest direction. Furthermore, \citet{Ohashi2022_L1489IRS} tentatively detect a ring-like structure at $r\sim90$\,au in the 1.3~mm dust continuum emission, which may be a hint of dust growth, or even planet formation.

% While this source is still embedded in the envelope \citep{Hogerheijde2000, Motte2001, Sai2022} and shows a bipolar outflow \citep{Tamura1991, Hogerheijde1998, Yen2014}, its high bolometric temperature indicates that this is a relatively evolved source (i.e., Class~I). Recent interferometric observations have unveiled a large Keplerian disk ($\sim$600--700\,au in radius) with a central stellar mass of $\sim$1.6\,$M_\odot$ \citep{Brinch2007, Yen2013, Yen2014, Sai2020}. Furthermore, the most recent ALMA observations have tentatively identified a ring-like structure in dust continuum emission \citep{Ohashi2022_L1489IRS}. L1489~IRS is thus one of the best sources to explore the planet formation in early evolutionary phases. 

% L1489~IRS is an ideal source to investigate the planet formation in young embedded disks as it harbors a well-established Keplerian disk and even shows a tentative ring-like structure in dust emission. However, the spatial resolution is limited to $\sim$$0\farcs3$ or $\sim$$40$--$50$\,au, which is insufficient to probe the disk substructure. Here, we present high-spatial-resolution ($\sim$$0\farcs1$ or $\sim$15\,au) observations of the dust continuum and molecular lines, \tco $J=2$--$1$, \ceo $J=2$--$1$, and SO $J_N = 6_5$--$5_4$, as part of the eDisk Large Program. 

In this paper, we present high spatial resolution ($\sim0\farcs1$) observations of dust continuum and molecular lines toward the embedded disk around L1489~IRS as part of the eDisk ALMA Large Program. This paper is structured as follows. We describe the observational details in Section \ref{sec:observation}, and the observational results (image maps and radial profiles) in Section \ref{sec:obs_results}. Section \ref{sec:analysis} presents the analysis of the dust continuum and CO isotopologue line data to constrain the physical structures of the disk. The results are discussed in \ref{sec:discussion}, and finally summarized in Section \ref{sec:summary}. 

% \item Description of planet formation in young disks, with particular focuses on ring structures observed in dust continuum
% \item Description of L1489 IRS
% \item Description of distance to L1489 IRS

\section{Observations and data reduction} \label{sec:observation}
L1489~IRS was observed as part of the eDisk Large Program (project code: 2019.1.00261.L, PI: N. Ohashi) and a dedicated Director’s Discretionary Time (DDT) program (project code: 2019.A.00034.S, PI: J. J. Tobin). Details of the observations and data reduction procedures are provided in \citet{eDisk_overview}. Here we summarize the key aspects specific to L1489~IRS. Our observations were made in a total of five executions, two of which used extended antenna configurations (C43-8) from the Large Program, and the other three used compact antenna configurations (C43-5) from the DDT program. The two executions with extended antenna configurations were conducted in 2021 August, while the three executions with compact antenna configurations in the DDT program were conducted in 2021 December and 2022 July. 
% The observational details including observing dates, number of antennas, integration times, baseline coverages and phase and flux calibrators, are listed in Table 5 of \citet{eDisk_overview}.

The eDisk program is designed to observe the dust continuum at 1.3~mm (or 225~GHz), as well as several molecular lines. The targeted molecular lines are listed in Table 2 of \citet{eDisk_overview}. In this specific first-look paper on L1489~IRS, we focused on three molecular lines: \tco $J=$~2--1, \ceo $J=$~2--1, and SO $J_N=$~6$_5$--5$_4$. In addition to these lines, several other lines ($^{12}$CO, DCN, H$_2$CO, and $c$-C$_3$H$_2$) are also detected, while SiO and CH$_3$OH lines are not detected (see Table 2 in \citet{eDisk_overview} for the complete list of the targeted transitions).
% We also show the observational results of other molecular lines in Appendix \ref{appendix:other_lines}.
% \textbf{The correlator setups include continuum spw and lines... The readers are referred to \citet{eDisk_overview} for further details.}

The initial calibration was performed by the ALMA observatory using the standard ALMA calibration pipeline version 2021.2.0.128. We then carried out self-calibration using Common Astronomy Software Applications (CASA) version 6.2.1 \citep{CASA}. 
% We basically follow the basic self-calibration procedure including a number of preprocessings detailed in \citet{eDisk_overview}. 
Prior to self-calibration, we imaged the continuum data of each execution block separately. We then found an emission peak on each image and aligned the peaks to a common phase center using the CASA task \texttt{fixvis} and \texttt{fixplanets}. For the aligned visibilities, we applied an amplitude rescaling by inspecting the azimuthally averaged visibilities (amplitude profile against baseline length) from different executions to correct for possible flux calibration uncertainties \citep[see][]{eDisk_overview}. 
% As a special note for L1489~IRS among the eDisk samples, the original phase center of the observations are significantly offset by $\sim$3$\arcsec$ from the actual source position. In this case, any shifts of phase center using \texttt{fixvis} and \texttt{fixplanets} will result in an incorrect primary beam pattern because those tasks shift the primary beam pattern as well to be centered on the shifted phase center. To compromise this issue, we used the visibilities which is not shifted but rescaled by the scaling factors estimated from shifted visbilities. While the alignment of phase center is necessary to correctly evaluate the rescaling factor at the same uv-distance, the derived rescaling factors can be directly applied to non-shifted visibilities. 

% After those preprocessing
We then performed four iterations of phase-only self-calibration and then two iterations of phase and amplitude self-calibration on the compact configuration continuum data. The data were then combined with the extended configuration data and self-calibrated together. For the combined data, we performed two iterations of phase-only self-calibration and no phase+amplitude self-calibration. The self-calibration solutions were then applied to the line data. Finally, the line data were continuum-subtracted using the CASA task \texttt{uvcontsub} in the visibility domain.

For the continuum data, we CLEANed down to 2$\times$ the RMS noise level using different Briggs robust parameters as shown in Appendix \ref{appendix:continuum_maps_different_robust}. \textrm{The \texttt{auto-multithresh} algorithm implemented in CASA was used to generate the CLEAN masks.} We adopted the image with \texttt{robust} $=1.0$ as the representative image considering the balance between spatial resolution and sensitivity. The resulting beam size and RMS noise level of the continuum image are $0\farcs105\times0\farcs078$ (PA $=12\arcdeg$) and 14\,\textmu Jy\,beam$^{-1}$, respectively (Table \ref{tab:cube_property}). We note that the beam size of the representative image is a factor of $\approx$2 larger than the typical spatial resolution of other eDisk continuum images ($\approx$0\farcs05); because the continuum emission of the L1489~IRS disk is faint, we need to trade off spatial resolution for better sensitivity. Additionally, to clearly show the extended, faint emission, we made another continuum image with \texttt{robust} $=2.0$ and \texttt{uvtaper} $=1000$\,k$\lambda$, which has a beam size and RMS noise level are 0$\farcs$221$\times$0$\farcs$179 (PA $=11\arcdeg$) and 18\,\textmu Jy\,beam$^{-1}$, respectively (Table \ref{tab:cube_property}). For the line data, we CLEANed down to 3$\times$ the RMS noise level with \texttt{robust} $=1.0$ (and \texttt{uvtaper} $=2000$\,k$\lambda$ as the default imaging parameter of eDisk line images, see \citealt{eDisk_overview}) for all lines presented in this paper. \textrm{As in the case of continuum imaging, the \texttt{auto-multithresh} algorithm implemented in CASA was used to generate the CLEAN masks.} The typical beam size and RMS noise level of the line images are 0\farcs1 and 2 mJy beam$^{-1}$ at a velocity channel width of 0.2 km\,s$^{-1}$, respectively. We assume an absolute flux calibration uncertainty of 10\%. \textrm{The maximum recoverable scale is $\sim2\farcs4$ \footnote{\textrm{Based on the Equation 7.7 in \url{https://almascience.nao.ac.jp/documents-and-tools/cycle8/alma-technical-handbook}.}}. } The detailed properties of the continuum image and line image cubes, as well as the corresponding spectroscopic data, are reported in Table \ref{tab:cube_property}\footnote{The self-calibration and imaging scripts for this source are available at \url{https://github.com/jjtobin/edisk}.}.

We note that there is an additional uncertainty in the intensity scale of the CLEANed images due to the ``JvM effect'' \citep{Jorsater1995, Czekala2021}: the inconsistency of intensity units between the CLEANed model image and the residual image makes the intensity in the resulting CLEANed image incorrect (see Figure 3 in \citealt{Czekala2021} for a comprehensive illustration). \citet{Czekala2021} introduced a method to correct for this effect (``JvM correction''). However, \textrm{while the JvM correction recovers the correct intensity scales in the restored images,} correcting the data in this way artificially manipulates the noise level \textrm{by rescaling the residual image} and may exaggerate the signal-to-noise ratio (S/N) \citep[]{Casassus2022}. \textrm{The modification of S/N could lead to the misinterpretation of the structure and extent of emission. Since we mainly focus on the morphology of the emission, we adopt the images without JvM correction. We exceptionally use the JvM-corrected image when we estimate the dust mass based on the flux density of the continuum emission in Section \ref{subsec:continuum}. Another exception is Section \ref{subsec:visfit}, where we compare the flux density estimated from the visibility analysis with that measured on the image plane, since the visibility analysis is not affected by the JvM effect.}

% Therefore, unless otherwise noted, we do not include this correction in the data used in the present work where we focus mainly on the morphology of the observed emission (see also Ohashi et al. submitted). 

\begin{deluxetable*}{cccccccc}
\label{tab:cube_property}
\tablecaption{Properties of the continuum and line images}
\tablehead{\colhead{} & \colhead{Frequency} & \colhead{${E_\mathrm{up}}^\dagger$$^\ddagger$} & \colhead{Spectral Resolution} & \colhead{Channel Width} & \colhead{Robust} & \colhead{Beam Size (PA)} & \colhead{RMS}\\ \colhead{} & \colhead{[GHz]} & \colhead{[K]} & \colhead{[km s$^{-1}$]} & \colhead{[km s$^{-1}$]} & \colhead{} & \colhead{ } & \colhead{[mJy beam$^{-1}$]}}
\startdata
continuum & 225 & --- & --- & --- & 1.0 & 0$\farcs$105$\times$0$\farcs$078 (12$\arcdeg$) & 0.014 \\
continuum (tapered) & 225 & --- & --- & --- & 2.0 & 0$\farcs$221$\times$0$\farcs$179 (11$\arcdeg$) & 0.018 \\
$^{13}$CO $J=$ 2--1 & 220.3986842$^\ddagger$ & 15.9 & 0.17 & 0.20 & 1.0 & 0$\farcs$13$\times$0$\farcs$11 (8.9$\arcdeg$) & 2.4 \\
C$^{18}$O $J=$ 2--1 & 219.5603541$^\ddagger$ & 15.8 & 0.17 & 0.20 & 1.0 & 0$\farcs$13$\times$0$\farcs$10 (13$\arcdeg$) & 1.7 \\
SO $J_N=$ $6_5$--$5_4$ & 219.9494420$^\ddagger$ & 35 & 0.17 & 0.20 & 1.0 & 0$\farcs$13$\times$0$\farcs$10 (15$\arcdeg$) & 2.1
\enddata
\tablenotetext{\dagger}{Upper state energy of the transition.}
\tablenotetext{\ddagger}{Taken from the Cologne Database for Molecular Spectroscopy \citep[CDMS;][]{CDMS1, CDMS2, CDMS3}.}
\end{deluxetable*}

\section{Observational results} \label{sec:obs_results}
\subsection{1.3 mm continuum}\label{subsec:continuum}

\begin{figure*}
\plotone{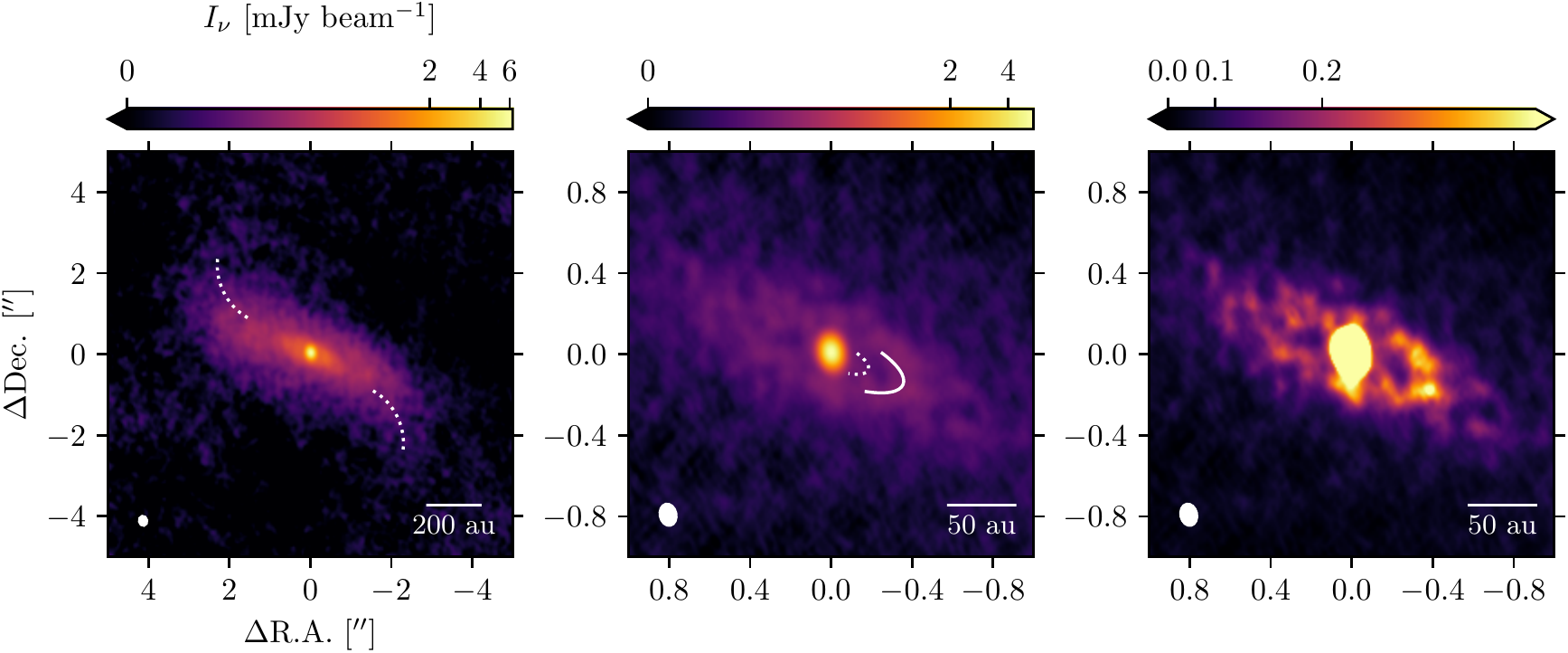}
% \epsscale{1.2}
% \plotone{continuum_gallery}
\caption{1.3\,mm continuum images of the disk around L1489 IRS. Left: Large-scale view of the disk with \texttt{robust} $=2.0$ and \texttt{uvtaper} $=1000$\,k$\lambda$. A non-axisymmetric feature appears in the outermost region as guided by the white dashed arcs. Center: Zoom-in view of the disk with \texttt{robust} $=1.0$. The locations of the ring and gap are indicated by white solid and dashed lines, respectively. The gap location is identified visually, while the ring location is based on the visibility analysis described in Section \ref{subsec:visfit}. For the left and center panels, the color scaling is stretched by arcsinh function to accentuate features with low surface brightness. Right: Same as the center panel, but with a limited color range stretched by sinh function to accentuate the non-axisymmetric feature at the ring. For each panel, the color scaling is saturated at 0.0, the beam size is represented by the white ellipse at the lower left, and the white scale bars are in the lower right.}
\label{fig:continuum}
\end{figure*}

Figure \ref{fig:continuum} shows the 1.3~mm dust continuum images of the L1489~IRS disk. The left panel of Figure \ref{fig:continuum} shows the large-scale view of the continuum emission with \texttt{robust} $=2.0$ and \texttt{uvtaper} $=1000$\,k$\lambda$, while the middle and right panels show the zoom-in view with \texttt{robust} $=1.0$. A disk-like structure elongated along the northeast to the southwest direction is detected, which is consistent with previous observations \citep{Yen2014, vantHoff2020, Sai2020, Tychoniec2021}. 
In the outermost region, non-axisymmetric, faint tails are also detected as indicated by the white dashed arcs in the left panel of Figure \ref{fig:continuum}. These tails are also seen in the continuum image in \citet{Tychoniec2021} and are consistent with the molecular line emission tracing the warped outer disk as well as the accretion flows \citep{Sai2020, Yen2014, vantHoff2020}. 

The zoom-in view of the continuum image (middle panel of Figure \ref{fig:continuum}) shows a central compact source at $\lesssim$0\farcs1, which is unresolved at the current angular resolution. The peak intensity is 5.4 mJy beam$^{-1}$, which corresponds to a brightness temperature of 21~K using the full Planck function. This component is spatially unresolved even in the higher resolution maps imaged with smaller robust parameters (Appendix \ref{appendix:continuum_maps_different_robust}), indicating that this component is quite compact ($\lesssim$$0\farcs04$).

Interestingly, an emission enhancement (i.e., a ring-like structure) is identified at a radius of $\sim0\farcs4$ from the central protostellar position is identified (middle panel of Figure \ref{fig:continuum}). The locations of the ring and the adjunct gap are marked by the white solid and dashed arcs, respectively, in the middle panel of Figure \ref{fig:continuum}. 
% \citet{Ohashi2022_L1489IRS} tentatively infer a similar ring-like structure at $\approx$$0\farcs64$ based on previous lower resolution data ($0\farcs24\times0\farcs16$). The higher resolution and better sensitivity of the present data exhibit the substructure more clearly. 
A non-axisymmetric brightness distribution (i.e., substructure in the azimuthal direction) at the ring is also observed. The right panel of Figure \ref{fig:continuum} shows the continuum image with an adjusted color scaling to clearly show the asymmetry. The western side of the ring is slightly brighter than its eastern counterpart. We will examine the properties of this ring-like structure in more detail in Section \ref{subsec:visfit} and discuss its origin in Section \ref{sec:discussion}.

To quantify the overall emission properties, we performed a two-component (i.e., central compact component and extended tenuous component) 2D Gaussian fit on the untapered continuum image using the \texttt{imfit} task in CASA. We note that due to the ring-like structure, the 2D Gaussian fit yields residuals at a level of $\approx$$8\sigma$ at the ring location. The position of the emission peak of the central compact component derived by the fit is $\alpha\mathrm{(ICRS)} = 04^\mathrm{h}04^\mathrm{m}43.080^\mathrm{s}$,  $\delta\mathrm{(ICRS)} = +26^\mathrm{d}18^\mathrm{m}56.12^\mathrm{s}$. This is used as the position of the central protostar in the following analysis. The deconvolved sizes of the compact and extended components are 0\farcs037 $\times$ 0\farcs024 (PA $=180\arcdeg$, $\sim$5.4\,au $\times$  $\sim$3.7\,au) and 3\farcs9 $\times$ 1\farcs3 (PA $=66\arcdeg$, $\sim$570\,au $\times$ $\sim$190\,au), respectively. The deconvolved size of the extended component is consistent with the previous observations \citep{Yen2014, Sai2020}. The inclination angle of the outer, extended disk is estimated from the deconvolved size to be $\sim$71$\arcdeg$ ($0\arcdeg$ for the face-on configuration) assuming a geometrically thin disk. The derived inclination angle is similar to the value measured by previous observations \citep[73$\arcdeg$;][]{Sai2020}. The total flux density originating from both components is 91 $\pm$ 9 mJy. The flux calibration uncertainty of 10\% is added in quadrature. The measured flux density is $\sim$53\% larger than that measured by \citet{Sai2020} ($\sim59$ mJy), which has been found to be due to the JvM effect \citep[][see also Section \ref{sec:observation}]{Jorsater1995, Czekala2021}.
% : the inconsistency of intensity units between the CLEANed model image and the residual image makes the flux density in the resulting CLEANed image incorrect (see Figure 3 in \citealt{Czekala2021} for a comprehensive illustration). To correct for this effect, we followed the methodology described in \citet[][``JvM'' correction]{Czekala2021} and 
Applying the JvM correction \citep{Czekala2021} resulted in \textrm{a JvM $\epsilon$ (the ratio of the CLEAN beam volume to the dirty beam volume) of $\sim0.26$ and} a similar flux density ($\sim$50\,mJy) to that derived by \citet{Sai2020}. 
% However, correcting the data in the manner proposed by \citet{Czekala2021} artificially manipulates the noise level and may exaggerate the S/N for point source emission \citep{Casassus2022}. Consequently, unless otherwise stated, we do not include this correction in the data used in the present work where we mainly focus on the morphology of the observed emission (see also Ohashi et al. submitted). 

Assuming that the continuum emission at $225\,\mathrm{GHz}$ is purely from the thermal dust emission and optically thin, the total dust mass can be estimated by
\begin{equation}
    M_\mathrm{dust} = \frac{F_\nu d^2}{\kappa_\nu B_\nu(T)},
\end{equation}
where $F_\nu$ is the flux density, $d=146$\,pc is the distance to the source, $\kappa_\nu$ is the dust mass opacity, and $B_\nu(T)$ is the Planck function of the blackbody radiation at the temperature of $T$. We adopted the flux density of $50$\,mJy from the JvM-corrected image, and the dust mass opacity of $\kappa_\nu = 2.3\,\,\mathrm{cm^2\,g^{-1}}$ based on \citet{Beckwith1990} with an assumption of the dust-to-gas mass ratio of 0.01. Adopting dust temperatures of $T = 20\,\mathrm{K}$, a typical value used in Class~II surveys \citep[e.g.,][]{Andrews2005, Ansdell2016, Tobin2020}, and $T = 58\,\mathrm{K}$ based on the prescription in \citet{Tobin2020},
\begin{equation}
    T = 43 \left(\frac{L_\mathrm{bol}}{1\,L_\odot}\right)^{1/4}\,\mathrm{K},
\end{equation}
with the bolometric luminosity of $L_\mathrm{bol} = 3.4\,L_\odot$ \citep{eDisk_overview}, we obtain disk dust masses of $31\,M_\oplus$ and $8.9\,M_\oplus$, respectively. We note that these estimates are lower limits given that the emission at the central compact component may be optically thick.

% These fundamental properties of continuum emission are reported in Table 4 in \citet{eDisk_overview} for comparison among the 19 eDisk sources. More detailed analysis on the visibility plane to derive the disk properties including substructures is presented in Section \ref{subsec:visfit}.

% Figure \ref{fig:continuum_radial_profile} shows the radial profiles of the continuum emission, a comprehensive 1D representation of radial substructures. The details of the generation and selection of the profiles are described in Appendix \ref{appendix:generation_of_profiles}. The ring-gap pair at $\sim$0\farcs2 and $\sim$0\farcs4 is clearly identified. In Figure \ref{fig:continuum_radial_profile}, we also show the profiles which are extracted from the east or west side only. Those profiles are overall consistent with the profile extracted from both side except at the ring and gap location: the east side presents lower intensity than the west side as indicated in the emission map (Figure \ref{fig:continuum}(c)) as well.

% \begin{figure}
% \plotone{figures/continuum_radial_profile.pdf}
% \caption{Radial profiles of continuum emission, extracted from the east (grey, dashed), west (grey, dotted), and both (black, solid) side of the disk. The continuum emission is averaged over a $\pm45\arcdeg$ wedge with respect to the disk major axis (see Appendix \ref{appendix:generation_of_profiles}). The color-shaded regions represent the 1$\sigma$ uncertainty of profiles. The vertical lines indicate the location of the ring and gap.}
% \label{fig:continuum_radial_profile}
% \end{figure}

\subsection{Molecular lines}\label{subsec:obs_results_line}
Figures \ref{fig:large_scale_maps} and \ref{fig:zoom_maps} show the molecular line maps of the velocity-integrated intensity (or ``zeroth moment''), velocity centroid, and peak brightness temperature with large-scale and zoom-in views, respectively. 
% The same set of maps for C$^{18}$O ($J=$ 2--1) and SO ($J_N=$ 6$_5$--5$_4$) are presented in Figure \ref{fig:C18O_maps} and Figure \ref{fig:SO_maps}, respectively.
% We present the velocity-integrated intensity (or ``zeroth moment'') maps, velocity centroid maps, and peak brightness temperature maps of $^{13}$CO ($J=$ 2--1) (Figure \ref{fig:13CO_maps}), C$^{18}$O ($J=$ 2--1) (Figure \ref{fig:C18O_maps}), and SO ($J_N=$ 6$_5$--5$_4$) (Figure \ref{fig:SO_maps}).
All the maps are generated using \texttt{bettermoments} \citep{bettermoments}. For the velocity-integrated intensity maps, we integrate only the emission above 2$\sigma$. For the velocity centroid maps and peak brightness temperature maps, we use the quadratic method implemented in \texttt{bettermoments} rather than traditional first/eighth moments.
% \footnote{The molecular line emission may also be affected by the JvM effect, and the values of the velocity-integrated intensity maps and the peak brightness temperature maps may thus have an uncertainty due to this effect. However, since we do not show any quantitative analysis based on the intensity of the emission, the results in the following sections are still hold.}. 
% We note that as the S/N of the molecular line emission is not so high that the detailed analysis on the channel maps is possible, we infer the emission morphologies based on the moment maps.
% The detailed method to generate these maps are described in Appendix \ref{appendix:map_generation_method}. The full channel maps and properties of those image cubes are presented in Appendix \ref{} and Table \ref{tab:cube_property}. \textbf{OTHER LINES SHOULD GO TO APPENDIX?}

\subsubsection{\texorpdfstring{$^{13}$CO}{}}
The maps of the $^{13}$CO $J=$ 2--1 emission are shown in Figures \ref{fig:large_scale_maps} and \ref{fig:zoom_maps} (top rows). The emission is detected above 3$\sigma$ in a velocity range from $-$0.2 km s$^{-1}$ to 15.6 km s$^{-1}$, where the systemic velocity is 7.38 km s$^{-1}$ as derived by the velocity structure analysis (Section \ref{subsec:SLAMfit}). The velocity-integrated intensity map with a large-scale view (top-left panel of Figure \ref{fig:large_scale_maps}) shows an elongated morphology along the major axis of the dust disk at $r\lesssim1\farcs5$, and the velocity centroid map (top-middle panel of Figure \ref{fig:large_scale_maps}) shows a rotation signature of the disk. In the outer region ($r\gtrsim1\farcs5$), the \tco emission shows an extended structure, suggesting that the \tco emission traces the envelope component.
% $^{13}$CO ($J=$ 2--1) emission also partially traces the infalling materials in the protostellar envelope, as presented in the map with large-scale view (Figure \ref{} in Appendix \ref{}). 
\citet{Sai2020} also observed the same transition and found a similar emission morphology (see Appendix C in their work). 

The velocity-integrated intensity map with a zoom-in view (top-left panel of Figure \ref{fig:zoom_maps}) shows a bright, double-peaked \tco emission at a spatial scale of $r\lesssim0\farcs5$. Interestingly, the peaks of the \tco emission are slightly shifted southward from the disk major axis. This would be evidence that the \tco emission is originating from the surface of the inclined disk. The brighter emission on the southern side indicates that the southern side is the far side of the disk, from which the emission of the warm disk surface reaches the observer directly \citep[e.g.,][]{Ruiz-Rodriguez2017, Lee2017, Villenave2020, eDisk_IRS7B_model}. This configuration of the disk is consistent with the outflow directions, where the red- and blue-shifted robes are located in the northern and southern sides, respectively \citep{Yen2014}.

The top-right panels of Figure \ref{fig:large_scale_maps} and \ref{fig:zoom_maps} show the peak brightness temperature maps. The peak brightness temperature reaches $\gtrsim$ 60~K at $r\lesssim0\farcs5$, again suggesting that the \tco emission traces the disk surface where the temperature is higher than the midplane due to the heating by the protostar or the accretion shock. The map also shows the higher brightness temperature on the southwestern (or redshifted) side, which may indicate a non-axisymmetric temperature structure of the disk. The brighter molecular line emission on the southwestern (or redshifted) side of the disk has also been previously observed in C$^{17}$O and H$_2$CO emission \citep{vantHoff2020}.

\subsubsection{\texorpdfstring{C$^{18}$O}{}}\label{subsubsec:C18O}
The maps of the C$^{18}$O $J=$ 2--1 emission are shown in Figures \ref{fig:large_scale_maps} and \ref{fig:zoom_maps} (middle rows). The emission is detected above 3$\sigma$ in a velocity range from 0.8 km s$^{-1}$ to 12.8 km\,s$^{-1}$. The velocity-integrated intensity map (middle-left panel of Figure \ref{fig:zoom_maps}) shows a lack of the \ceo emission in the innermost region ($r\lesssim0\farcs2$). In the intermediate region ($0\farcs2\lesssim r\lesssim 1\farcs5$), an elongated disk-like emission exists. While the \ceo emission shows the double-peaked morphology at $r\approx0\farcs5$ similar to that of \tco, no significant shifts of the peaks from the major axis of the disk are seen, in contrast to the velocity-integrated intensity map of \tco (the top-left panel of Figure \ref{fig:zoom_maps}). This indicates that the \ceo emission mainly traces the gas near the disk midplane. Also, the radii of the \ceo peaks are slightly larger than those of the \tco peaks, which is more clearly presented in the radial intensity profiles in Section \ref{subsec:radial_profiles}.

In the outer region ($1\farcs5 \lesssim r\lesssim 4\arcsec$), a warped structure is seen, where the redshifted part is curved toward southwest and the blueshifted part toward northeast (Figure \ref{fig:large_scale_maps}). This feature was first found by \citet{Sai2020}, where they interpreted it as a misaligned Keplerian disk based on the velocity structure analysis. Furthermore, an extended redshifted emission is present in the outermost region ($r\gtrsim4\arcsec$) of the southwestern (redshifted) side (see the middle-left panel of Figure \ref{fig:large_scale_maps}), consistent with the infalling accretion flow observed in \citet{Yen2014}. The blushifted conuterpart of this component is also seen as a slight extension to the north on the northeastern side.
% The velocity-integrated intensity map (the middle-left panel of Figure \ref{fig:large_scale_maps}) shows an elongated disk-like structure similar to that seen in \tco.
% C$^{18}$O also partially traces the infalling materials as indicated by the large-scale map (Figure \ref{} in Appendix \ref{}). 
% The overall morphology of the emission, including the outer, warped disk, is consistent with the previous observations by \citet{Sai2020}. 
% In contrast to $^{13}$CO, C$^{18}$O does not show the ``V''-shaped morphology, but show the elliptical disk structure, indicating the C$^{18}$O emission mainly traces the gas in the disk midplane. 
% While the C$^{18}$O emission shows the double-peaked morphology at $r\approx0\farcs5$ similar to that of \tco, no significant shifts of the peaks from the disk major axis are seen, unlike the velocity-integrated intensity map of \tco (Figure \ref{fig:zoom_maps}). This indicates that the \ceo emission mainly traces the gas near the disk midplane.
% which is consistent with the previous observations by \citet{Sai2020}. 
% Also, the radii of the \ceo peaks are slightly larger than the \tco peak, which is more clearly presented in the radial intensity profiles in Section \ref{subsec:radial_profiles}.

The centroid velocity map (central panel of Figure \ref{fig:zoom_maps}) shows the highest velocity ($\pm$3--5 km\,s$^{-1}$ with respect to the systemic velocity) along the major axis of the disk, suggesting that the high-velocity component of the \ceo emission traces the gas in the rotating disk. A similar velocity structure was observed by \citet{Sai2020}. In the innermost region ($\lesssim$0\farcs2), the map lacks the highest velocity component, which is due to the absence of the disk emission (possibly caused by the actual gas absence and/or the continuum over-subtraction; see Section \ref{subsec:discussion_dust_ring}), as seen in the velocity-integrated intensity map (middle-left panel of Figure \ref{fig:zoom_maps}). 
% The lower velocity component traces the emission from the outer disk and envelope.
% The C$^{18}$O emission also presents a redshifted protrusion in the SE side of the disk, indicative of the asymmetric infall motion as also suggested by $^{13}$CO emission (Figure \ref{fig:C18O_maps}(b)) as well as previous observations \citep{Yen2014, Sai2020}. 

The peak brightness temperature map (the middle-right panel of Figure \ref{fig:zoom_maps}) also suggests that the \ceo emission originates from the disk midplane, as the peak emission components are located close to the disk major axis. Similar to the velocity-integrated intensity map (the middle-left panel of Figure \ref{fig:zoom_maps}), the peak brightness temperature also shows a central dip at $r \lesssim 0\farcs4$ and a ring-like structure with a radial peak at $r\sim0\farcs4$. In addition, the southwestern (or redshifted) side of the disk is slightly brighter than the northeastern side, consistent with the distributions of the \tco peak brightness temperature structure.
% The peak brightness temperature is thus used as the lower limit of the physical temperature in the disk midplane (assuming the C$^{18}$O emission is not completely optically thick). We also identify the local decrease and enhancement of the peak brightness temperature at $\sim$1$\arcsec$ and $\sim$2$\arcsec$ SE, respectively, which are also tentatively present in the velocity-integrated intensity map (Figure \ref{fig:C18O_maps}(a)) and more clearly identified in the radial intensity profiles (see Section \ref{subsec:radial_profiles}). These features are consistent with the C$^{18}$O dip observed in \citet{Sai2020}.

% \begin{figure*}
% \plotone{figures/C18O_maps_v3.pdf}
% \caption{Same as Figure \ref{fig:13CO_maps}, but for C$^{18}$O ($J=$ 2--1). The contours in the bottom-left panel start from $10\sigma$, followed by a step of $20\sigma$, where $\sigma=0.57$\,mJy\,beam$^{-1}$\,km\,s$^{-1}$.}
% \label{fig:C18O_maps}
% \end{figure*}

\subsubsection{SO}\label{subsubsec:SO}
\begin{figure*}
\plotone{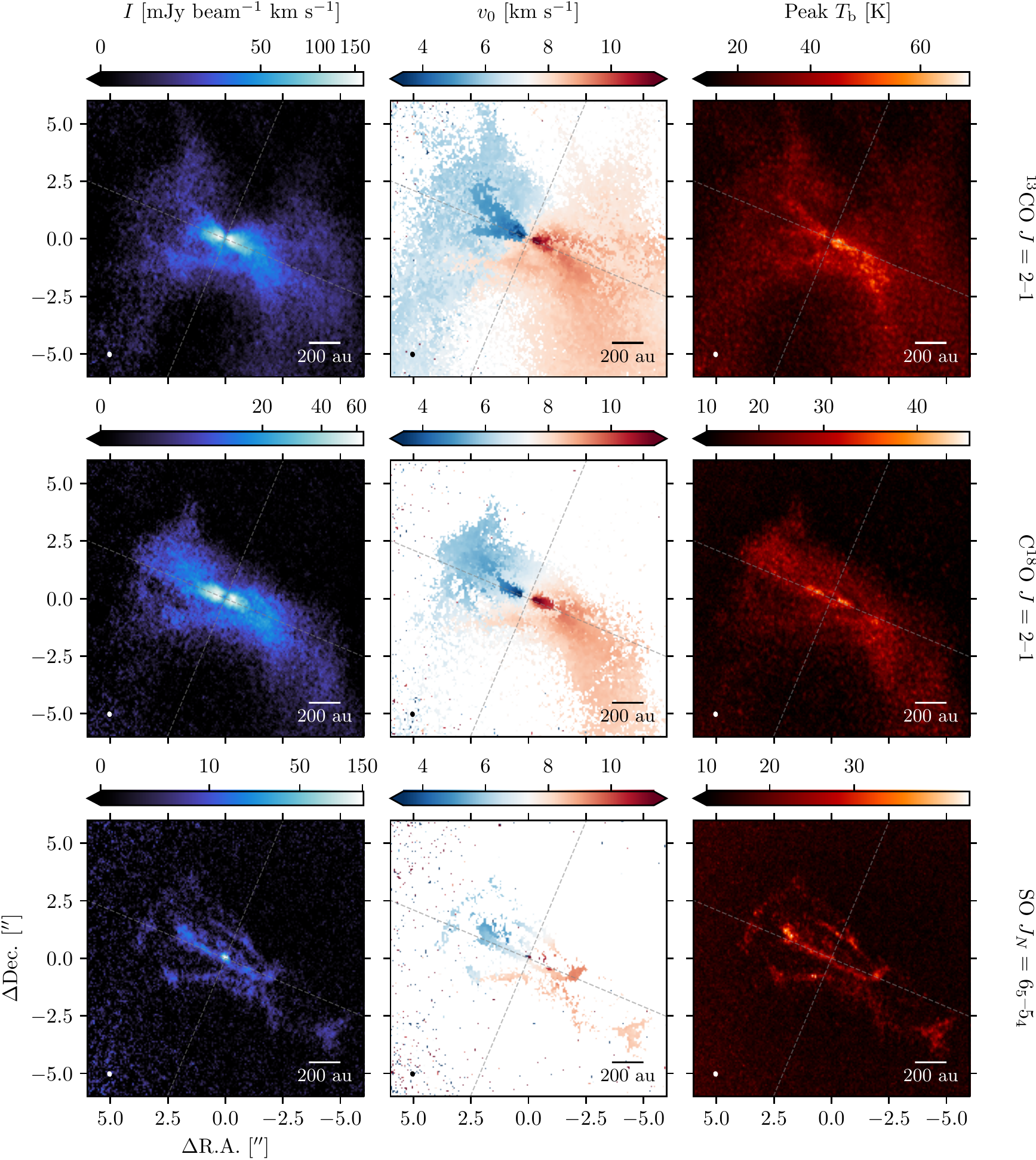}
\caption{Velocity-integrated intensity maps (left column), velocity centroid maps (center column), and peak brightness temperature maps (right column) of \tco $J=$ 2--1 (top row), \ceo $J=$ 2--1 (center row), and SO $J_N=$ 6$_5$--5$_4$ (bottom row). The velocity centroid and peak brightness temperature maps are generated by the quadratic method implemented in \texttt{bettermoments}. For the velocity-integrated intensity maps, only emission larger than 2$\sigma$ has been integrated. The color scaling is stretched by arcsinh function and saturated at 0.0. For the peak brightness temperature maps, the conversion from intensity to brightness temperature has been done using the full Planck function. The color scaling is stretched by sinh function and saturated at 10~K for visual clarity. For all panels, the synthesized beam is shown in the bottom left, while the scale bar in the bottom right indicates 200~au. The gray dashed lines indicate the direction of the major and minor axis of the dust disk (PA $=67\arcdeg$ as derived by the visibility analysis described in Section \ref{subsec:visfit}).}
\label{fig:large_scale_maps}
\end{figure*}

\begin{figure*}
\plotone{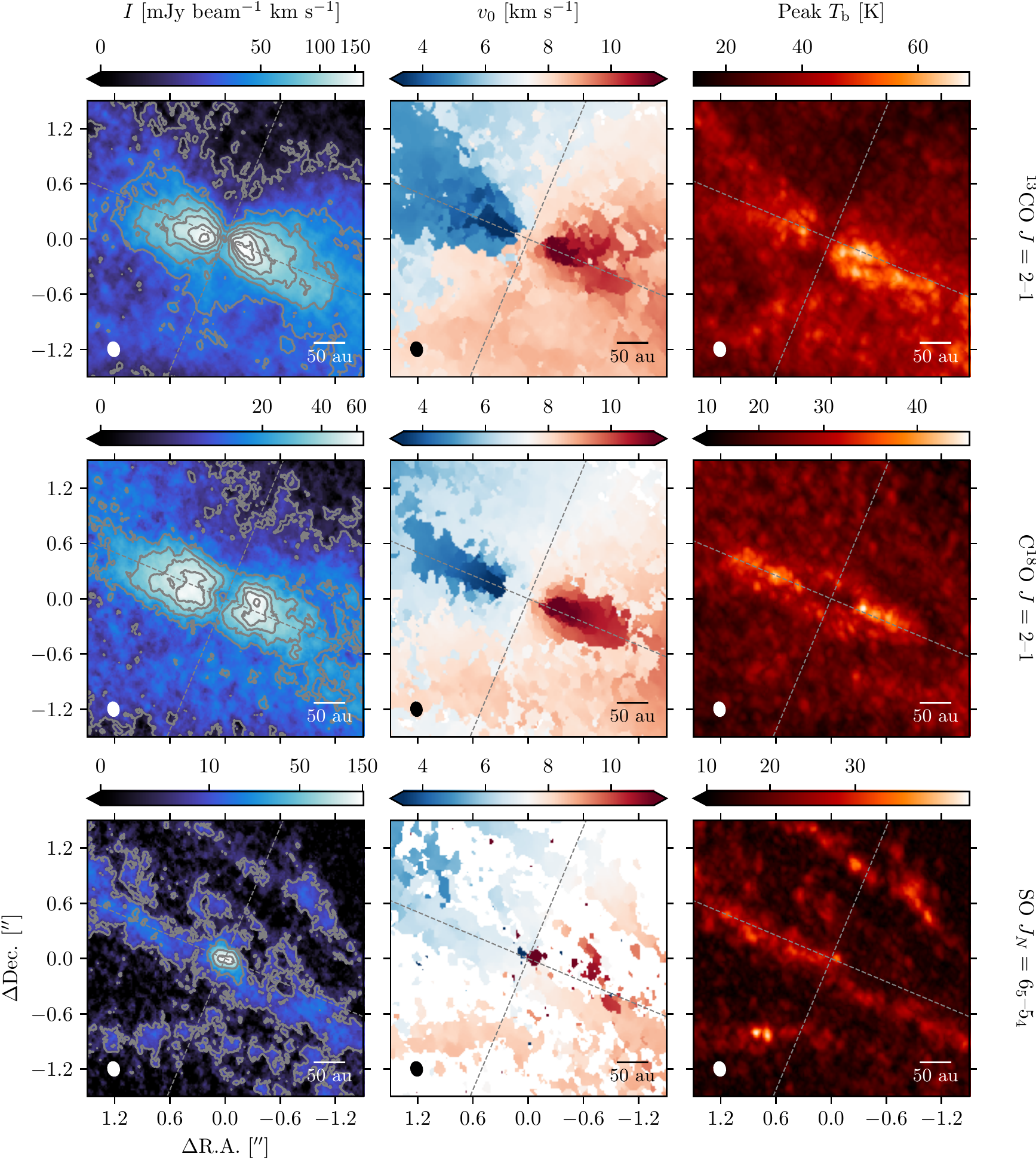}
\caption{Same as Figure \ref{fig:large_scale_maps}, but with zoom-in views. The contours in the velocity-integrated intensity maps (left column) are [10, 30, 50, 70, 90, 110, 130]$\sigma$ ($\sigma = 1.2$~mJy~beam$^{-1}$~km~s$^{-1}$), [10, 30, 50, 70, 90, 110]$\sigma$ ($\sigma = 0.57$~mJy~beam$^{-1}$~km~s$^{-1}$), and [10, 30, 100, 200]$\sigma$ ($\sigma = 0.61$~mJy~beam$^{-1}$~km~s$^{-1}$) for \tco, \ceo, and SO, respectively.} 
\label{fig:zoom_maps}
\end{figure*}

The maps of the SO ($J_N=$ 6$_5$--5$_4$) emission are shown in  Figure \ref{fig:large_scale_maps} and \ref{fig:zoom_maps} (bottom rows). The emission is detected above 3$\sigma$ over a wide velocity range from $-$3.8 km\,s$^{-1}$ to 19.2\,km\,s$^{-1}$. Overall, the morphology of the SO emission is different from that of \tco and \ceo. The velocity-integrated intensity map with the zoom-in view (bottom-left panel of Figure \ref{fig:zoom_maps}) exhibits a prominent compact emission in the innermost region ($r\lesssim0\farcs2$) and the diffuse extended emission in the outer disk, while the diffuse SO emission lacks the high-velocity component at $r \approx 0\farcs2$--$1\arcsec$ seen in \tco and \ceo (see the middle column of Figure \ref{fig:zoom_maps}). The central compact emission is marginally spatially resolved and shows an elongated structure along the east-west direction. Interestingly, this emission component shows a velocity gradient along the elongation direction over the extremely high velocities of $\pm$10--12 km s$^{-1}$ with respect to the systemic velocity (bottom-middle panel of Figure \ref{fig:zoom_maps}). The origin of this emission is discussed in detail in Section \ref{subsec:discussion_SO_emission}. The outer diffuse emission shows a complex structure where the emission is mainly distributed along the major axis of the disk as well as in the regions relocated from the axis (bottom-left panel of Figure \ref{fig:large_scale_maps}). It also shows the velocity gradient along a direction similar to the major axis of the disk, partially tracing the rotating disk (bottom-middle panel of Figure \ref{fig:large_scale_maps}).
% , which indicates that the SO emission is not originating from the disk. 
In addition, an SO emission clump is detected on the southwestern side of the disk ($r\approx6\arcsec$), consistent with previous observations \citep{Yen2014}.
% The redshifted protrusion observed in $^{13}$CO and C$^{18}$O is also identified. 

The peak brightness temperature map of the SO emission (the bottom-right panels of Figures \ref{fig:large_scale_maps} and \ref{fig:zoom_maps}) is highly intriguing. The multiple local enhancements of the peak brightness temperature are identified in the outer region. They may indicate the local enhancement of the gas temperature due to the accretion shock. On the other hand, the central compact emission shows lower peak brightness temperatures. This will be discussed in more detail in Section \ref{subsec:discussion_SO_emission}.

% \begin{figure*}
% \plotone{figures/SO_maps_v3.pdf}
% \caption{Same as Figure \ref{fig:13CO_maps}, but for SO ($J_N=$ 6$_5$--5$_4$). The contours in the bottom-left panel corresponds to [10, 30, 100, 200]$\sigma$, where $\sigma=0.61$\,mJy\,beam$^{-1}$ km\,s$^{-1}$.}
% \label{fig:SO_maps}
% \end{figure*}

\subsection{Radial intensity profiles}\label{subsec:radial_profiles}
\begin{figure*}
% \centering
\plotone{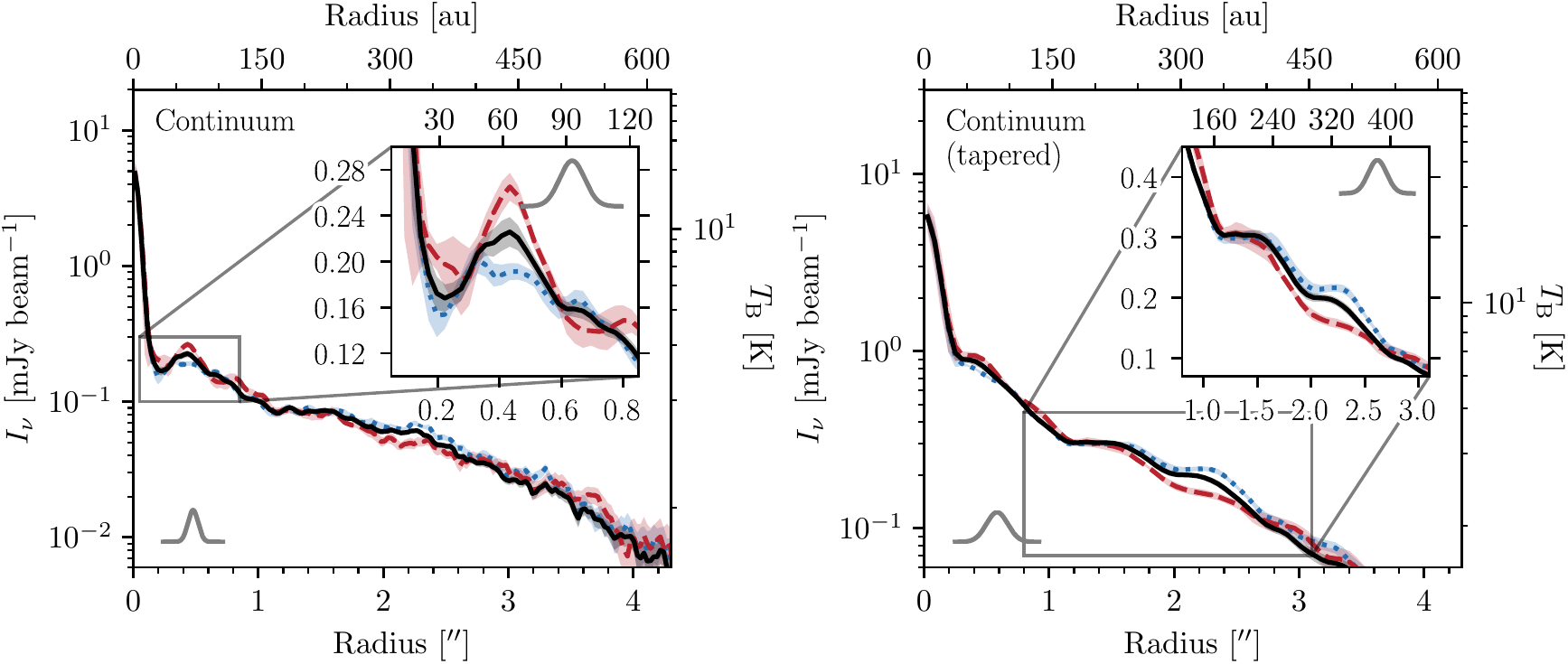}
\caption{Deprojected, azimuthally averaged radial intensity profiles of the 1.3\,mm continuum. The left panel presents the radial profile of the image with \texttt{robust} $=1.0$, while the right panel shows the radial profile of the \texttt{robust} $=2.0$ tapered image. The dashed red and dotted blue lines represent the profiles extracted from the southwestern (redshifted) side and the northeastern (blueshifted) side of the disk, respectively. The solid black line shows the profile averaged over both sides of the disk. Those profiles are computed by averaging over $\pm45\arcdeg$ wedges with respect to the disk major axis (PA $=67\arcdeg$). The shaded regions indicate 1$\sigma$ scatter at each radial bin (see Appendix \ref{appendix:generation_of_profiles}). In the inset of each panel, the zoom-in views of the innermost ring (left panel) and outer shoulders (right panel) are shown, respectively. The Gaussian profile in the lower left corner of each panel and the upper right corner of the inset indicates the width of the major axis of the synthesized beam. The right vertical axis indicates the brightness temperature, converted from the intensity using the full Planck function.}
\label{fig:radial_profiles_cont}
\end{figure*}

\begin{figure*}
% \centering
\plotone{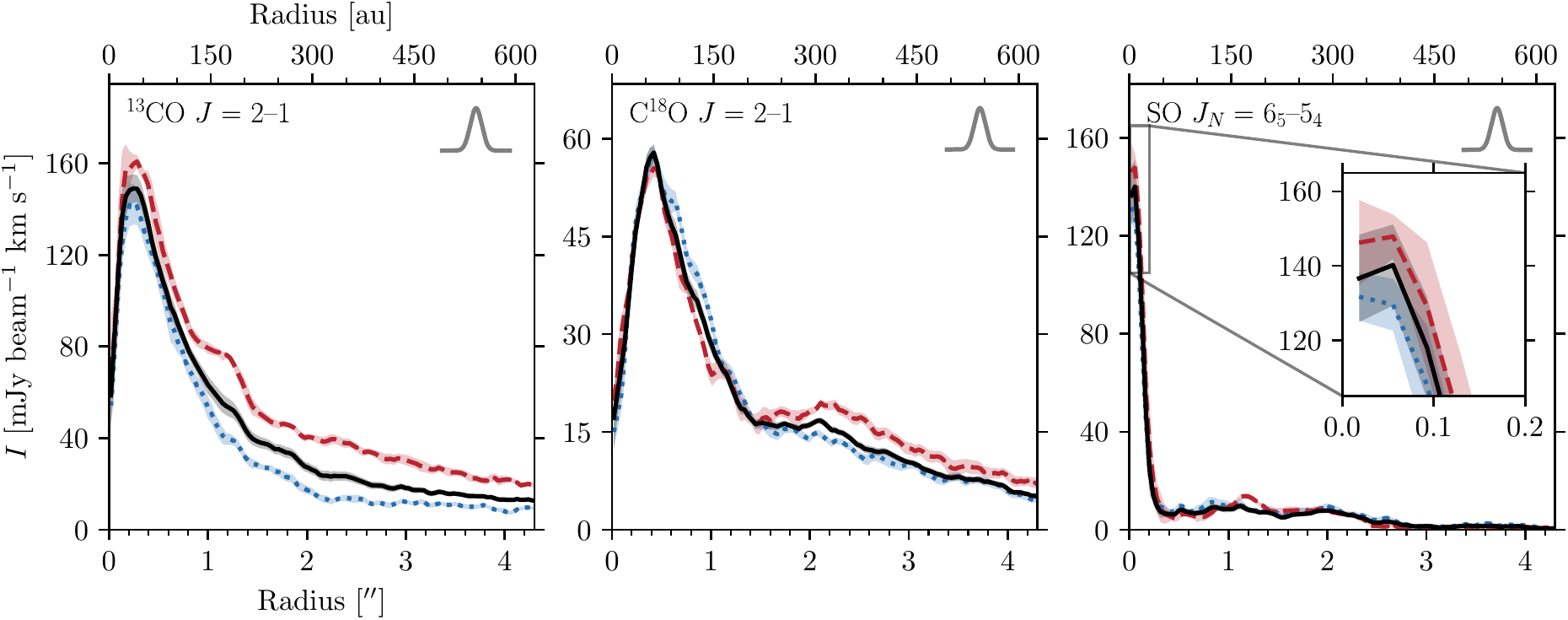}
\caption{Deprojected, azimuthally averaged radial intensity profiles of $^{13}$CO $J=$ 2--1 (left), C$^{18}$O $J=$ 2--1 (middle), and SO $J_N=6_5$--$5_4$ (right). While the dashed red and dotted blue lines represent the profiles extracted from the southwestern (redshifted) side and the northeastern (blueshifted) side of the disk, respectively, the solid black line shows the profile extracted from both sides of the disk. Those profiles are computed by averaging over $\pm45\arcdeg$ wedges with respect to the disk major axis (PA $=67\arcdeg$). The shaded regions indicate 1$\sigma$ scatter at each radial bin (see Appendix \ref{appendix:generation_of_profiles}). The Gaussian profile in the upper right corner of each panel indicates the width of the major axis of the synthesized beam. The inset of the right panel shows the slight cetral depression of the SO emission.}
\label{fig:radial_profiles_line}
\end{figure*}

% show and explain radial intensity profiles of continuum and lines
% We present the radial intensity profiles of continuum (Figure \ref{fig:radial_profiles_cont}) and molecular lines of $^{13}$CO ($J=$ 2--1), C$^{18}$O ($J=$ 2--1), and SO ($J_N=$ 6$_5$--5$_4$) (Figure \ref{fig:radial_profiles_line}), 1D comprehensive representations of radial substructures.
Figures \ref{fig:radial_profiles_cont} and \ref{fig:radial_profiles_line} show the radial intensity profiles of the continuum and molecular lines, respectively. The methodology used to generate these radial profiles is described in detail in Appendix \ref{appendix:generation_of_profiles}. Briefly, we averaged the emission over certain radial bins within the limited wedge of $\pm45\arcdeg$ with respect to the disk major axis. In addition to the profiles averaged over both sides of the disk, we calculated the profiles extracted from each southwestern (redshifted) side and northeastern (blueshifted) side of the disk to show the azimuthal variations. The radial and azimuthal substructures for the continuum and each line emission are described below. A more detailed comparison of the radial profiles will be made in Section \ref{sec:discussion}.

\textit{Continuum} --- Figure \ref{fig:radial_profiles_cont} shows that the continuum emission consists of the centrally peaked compact component and the tenuous extended disk emission. In addition to these components, a gap-ring pair is identified at $\sim$0\farcs2 (or 30\,au) and $\sim$0\farcs4 (or 60\,au) radii from the disk center. The profile extracted from the southwestern side is brighter at the ring position than that extracted from the northeastern side by 0.08 mJy~beam$^{-1}$ (i.e., $\approx10\sigma$ significance), which can be seen more clearly in the inset panel of the left panel of Figure \ref{fig:radial_profiles_cont}. In the radial profile of the \texttt{robust} $=2.0$ tapered image (the right panel of Figure \ref{fig:radial_profiles_cont}), two shoulder-like structures are tentatively identified at $\approx$1\farcs5 (or 220\,au) and $\approx$2\farcs3 (or 340\,au). The profiles also show slight variations between the southwestern side and the northeastern side at those shoulder-like substructures. The northeastern side is brighter than the southwestern side.

\textit{$^{13}$CO $J=$ 2--1} --- The radial intensity profile extracted from both sides peaks at $\sim$0\farcs2 (or 30\,au) radii. Additionally, a subtle shoulder is identified at $\sim$1\farcs3 (or 190\,au) radii. This feature is also seen in the profiles extracted from the southwestern and northeastern sides of the disk, although the feature in the northeastern side is subtle. Overall, the southwestern side is brighter than the northeastern side, while the difference is more significant in the outer radii ($\gtrsim0\farcs8$).   

\textit{C$^{18}$O $J=$ 2--1} --- The peak of the C$^{18}$O intensity profile is slightly shifted outward ($\approx$0\farcs4 or 60\,au) compared to the peak of $^{13}$CO. From this peak, the \ceo profile shows a steep decline until $r\approx$1\farcs4 (or 200\,au). At $r\approx$2\farcs2, the \ceo intensity profile from the southwestern side shows a slight enhancement, which is brighter than the emission on the northeastern side. While the \ceo profiles show the difference between the southwestern and northeastern sides of the disk in the outer region ($r\gtrsim1\farcs5$ or 220\,au), the profiles are consistent in the inner disk ($r\lesssim1\farcs5$). 
% This feature corresponds to the local emission enhancement at $\sim$2$\arcsec$ SW in the maps (Figure \ref{fig:C18O_maps}), which is also observed in \citet{Sai2020}. 
% \textbf{On the other hand, the dip in the east side at $\sim$2$\arcsec$ seen in \citet{Sai2020} is not identified in our high-resolution data. NEED TO CHECK TAPERED MAP?}

\textit{SO $J_N=$ 6$_5$--5$_4$} --- The SO emission is centrally peaked at $r\lesssim$ 0\farcs2, although the innermost bin shows a slight central depression (the inset of the right panel of Figure \ref{fig:radial_profiles_line}). The extent of the central compact component is up to $\approx$0\farcs2. In addition to the central compact component, there is a tenuous diffuse emission as also seen in the maps (Figure \ref{fig:large_scale_maps}). There is no major variation between the southwestern and northeastern sides, although the 2D emission distributions are complex (Figure \ref{fig:large_scale_maps}).  

% \begin{figure*}
% \centering
% \includegraphics[width=\hsize]{figures/radial_emission_profile.pdf}
% \caption{Deprojected, azimuthally averaged radial intensity profiles of continuum, $^{13}$CO ($J=$ 2--1), C$^{18}$O ($J=$ 2--1), and SO ($J_N=6_5$--$5_4$). While for the continuum profile (the leftmost panel) we show it in logarithmic scale, we show the line profile in linear scale. While the dotted and dashed grey lines represent the profiles extracted from west and east side of the disk, respectively, the solid black line shows the profile extracted from the both (west and east) side of the disk. Those profiles are computed by averaging over $\pm45\arcdeg$ wedges with respect to the disk major axis (PA $=67.2\arcdeg$). The shaded regions indicate 1$\sigma$ uncertainty at each radial bin (see Appendix \ref{appendix:generation_of_profiles}). For continuum profile (leftmost panel), we show the zoom-in view of the ring-gap pair at $\sim$0\farcs2--0\farcs4 radii in linear scale in the inset axis for visual clarity. The Gaussian profile in the lower left corner (continuum panel) or upper right corner (continuum inset axis and line panels) indicate the width of the major axis of the synthesized beam.}
% \label{fig:radial_profiles}
% \end{figure*}

% \begin{figure*}
% \epsscale{1.2}
% \plotone{map_gallery_large}
% \end{figure*}

% \begin{figure*}
% \epsscale{1.2}
% \plotone{map_gallery_zoom}
% \end{figure*}

\section{Analysis}\label{sec:analysis}
We performed two simple modeling approaches to the observed continuum and line emission: (1) an analytical model fit to the observed continuum emission in the visibility domain to characterize the radial substructures and geometry of the dusty disk (Section \ref{subsec:visfit}) and (2) a rotation curve fit to the position-velocity (PV) diagram of the \ceo emission to investigate the nature of the disk rotation (Section \ref{subsec:SLAMfit}).

\subsection{Visibility analysis of the dust continuum emission}\label{subsec:visfit}
To precisely characterize the dust disk geometry and morphology, we conduct an analytic model fit to the observed dust continuum data in the visibility domain. Visibility analysis has two advantages against the image analysis. First, the uncertainties associated with imaging can be avoided. Second, visibility can probe the structures on scales smaller than the beam size of the image, although the smallest scale that can be probed is limited by the maximum baseline length. We only consider the axisymmetric component and the innermost dust ring, given that the non-axisymmetric components and outer shoulder-like structures are rather faint. We use a Gaussian ring model with an intensity distribution given by 
\begin{equation}
    I_{\nu, \mathrm{r}}(r) = I_\mathrm{\nu, r}\exp\left(- \frac{(r - r_{0, \mathrm{r}})^2}{2w_\mathrm{r}^2}\right),
\end{equation}
where $I_{\nu, \mathrm{r}}$ is the peak intensity of the ring component, $r_{0, \mathrm{r}}$ is the radius of the ring center, and $w_\mathrm{r}$ is the width of the ring. The radial coordinate $r$ is the disk coordinate deprojected with the position angle (PA) and the inclination angle ($i$).
% The disk coordinate projected onto the sky is defined as
% \begin{align}
%     x^\prime &= x\sin\mathrm{PA} + y\cos\mathrm{PA}, \\
%     y^\prime &= x\cos\mathrm{PA} - y\sin\mathrm{PA}, \\
%     r &= \sqrt{x^{\prime 2} + \left(\displaystyle \frac{y^\prime}{\cos i}\right)^2}, \\
%     \phi &= \arctan\frac{y^\prime}{x^\prime \cos i}
% \end{align}
% where $(x, y)$ is the offset from the phase center of the observation, PA is the position angle measured from north to east, and $i$ is the inclination angle of the disk (0$\arcdeg$ for face-on). 
The central compact component is given by
\begin{equation}
    I_{\nu, \mathrm{c}}(r) = I_\mathrm{\nu, c}\exp\left(- \frac{r^2}{2w_\mathrm{c}^2}\right).
\end{equation}
We add the extended disk component as a Gaussian and it is similarly given by 
\begin{equation}
    I_{\nu, \mathrm{e}}(r) = I_\mathrm{\nu, e}\exp\left(- \frac{r^2}{2w_\mathrm{e}^2}\right).
\end{equation}
The full model is then given by the sum of all the components,  
\begin{equation}
    I_\nu(r) = I_{\nu, \mathrm{r}}(r) + I_{\nu, \mathrm{c}}(r) + I_{\nu, \mathrm{e}}(r).
\end{equation}

We directly fit this model to the observed visibilities. We used the flux density ($F_{\nu, *}$) instead of the peak intensity ($I_{\nu, *}$) for the parameters in practice. First, the model image is generated in the image plane, and then Fourier-transformed by the \texttt{GALARIO} code \citep{GALARIO} onto the observed sampling of the $(u, v)$-plane. We also consider the disk center offset from the phase center of the observations, resulting in additional parameters of $(x_0, y_0)$. In total, we consider 11 parameters $\hat{\theta} = \{F_{\nu, \mathrm{r}}, w_\mathrm{r}, r_{0, \mathrm{r}}, F_{\nu, \mathrm{c}}, w_\mathrm{c}, F_{\nu, \mathrm{e}}, w_\mathrm{e}, \mathrm{PA}, i, x_0, y_0\}$, and search the parameter space with the Markov Chain Monte Carlo (MCMC) method using the \texttt{emcee} package \citep{emcee}. 
% The adopted bound for uniform priors and initial position of walkers are listed in table \ref{}. We determine them from visual inspection on the image except for initial positions of PA and $i$, where we used the nominal value in the literature \citep[69\arcdeg and 73\arcdeg, respectively;][]{Sai2020}. 
The parameter space is sampled by 200 walkers with 1000 steps, and the initial 700 steps are discarded as burn-in.  

Table \ref{tab:visibility_fit_results} summarizes the resultant parameters of the fit. We find a PA of $67\arcdeg$ and an inclination of $72\arcdeg$, both of which are consistent with the 2D Gaussian fit on the image plane (see Section \ref{sec:obs_results}). The ring radius is estimated to be $0\farcs38$ (or 56\,au), which is consistent with the peak of the radial intensity profile (Figure \ref{fig:radial_profiles_cont}). The ring width is $0\farcs24$ (or 35\,au). The size of the central compact component is derived to be $0\farcs0073$ (or 1\,au), which is smaller than the deconvolved size derived by the 2D Gaussian fit on the image plane (see Section \ref{sec:obs_results}). This indicates that the central component is indeed very compact and not spatially resolved at the current resolution. The size of the extended component ($1\farcs58$ in the standard deviation of Gaussian, or $3\farcs72$ in FWHM) is comparable to that derived by the 2D Gaussian fit. The total flux density (summed up all the components) of 52\,mJy is similar to the value derived from the JvM-corrected image (which is expected to recover the correct flux scale; see Section \ref{sec:obs_results}).

\begin{figure}
\plotone{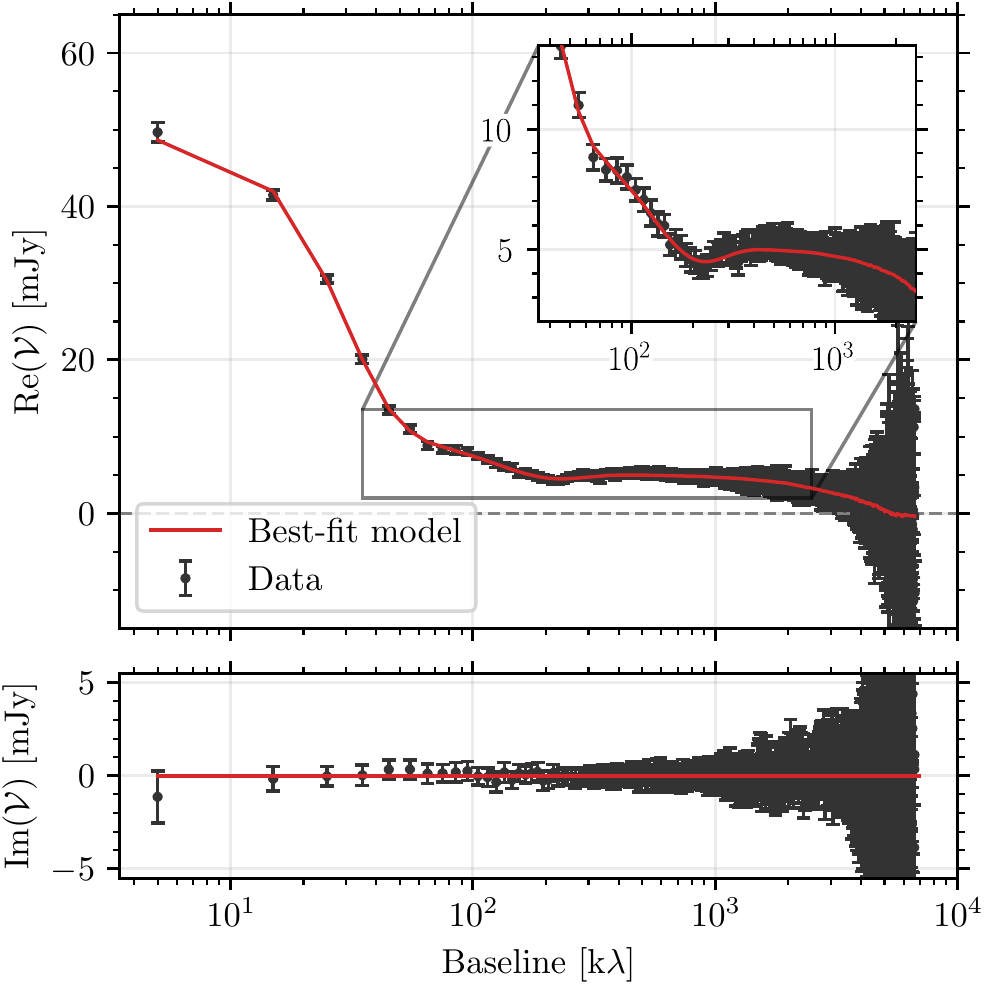}
\caption{Comparison of the deprojected, azimuthally averaged observed visibilities (black) and the best-fit model (red). The upper and lower panels show the real and imaginary parts, respectively. The zoom-in view of the long-baseline regime is shown in the inset in the upper panel.}
\label{fig:vis_profile}
\end{figure}

The comparison of the best-fit model (i.e., the model that maximizes the likelihood) with the observed visibilities (deprojected and azimuthally averaged) is shown in Figure \ref{fig:vis_profile}. Overall, the observed visibilities are reproduced well by the model. Importantly, a dip at $\sim$200\,k$\lambda$ (the inset of Figure \ref{fig:vis_profile}) clearly indicates the presence of a ring-like structure. In the shortest-baseline bin, a slight discrepancy between the observations and the model in both real and imaginary parts exists, although it is within the uncertainty. The discrepancy in the real part indicates that additional emission component(s) in the largest spatial scale (which can be traced by the present observations) may be required to better reproduce the observations. The discrepancy in the imaginary part in the shortest-baseline bin suggests that the additional component(s) would be non-axisymmetric with respect to the phase center. The potential presence of a large-scale, non-axisymmetric structure is consistent with the faint tails in the outermost region seen in the dust continuum image (the left panel of Figure \ref{fig:continuum}), which are not included in the current model.

\begin{deluxetable*}{lClC}
\footnotesize
\tablecaption{Best-fit parameters of the visibility fit}
\label{tab:visibility_fit_results}
\tablehead{\colhead{Parameter} & \colhead{Unit} & \colhead{Description} & \colhead{Value$^\dagger$}}
\startdata
$F_{\nu, \mathrm{c}}$ & \mathrm{mJy} & Flux density of the central compact component & 5.05^{+0.04}_{-0.04} \\
$w_\mathrm{c}$ & \arcsec & Size of the central compact component & 0.00722^{+0.00024}_{-0.00024} \\
$F_{\nu, \mathrm{r}}$ & \mathrm{mJy} & Flux density of the ring component & 5.39^{+0.22}_{-0.22} \\
$w_\mathrm{r}$ & \arcsec & Width of the ring & 0.238^{+0.016}_{-0.016} \\
$r_{0, \mathrm{r}}$ & \arcsec & Radius of the ring location & 0.384^{+0.018}_{-0.019} \\
$F_{\nu, \mathrm{e}}$ & \mathrm{mJy} & Flux density of the extended component & 41.4^{+0.4}_{-0.4} \\
$w_\mathrm{e}$ & \arcsec & Size of the extended component & 1.58^{+0.02}_{-0.02} \\
$\mathrm{PA}$ & \arcdeg & Position angle of the disk measured from north to east & 67.0^{+0.3}_{-0.3} \\
$i$ & \arcdeg & Inclination angle of the disk (0$\arcdeg$ for face-on) & 72.0^{+0.2}_{-0.3} \\
$x_0$ & \arcsec & Source offset from the phase center along Right Ascension & 0.00881^{+0.00010}_{-0.00010} \\
$y_0$ & \arcsec & Source offset from the phase center along Declination & 0.00101^{+0.00015}_{-0.00016} \\
\enddata
\tablenotetext{\dagger}{The uncertainties correspond to the 16th and 84th percentiles of the posterior distributions.}
\end{deluxetable*}

\subsection{Velocity structure traced by the \texorpdfstring{\ceo}{} emission}\label{subsec:SLAMfit}
\begin{figure}
\plotone{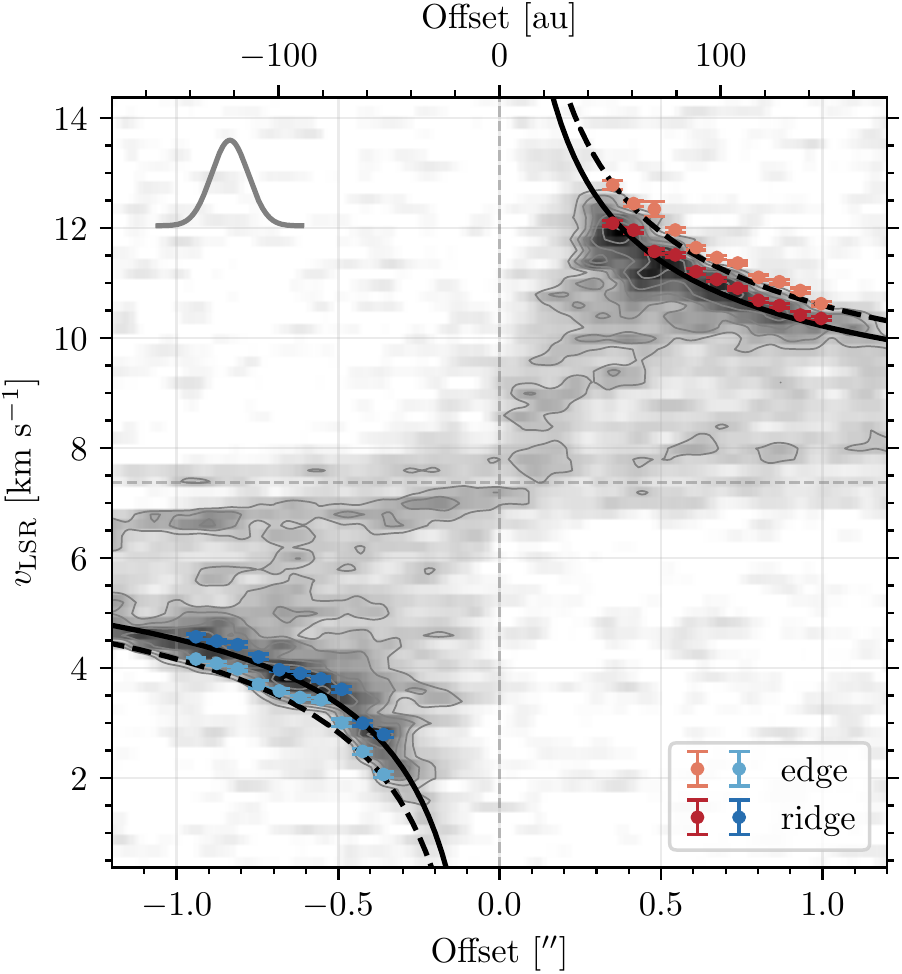}
\caption{PV diagram of the \ceo~$J=2$--1 emission along the disk major axis (PA$=67\arcdeg$). The contours are drawn from 5$\sigma$ to 15$\sigma$ in steps of 2$\sigma$, where $\sigma=1.7$\,mJy beam$^{-1}$. The derived data points from the PV diagram by SLAM are indicated by red/blue circles (ridge method) and orange/light blue circles (edge method). The black solid and dashed lines represent the rotation curve with the best-fit parameters listed in Table \ref{tab:pvfits_C18O_results} for data points derived from the ridge and edge methods, respectively. The systemic velocity derived from both fits (averaged the best-fit values; 7.375 km s$^{-1}$) is indicated by the horizontal grey dashed line. The Gaussian profile in the upper left corner indicates the size of the major axis of the synthesized beam.}
\label{fig:C18O_PVdiagram}
\end{figure}

To investigate the velocity structure of the disk and infer the dynamical stellar mass, we use the \ceo emission since the emission morphology suggests that it traces the rotating gas near the midplane (see the middle-left panel of Figure \ref{fig:zoom_maps}). \citet{Sai2020} found a large Keplerian disk extended to $r\sim600$\,au with a warped structure at $r\sim200$\,au by conducting the power-law fitting to the data points derived from the position-velocity (PV) diagram of \ceo~$J=$ 2--1, and estimated the central stellar mass of $1.64\pm0.12\,M_\odot$ \citep[see also][]{Yen2014}. We here independently verify the Keplerian rotation of the disk and estimate the central stellar mass using the Spectral Line Analysis/Modeling (\texttt{SLAM}) code \citep[][]{SLAM}. We follow the methodology described in \citet{eDisk_overview}. First, the \ceo PV diagram along the major axis of the dusty disk (PA $=67\arcdeg$ as derived by the visibility analysis; see Section \ref{subsec:visfit}) is generated and shown in Figure \ref{fig:C18O_PVdiagram}. The representative data point pairs $(r, v)$ are extracted from the diagram, to which a power-law function ($v \propto r^{-p}$) is subsequently fitted. To extract the representative data points, we identified a peak velocity channel for each pixel of the positional axis and subsequently fitted a Gaussian to the adjacent two channels (i.e., in total three channels). This is referred to as the ridge method \citep[e.g.,][see also \citealt{Aso2020}]{Yen2013, Lee2018, Sai2020}. As a complementary method, we extracted the points where the emission reaches 5$\sigma$ for each pixel of the positional axis, which is referred to as the edge method \citep[e.g.,][see also \citealt{Aso2020}]{Seifried2016, Alves2017}. The two methods result in different estimates of the central stellar mass: the ridge method can underestimate it and the edge method can overestimate, depending on the spatial resolution \citep{Maret2020}. The difference between the central stellar masses derived from these methods is thus adopted as a systematic uncertainty. In addition, we only used $r=\pm$0\farcs3--1\farcs0 region to avoid the effects of spatial resolution as described in Appendix of \citet{Aso2015} and substructures (i.e., the central depression and the warped structure at $r\gtrsim200$\,au). The extracted representative data points are shown by red and blue points in Figure \ref{fig:C18O_PVdiagram}. We verified that these data points follow Keplerian rotation by conducting a power-law fit to them with SLAM. The best-fit power-law index $p$ are $0.480^{+0.008}_{-0.008}$ and $0.495^{+0.008}_{-0.008}$ for the ridge and edge methods, respectively; the index $p\approx0.5$ is consistent with Keplerian rotation. To infer the central stellar mass, we fit the Keplerian rotation model to these data points:
\begin{equation}
    v_\mathrm{los} = \pm\sqrt{\frac{GM_\star}{r}}\sin(i) + v_\mathrm{sys},
\end{equation}
where $v_\mathrm{los}$ is the line-of-sight component of the rotation velocity, $G$ is the gravitational constant, $M_\star$ is the central stellar mass, $r$ is the radius, $i$ is the disk inclination angle, and $v_\mathrm{sys}$ is the systemic velocity. The free parameters are $M_\star$ and $v_\mathrm{sys}$. The disk inclination angle is fixed to 72$\arcdeg$ as derived from the visibility analysis described in Section \ref{subsec:visfit}. To search the parameter space, the MCMC algorithm implemented in the \texttt{emcee} package \citep{emcee} is adopted. We ran 1000 steps with 200 walkers and discarded the initial 500 steps as burn-in. The convergence was tested by auto-correlation analysis. The best-fit values (median of the posterior distributions) and uncertainties (16th and 84th percentiles) are reported in Table \ref{tab:pvfits_C18O_results} for both ridge and edge methods. 

While the derived systemic velocities are consistent between the two methods, the best-fit stellar masses are significantly different (Table \ref{tab:pvfits_C18O_results}). The value from the edge method is larger than that from the other method, which is natural as the edge method tends to trace a higher velocity than the ridge method at each pixel. Thus, the derived stellar mass including the systematic uncertainty is $1.5$--$1.9\,M_\odot$. This is consistent with the previous estimates of $1.64\pm0.12\,M_\odot$ by \citet{Sai2020}, where they used the ridge method and the data points from the outer warped disk as well.

% The larger value by ``edge'' method is considered as the upper limit. The fiducial stellar mass is slightly lower than the value derived by \citet[][$1.64\pm0.12\,M_\odot$]{Sai2020}, but consistent within $\sim$2$\sigma$ significance. The systemic velocity is slightly larger than the value in \citet[][$7.22\pm0.09$ km s$^{-1}$]{Sai2020}

\begin{deluxetable}{cCC}
\footnotesize
\tablecaption{Results of the Keplerian fits to \ceo PV diagram}
\label{tab:pvfits_C18O_results}
\tablehead{\colhead{Method} & \colhead{$M_\star$} & \colhead{$v_\mathrm{sys}$} \\
\colhead{} & \colhead{[$M_\odot$]} & \colhead{[km s$^{-1}$]}}
\startdata
ridge & 1.498^{+0.008}_{-0.008} & 7.372^{+0.009}_{-0.009} \\
edge & 1.911^{+0.010}_{-0.009} & 7.377^{+0.010}_{-0.010}
\enddata
\tablecomments{The uncertainties correspond to the 16th and 84th percentiles of the posterior distributions.}
\end{deluxetable}

\section{Discussion} \label{sec:discussion}
% \subsection{Comparison of the radial intensity profiles}
% compare the radial intensity profiles of dust /molecular line emission

% We compare the radial intensity profiles of continuum, \tco ($J=$ 2--1), \ceo ($J=$ 2--1), and SO ($J_N=$ 6$_5$--5$_4$) emission in Figure \ref{}. 

% \subsection{Possible origins of dust and moleculer substructures}
% discuss the possible origin of dust/molecular gap/ring. Mostly about inner most dust ring, but will be the relation between outer dust ring and C18O dip
% Our key results of this work is the findings of the disk substructures in both dust and molecular line emission. In this section, we will discuss the possible origins of those substructures.
% \subsubsection{Radial substructures}\label{subsubsec:discussion_radial_substructures}
\subsection{Origin of the dust ring}\label{subsec:discussion_dust_ring}
As described in Section \ref{sec:obs_results}, our observations have detected several radial substructures in both dust continuum and molecular line emission. Various mechanisms have been suggested as the origin of ring/gap structures in protoplanetary disks: e.g.,  planet-disk interaction \citep[e.g.,][]{Zhang2018}, modification of dust grain properties at the snowlines of volatiles \citep{Zhang2015, Okuzumi2016}, disk winds \citep[e.g.,][]{Takahashi2018}, and dust growth \citep{Ohashi2021}. 

% The most intriguing explanation for the origin of the dust ring at $\approx57$\,au would be the depletion of both dust and gas, which might be carved by formed planets. 
If the dust ring at $\sim$$56$\,au is carved by a planet, gas will also show a ring-gap structure at a similar radius. To explore the gas and dust structures around the dust ring, we compare the radial profiles of the dust continuum and line emission in the innermost region in Figure \ref{fig:radprof_comparison_inner_outer} (left panel). Interestingly, the dust ring and the radial peak of \ceo intensity profile coincide with each other. Because the \ceo emission is likely to trace the region near the midplane based on the emission morphology on the velocity-integrated emission map (Figure \ref{fig:zoom_maps}), this coincidence may indicate the presence of a ring-like structure in both dust and gas. In the adjacent dust gap (at $\approx$30\,au), the \tco emission is bright, which is likely due to its optically thick emission at the disk surface as indicated by the morphology on the velocity-integrated intensity map (Figure \ref{fig:zoom_maps}). While the dust and \ceo intensity profiles at $r\gtrsim30$\,au show similar behavior, the \ceo intensity profile monotonically decreases inwards at $r\lesssim30$\,au, where the dust intensity profile shows a rapid increase. The \tco intensity profile also shows a rapid decrease at $r\lesssim20\,$au. These are likely due to the continuum over-subtraction \citep[e.g.,][]{Weaver2018}, which makes it difficult to infer the presence/absence of the gas gap, while the coincidence of the dust ring and the radial peak of the \ceo emission is likely real. 
% Although it is also unclear if a planet that is massive enough to carve a gap will be already formed in Class~I stage, 

If we assume that the gas disk has a gap with the same depth as the dust gap and that it is carved by a formed planet, the mass of the planet $M_\mathrm{p}$ can be estimated using 
% However, it is difficult to infer the presence/absence of the gas gap due to continuum oversubtraction \citep[e.g.,][]{Weaver2018}, which possibly causes the monotonic decrease of the line intensity profiles toward the disk center at $\lesssim$30\,au, where the continuum intensity profile shows a rapid increase. Also, it is unclear if massive bodies that are enough to carve a gap will be already formed in Class~I stage. If we assume that the dust continuum gap is carved by a formed planet, the mass of the planet $M_\mathrm{p}$ can be estimated using 
\begin{equation}\label{eq:planet_mass}
    M_\mathrm{p} = 5 \times 10^{-4}\left(\frac{1}{\Sigma_\mathrm{p}/\Sigma_0}-1\right)^{1/2}\left(\frac{h_\mathrm{p}}{0.1}\right)^{5/2}\left(\frac{\alpha}{10^{-3}}\right)^{1/2}\,M_\star,
\end{equation}
where $\Sigma_\mathrm{p}/\Sigma_\mathrm{0}$ is the surface density contrast (i.e., gap depth), $h_\mathrm{p}$ is the disk aspect ratio (i.e., the ratio of the scale height to the disk radius), $\alpha$ is the turbulence parameter, and $M_\star$ is the central stellar mass \citep{Kanagawa2015}. If we adopt the contrast of 0.64 as measured on the dust intensity profile (Figure \ref{fig:radial_profiles_cont}; assuming the emission is optically thin) as $\Sigma_\mathrm{p}/\Sigma_\mathrm{0}$, and \textrm{assume $h_\mathrm{p} = 0.05$--$0.1$, $\alpha=10^{-4}$--$10^{-2}$, and the central stellar mass of $1.5$--$1.9\,M_\odot$ as derived in Section \ref{subsec:SLAMfit}, the mass of the planet (if present) would be $\sim$0.033--2.4\,$M_\mathrm{J}$.}

Another possible mechanism to explain the innermost ring/gap structures in the dust emission is the dust growth front, which is recently proposed by \citet{Ohashi2021}. They calculated the time evolution of the dust size distribution and surface density profile considering the grain growth via coagulation and radial drift of dust grains and found that a ring-like structure in the surface density profile is formed at the dust growth front (or pebble production line; \citealt{Lambrechts2014}). Outside the dust growth front, the dust particles have not grown and remain in their initial states (\textmu m size), while inside the growth front they have grown to mm size or larger and radially drifted, resulting in a local maximum of the surface density profile (i.e., ring-like structure) at the growth front. The expected radial location of the ring matched well the observed dust ring locations in young protostellar disks \citep{Ohashi2021, Ohashi2022_L1489IRS}, suggesting that the dust growth front mechanism could be a possible explanation for the origin of dust rings particularly in young disks.
% The grown dust grains will radially drift inward by the inside-out manner and make a single ring structure in dust surface density profile. 
\citet{Ohashi2021} constructed an equation that predicts the radial location of the ring formed by this mechanism based on the radiative transfer modeling. The important parameters that control the location of the dust ring ($R_\mathrm{c}$) are the disk age ($t_\mathrm{disk}$), stellar mass ($M_\star$), and dust-to-gas mass ratio ($\zeta_\mathrm{d}$):
\begin{equation}
    R_\mathrm{c} = 56\, \left(\frac{M_\star}{M_\odot}\right)^{1/3}\left(\frac{\zeta_\mathrm{d}}{0.01}\right)^{2/3}\left(\frac{t_\mathrm{disk}}{0.1\,\mathrm{Myr}}\right)^{2/3}\,\mathrm{au}.
\end{equation}
If we adopt the stellar mass of $1.5$--$1.9\,M_\odot$ as derived in Section \ref{subsec:SLAMfit}, $\zeta_\mathrm{d} = 0.01$, and $t_\mathrm{disk} = $ (3--8) $\times 10^4\,\mathrm{yr}$ \citep{Sai2022}, the expected location of the dust ring is 28--59\,au. 
% The age of the L1489~IRS is longer by a factor of $~$2--3 if we consider the assumption of modeling in \citet{Sai2022}, and if we adopt , 
The ring location of 56\,au found by the present observations is consistent with this expectation. \citet{Ohashi2022_L1489IRS} also tentatively detected the ring-like structure at $\sim$90\,au in the L1489~IRS disk; the difference with our result is likely due to their coarser spatial resolution. If the dust ring is produced by the dust growth front mechanism, dust grains in the inner $\lesssim$60 au has already grown to millimeter size or even larger \citep{Ohashi2021}. To confirm that grain growth indeed has taken place, higher resolution, multi-band observations are needed to measure the spectral indices across the dust ring.

% the mass of the planet will be $\sim$0.5\,$M_\mathrm{J}$ based on the prediction by \citet{Kanagawa2015}. Here we use the dust surface density contrast of 0.5 measured in the dust surface density profile in Figure \ref{} and assume that the aspect ratio of the disk is 0.1 and a turbulence parameter of $\alpha=10^{-3}$ (see Equation (7) in \citealt{Kanagawa2015}).

% Inside the dust ring, the molecular line profiles show the inner depressions and radial peaks at different radii depending on molecules. The inner depression is most likely an artifact due to continuum oversubtraction. The different peak locations reflect the temperature of the region (or disk height) traced by each line. While SO traces the highest temperature region (i.e., uppermost disk surface), \ceo originates from the lowest temperature region (i.e., near the midplane), and \tco from the intermediate. The higher sublimation temperature of SO molecules also support this interpretation. Molecular line emission originated from higher temperature regions is less affected by continuum oversubtraction, and thus show the inner depression at only the innermost region. 
% This effect also complicates the interpretation of the coincidence of dust and \ceo rings (see above).

The observed ring/gap structure may also be formed by the inner misaligned disk. We will discuss this possibility in Section \ref{subsec:warp_discussion}.

% The final, but  possiblity for the origin of the dust gap/ring is the disk warp inside 
\subsection{Outer disk structures}
\begin{figure*}
\plotone{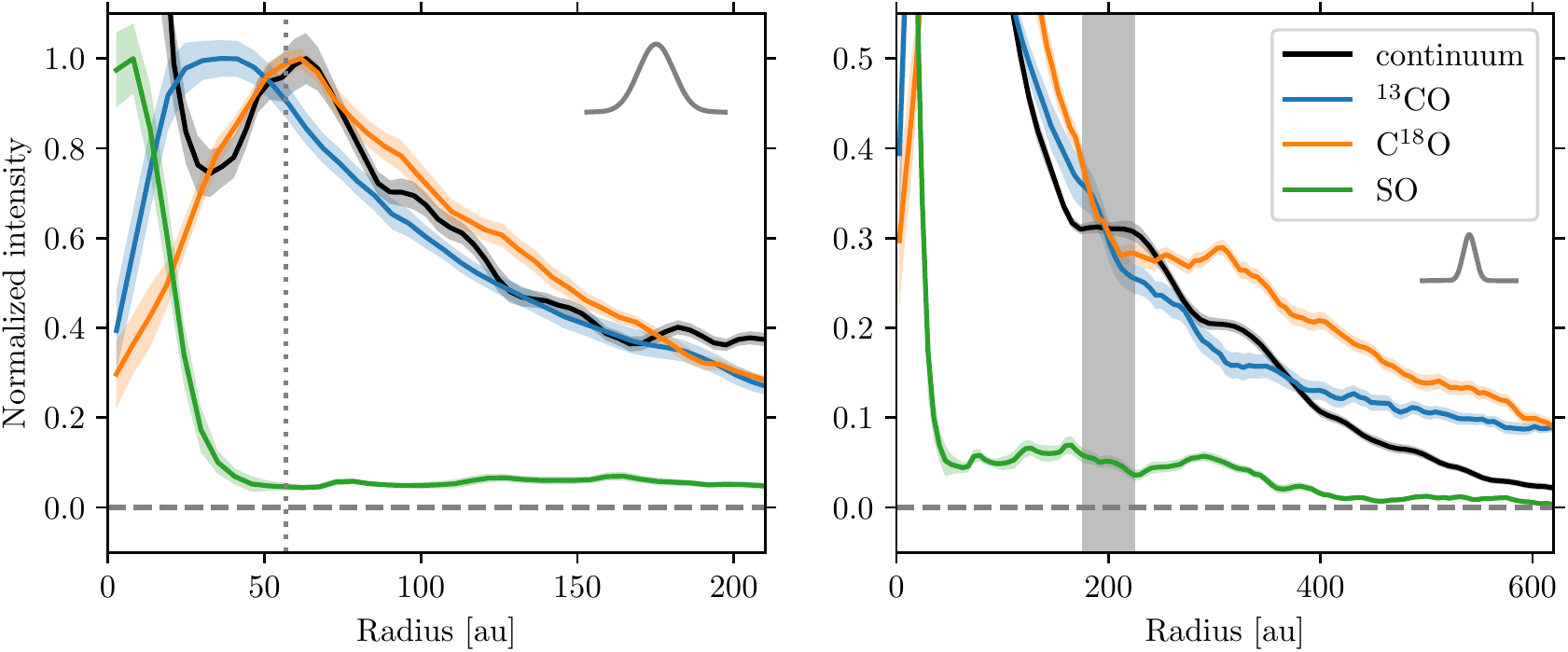}
\caption{Comparison of radial intensity profiles of continuum, \tco, \ceo, and SO. While the left panel shows the profiles in the innermost region, the right panel  presents the outer region. The profiles of line emission are normalized by the radial peaks, while the continuum profile is normalized by the intensity at the peak of the ring. The horizontal dashed line represents the zero-intensity levels. In the left panel, the vertical dotted line indicates the position of the innermost dust ring, while the vertical grey-shaded region in the right panel marks the position of the CO snowline inferred by \citet{Sai2020}.}
\label{fig:radprof_comparison_inner_outer}
\end{figure*}

In addition to the dust ring at 56\,au, we identified two shoulder-like structures in the outer disk in the dust continuum radial profile (Figure \ref{fig:radial_profiles_cont}). 
% Figure \ref{fig:radprof_comparison_inner_outer}(b) present the radial continuum intensity profile derived from additionally generated image with a Briggs robust parameter of 2.0 and taper of $1000\mathrm{k\lambda}$ in the $(u,v)$ domain,  
The comparison of the radial intensity profiles of the dust continuum and line emission in the outer region is shown in the right panel of Figure \ref{fig:radprof_comparison_inner_outer}. 
The inner shoulder at $\sim$220\,au coincides with the CO snowline location based on the radiative transfer modeling by \citet{Sai2020}.
% which implies that this substructure may have the snowline origin \citep{Zhang2015, Okuzumi2016}. 
The rapid inward increase of the \ceo profiles at the shoulder also indicates the sublimation of CO molecules. Similar behaviors of CO line intensity profiles have been observed in Class~II disks as well \citep[e.g.,][]{Zhang2021_MAPSV}. While the coincidence between the dust shoulder and the CO snowline may imply the change of the dust property to form the substructure, it is also possible that the rapid increase of the radial profiles of the CO isotopologues could simply be due to the gas density enhancement. We also note that the shoulder radius of $\sim200$\,au coincides with the warped structure as revealed by the \ceo emission (Section \ref{subsubsec:C18O}; see also \citealt{Sai2020})
% corresponds to the connecting region \textbf{of the warped inner and outer disks} as revealed by \citet{Sai2020}. 
% More detailed analysis would be required to determine the origin of this substructure.

The outer shoulder at $\sim$340\,au is located in the outer warped disk. At the same location, the \ceo emission, which traces near the disk midplane, exhibits a subtle emission enhancement. Similar enhancements of the line emission have been observed in \ceo $J=$ 2--1 by \citet{Sai2020} and C$^{17}$O $J=$ 2--1 by \citet{vantHoff2020}. 
% On the other hand, the \tco emission, which probably traces the disk surface, does not show any significant enhancement at the \ceo ring. 
This may imply that CO depletion is much less significant outside $\sim$340 au, which we call the secondary CO snow line. As the disk column density decreases with radius, photodesorption due to penetrating UV radiation could keep CO in the gas phase outside a certain radius. A similar re-enhancement of the line emission at an outer radius is also observed in several disks \citep{Dutrey2017, Flores2021, eDisk_IRAS04302}. In Class~II disks, similar secondary CO snow lines have also been suggested by the double emission rings of DCO$^+$, which is expected to trace the gas-phase CO \citep{Oberg2015, Cataldi2021}. 

\subsection{Azimuthal asymmetry}\label{subsubsec:discussion_azimuthal_aymmetry}
In addition to the radial substructures, the dust continuum and molecular line emission also show azimuthal asymmetries (Section \ref{sec:obs_results}).
% We also detected the azimuthal asymmetry in the dust continuum and several molecular lines.
As for the dust continuum, the southwestern side of the disk is brighter than the northeastern side in the dust ring (see the right panel of Figure \ref{fig:continuum}). Non-axisymmetric dust emissions have been observed in a handful of disks of both Class 0/I and Class~II \citep[e.g.,][]{vanderMarel2013, Sheehan2020}, at least part of which is considered to be non-axisymmetric dust traps, e.g., vortices. In the L1489~IRS disk, the ring at $\sim56$\,au may thus have the dust trap (i.e., gas pressure maximum) on the southwestern side. Alternatively, the azimuthal variance of the temperature can also lead to the asymmetry of the dust emission, which could be caused by an inner misaligned disk (see Section \ref{subsec:warp_discussion}).

The L1489~IRS disk shows the asymmetry of the dust continuum emission along the major axis (see the left panel of Figure \ref{fig:continuum}). This is in contrast to several other eDisk sources which exhibit asymmetries along the disk minor axis (see \citealt{eDisk_overview} for a gallery of the continuum images). The latter is evidence that the dust grains are flared but not settled to the disk midplane; brighter emissions from the far side of the disk and fainter emissions from the near side of the disk are observed as a natural consequence of an flared optically thick disk with a radial temperature gradient \citep[decreasing temperature as a function of radius;][]{Ohashi2022_L1527IRS, eDisk_IRAS04302}. 

The lack of brightness asymmetry along the disk minor axis in the L1489~IRS disk could be due to its optically thin emission. In the case of the optically thin emission, the entire columns are seen at both near and far sides, resulting in a symmetric emission distribution along the minor axis.
% In the case of the optically thin emission, while we can directly observe the disk midplane, 
% the contribution of the flared dust to the observed emission is smaller than in an optically thick disk depending on the disk inclination.
The low brightness temperature of the dust emission in the outer disk ($\lesssim4$\,K, see Figure \ref{fig:radial_profiles_cont})  indeed suggests that the dust emission is likely optically thin except for that originating from the spatially-unresolved innermost region ($\lesssim0\farcs1$). 
% In addition, observations of young Class~0/I edge-on disks suggests that the dust in Class~0/I disks are significantly flared from the midplane \citep{Lee2017, eDisk_IRAS04302, eDisk_IRS7B_model}. Since L1489~IRS is classified as a Class~I protostar based on its spectral energy distribution (SED; \citealt{eDisk_overview}), it is not surprising even if the dust in the L1489~IRS disk is indeed flared.

It is also possible that the dust in the L1489~IRS disk is settled to the midplane, which causes the lack of asymmetry along the minor axis of the disk. In contrast to young Class~0/I disks, observations of Class~II disks indicate that the large dust grains are settled well to the midplane \citep[e.g.,][]{Kwon2011, Pinte2016, Villenave2020, Villenave2022}. This suggests that the dust settling occurs during the evolution from the Class~I phase to the Class~II phase. It has also been speculated that L1489~IRS is in the later evolutionary stage of the Class~I phase based on its relatively small disk mass compared to other Class~I disks \citep{Sai2020}. Therefore, L1489~IRS may be in transition between the Class~I and Class~II phases, where the dust settling has started. A similar lack of asymmetry along the minor axis is also observed in the disk around V883~Ori, which is in transition from the Class~I phase to the Class~II phase \citep{Cieza2016, Lee2019}.

% Several eDisk sources, in particular younger ones (i.e., Class~0), show the asymmetry along the disk minor axis. This is an indication of the flared dust distribution, from which we observe the brighter emission in the far side of the disk (see e.g., Figure 18 in \citealt{Ohashi2022_L1527IRS} for a schematic figure). L1489~IRS does not show any clear asymmetry along the disk minor axis, suggesting that the dust is not flared but settled to the disk midplane. Older sources (i.e., Class~I sources) in eDisk sample tends not to show the asymmetry along the minor axis, possibly indicating the dust settling occurs in the middle of Class~0 and Class~I. Several recent studies also suggest that the mm dust grain traced by (sub)millimeter observations in young embedded sources are flared, while they are settled in evolved sources such as Class~II by observing edge-on disks \citep[e.g.,][]{Villenave2020, Villenave2022}.

In addition to the dust continuum emission, the molecular line emissions also show azimuthal asymmetries (Figure \ref{fig:large_scale_maps} and \ref{fig:zoom_maps}). The slight shift of the emission peaks from the disk major axis to the southern side of the disk on the \tco velocity-integrated emission map (Figure \ref{fig:zoom_maps}) is clear evidence that \tco emission traces a significantly higher layer above the disk midplane (so-called ``emission surface'')\footnote{We note that we tried to estimate the emission surface of \tco using the \texttt{disksurf} package \citep{Pinte2018, disksurf}, but no reasonable estimates were obtained due to the contamination from the envelope emission.}. The direction of the shift indicates that the southern side is the far side of the disk, which is consistent with the outflow configuration where the southern lobe is blueshifted \citep{Yen2014}. Similar molecular line structures elevated from the midplane have been observed in a number of disks \citep[e.g.,][]{Law2021_MAPSIV, Law2022, eDisk_IRAS04302}. Furthermore, the \tco and \ceo emissions show higher brightness temperatures on the southwestern side of the inner ($r\lesssim0\farcs5$) disk (Figure \ref{fig:zoom_maps}). These asymmetries, as well as the non-axisymmetric brightness distribution of the dust continuum emission, may be explained by the warped disk structure (see Section \ref{subsec:warp_discussion}).

% The \tco and \ceo emission also show azimuthal asymmetries in the outer disk ($\gtrsim$100\,au) as shown in Figure \ref{fig:radial_profiles_line}. They are brighter on the redshifted (southwestern) side than on the blueshifted (northeastern) side. A similar trend is observed in other molecular lines at a similar spatial scale as well \citep[e.g., C$^{17}$O and H$_2$CO;][]{vantHoff2020}. One possible explanation could be that those molecular lines partially trace the infalling envelope as in the outer region, where the material infalls to the disk in an asymmetric manner. For example, \citet{Yen2014} demonstrated that the large-scale tail of the \ceo ($J=$ 2--1) emission can be explained by an infalling flow, which is brighter in the redshifted side. On the other hand, in the innermost region ($\lesssim$100\,au), the radial intensity profile is similar between both sides, indicating those molecular lines trace the axisymmetric disk on that spatial scale.

\textrm{\subsection{Gas disk size and dust disk size}
Disk size is a key to undestanding disk evolution processes. Submm/mm Observations of Class~II disks have revealed that the extent of the bright $^{12}$CO line emission is generally larger than that of the dust continuum emission \citep[e.g.,][]{Andrews2012, deGregorioMonsalvo2013, ansdell2018, Long2022}. This difference has been interpreted as a consequence of grain growth and radial drift, although the difference could also be explained by the larger optical depth of the $^{12}$CO line emission than that of the dust continuum emission \citep[e.g.,][]{Panic2009, Facchini2017, Trapman2019}.}

\textrm{For the L1489~IRS disk, while the dust disk radius derived from the visibility analysis is $\sim$230\,au, the gas disk radius was estimated to be $\sim$600\,au from the rotation curve analysis of the \ceo emission \citep{Sai2020}. This factor of $\sim$2--3 difference in the size of the gas and dust disks is consistent with the averaged ratio of the gas disk radii to the dust disk radii in samples of Class~II disks \citep{ansdell2018, Long2022}, suggesting that the dust growth and radial drift have already been occurred in the L1489~IRS disk. This is also consistent with the ring-like structure of the dust emission (see Section \ref{subsec:discussion_dust_ring}) and suggests that the L1489 IRS could be in transition between the Class I phase to the Class II phase (see also Section \ref{subsubsec:discussion_azimuthal_aymmetry}).}

\textrm{However, it may be also possible that the difference in the radii of the gas and dust disks is due to the difference in the optical depths of the gas and dust emission \citep{Panic2009, Facchini2017}. The brightness temperature of the dust continuum emission in the outer region of the L1489 IRS disk ($\sim$ a few K; see Figure \ref{fig:radial_profiles_cont}) is lower than that of the \ceo emission ($\sim$ a few tens of K; see Figure \ref{fig:large_scale_maps}). This may indicate the difference in the optical depths, although it is also possible that the dust emission and the \ceo emission trace the different vertical layers of the disk that should have different temperatures.} 

% \textrm{Alternatively, a steep truncation of the radial intensity profile of the dust emission could be a better indicator of the growth and radial drift of dust grains than the difference in the sizes of the gas and dust disks \citep{Facchini2017}. Deeper observations of the dust continuum emission would be useful to constrain the outer edge of the dust disk and probe the dust growth in the L1489~IRS disk.}

\subsection{Origin of the SO emission: accretion shocks and a warm inner disk?}\label{subsec:discussion_SO_emission}
As described in Section \ref{subsubsec:SO}, the nature of the SO emission is highly complex. Outside the radius of $\sim$30\,au ($r\gtrsim0\farcs2$), diffuse SO emission has been observed (Figure \ref{fig:large_scale_maps}). The non-uniform diffuse SO emission suggests the localized enhancement of the SO column density rather than a smooth distribution of SO within the disk. In addition, the localized enhancements of the SO brightness temperature (Figure \ref{fig:large_scale_maps}) indicate the elevated gas temperature. In particular, the SO brightness temperature enhancement is prominent at $r\sim300$\,au on the disk major axis, and this radius roughly coincides with the landing point of the accretion flow observed in the \ceo emission. \citet{Yen2014} reproduced the observed accretion flow with parabolic trajectories assuming a landing point (or centrifugal radius) of $\sim$300\,au. It is thus likely that the origin of the diffuse SO emission is an accretion shock onto the protostellar disk caused by the infalling material. At the interface between the infalling material and the protostellar disk, slow shocks can occur; SO molecules are released into the gas phase due to the aerodynamic heating of dust grains in the post shock gas and/or formed from other desorbed molecules \citep[e.g.,][]{Aota2015, vanGelder2021}. In the Class~0 protostar L1527~IRS, the ring-like distributions of the SO emission have been interpreted as an indication of accretion shocks at the interface between the envelope and disk \citep[][but see also \citealt{eDisk_L1527}]{Ohashi2014, Sakai2014, Sakai2017}. 
% Although L1489~IRS is a Class~I protostar, the mass accretion from the envelope is still ongoing as indicated by the infalling accretion flow observed in \ceo \citep{Yen2014}. Indeed, an SO emission clump at $\sim5\arcsec$ southwest of the protostar spatially coincides well with the \ceo infalling flow (Figure \ref{fig:large_scale_maps}; see also \citealt{Yen2014}). 
SO emissions originating from the accretion shock of the late-stage infalling streamer are also observed in Class~I/II disks, DG Tau and HL Tau, as well \citep{Garufi2022}. \textrm{Alternatively, the outer diffuse SO emission may originate from non-thermally desorbed SO molecules (e.g. photodesorption). However, the non-uniform distribution of the SO emission would be better explained by accretion shock.}
% The explanation for the outer diffuse SO emission could be the release of the SO molecules due to the dust sputtering associated with accretion shocks. The ring-like distributions of the SO emission at the disk outer edge (or the centrifugal barrier) have been observed in other sources as well \citep[e.g.,][]{Sakai2014, Sakai2016, Sakai2017, Garufi2022}. The spotty local enhancements and partial ring-like distribution of the brightness temperature at the disk outer edge ($\sim$2\arcsec or $\sim$290\,au; Figure \ref{fig:SO_maps}(c)), suggesting the emission from shocked materials, also support this interpretation. The ring-like distribution is consistent with the expectation from a PV diagram analysis in \citet{Yen2014}. In addition to the partial ring-like brightness temperature distribution, a protrusion toward the NW side of the disk from the centrally peaked emission along the disk major axis has been observed (see Figure \ref{fig:SO_maps}). This might be explained as that the released SO molecules at the outer disk edge by the accretion shock drifted toward the center, similar to the interpretation of the similar morphology of SO emission in HL~Tau disk \citep{Garufi2022}.

In the innermost region ($r\lesssim0\farcs2$), on the other hand, a compact, prominent SO emission exists (Figure \ref{fig:zoom_maps}). This emission component is distinct from the outer diffuse emission in terms of the spatial distribution and the brightness, suggesting a separate origin. The comparison of the radial intensity profiles (Figure \ref{fig:radprof_comparison_inner_outer}) clearly shows that the SO emission is distributed in inner regions ($\lesssim30$\,au) compared to \tco and \ceo. Given the usual temperature structure of the disk (i.e., the higher temperature at a smaller radius), the compact SO emission traces the higher temperature region than that traced by CO isotopologue lines. \citet{Sai2020} calculated the temperature structure of the L1489~IRS disk by radiative transfer modeling. The midplane dust temperature at the SO emitting radii ($\lesssim30$\,au) is $\gtrsim40$\,K (see Appendix \ref{appendix:disk_model}), which is comparable to or higher than the sublimation temperature of SO \citep[$\sim50$\,K; e.g.,][]{Sakai2014, Aota2015, Miura2017}, and the disk surface would be even warmer. The compact SO emission thus likely originates from the thermal sublimation of SO or its precursor (e.g., H$_2$S) at the warm inner region. The bright emission also requires abundant SO gas, which could be naturally explained by thermal sublimation. Although the relatively low peak brightness temperatures of SO in the innermost region (see Figure \ref{fig:zoom_maps}) seemingly contradicts this interpretation, the brightness temperatures at the innermost region may be reduced due to the beam dilution effect. Additionally, the SO emission in the innermost region could be optically thin, resulting in lower peak brightness temperature than the actual gas temperature.

\textrm{While a bright, compact SO emission is detected in the innermost region, a CH$_3$OH transition with an upper state energy of 45\,K, which is observed in the same spectral setup, is not detected. In the Class~II disks around IRS~48 and HD~100546, SO emission and bright emission from complex organic molecules including CH$_3$OH have been detected co-spatially \citep{Booth2021_IRS48, Booth2021_HD100546, Booth2023, Brunken2022}, indicating a hot-corino-like chemistry in the warm ($\gtrsim100$\,K) region. The detection of SO and the non-detection of CH$_3$OH in the L1489~IRS disk may indicate that the temperature is not high enough for CH$_3$OH ice to thermally sublimate: while the sublimation temperature of SO is $\sim50$\,K \citep[e.g.,][]{Sakai2014, Aota2015, Miura2017}, a higher temperature ($\sim100$\,K) is required for CH$_3$OH ice to sublimate. Alternatively, the emitting region size of CH$_3$OH molecules may be smaller than that of SO and the CH$_3$OH emission is thus beam-diluted and/or absorbed by the bright continuum emission. Higher angular resolution observations at lower frequencies, where the dust continuum emission is expected to be more optically thin, would be a key to revealing the chemistry in the innermost region of the L1489~IRS disk.}

% , where the continuum emission is significant, cannot be used directly as the proxy of temperature. If we consider the continuum brightness temperature at the peak ($\approx21$\,K), the peak brightness temperature of the SO plus dust emission is $\approx50$\,K, consistent with the sublimation temperature of SO. \textbf{Additionally, the SO emission is only marginally spatially resolved, and the peak brightness temperature can thus be reduced due to the beam dilution effect.}

\subsection{Warped disk structures}\label{subsec:warp_discussion}

\begin{figure*}
\plotone{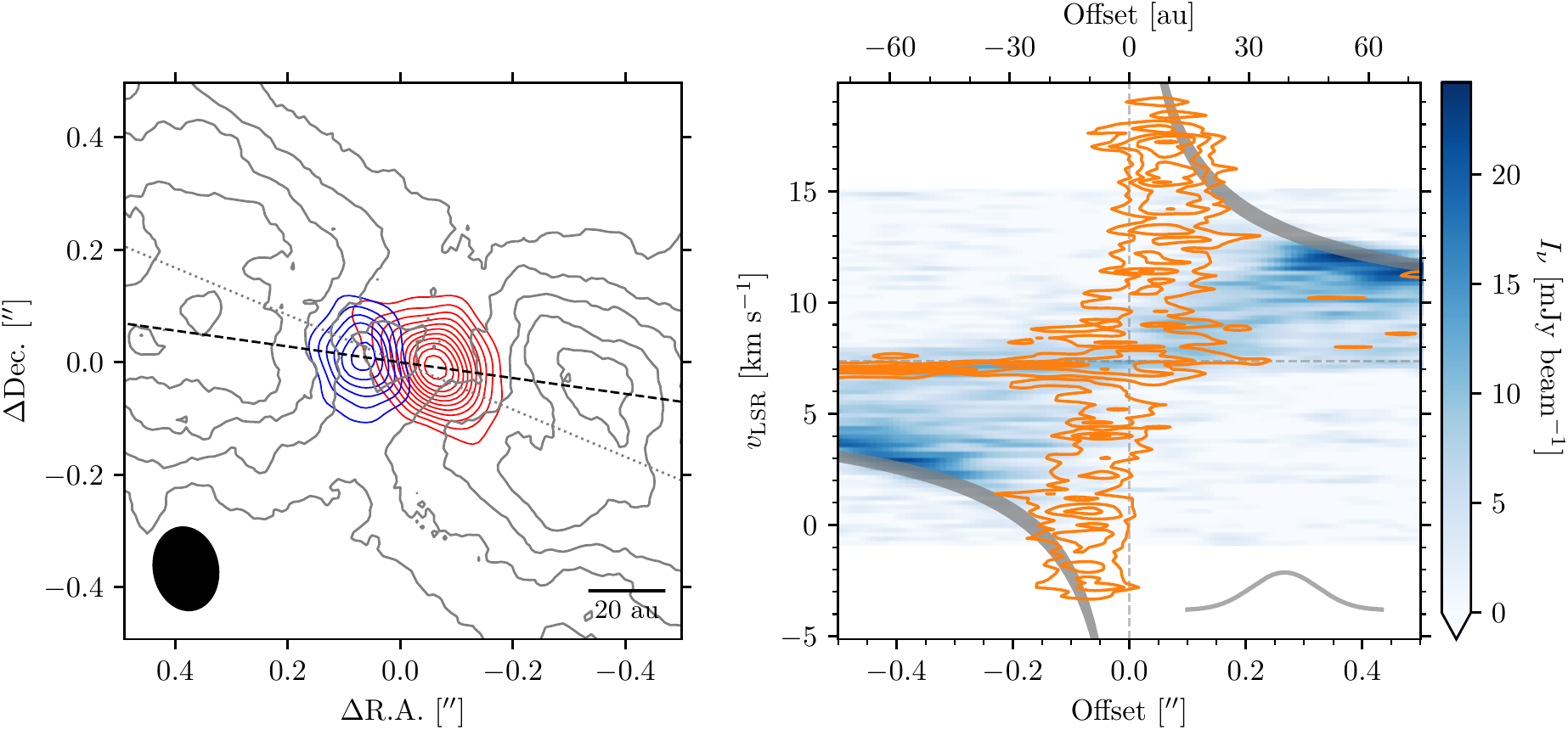}
\caption{Left: Velocity-integrated intensity map (red and blue contours for red- and blue-shifted emission respectively) of the SO emission for the highest velocity component ($\pm$8--12\,km s$^{-1}$ with respect to $v_\mathrm{sys}$), overlayed on the \ceo velocity-integrated intensity map (grey contours; same as the middle-left panel of Figure \ref{fig:zoom_maps}). The contours of the SO emission start from 5$\sigma$ followed by 2$\sigma$ steps, where $\sigma=1.9$\,mJy\,beam$^{-1}$\,km\,s$^{-1}$. The contours of the \ceo emission indicate [20, 30, 40, 50, 60, 70]$\sigma$, where $\sigma=1.0$\,mJy\,beam$^{-1}$\,km\,s$^{-1}$. While the grey dotted line marks the dust disk PA of 67\arcdeg estimated by visibility analysis (see Section \ref{subsec:visfit}), the black dashed line indicates a slightly tilted PA of 82\arcdeg, which approximately intercepts the peak location of the red- and blue-shifted SO emission lobes. The beam is shown in the lower left. The scale bar in the lower-right indicates 20\,au scale. Right: PV diagrams of the \ceo emission (color) along a position angle of $67\arcdeg$ and the SO emission (orange contour) along a position angle of $82\arcdeg$. The contours indicate [3, 5, 7, 9]$\sigma$, where $\sigma=2.1$\,mJy\,beam$^{-1}$. The gray-shaded curves indicate the Keplerian velocity assuming the stellar mass derived from the PV diagram fits (Section \ref{subsec:SLAMfit}). The shaded region indicates the uncertainty owing to the range of the stellar mass ($1.5$--$1.9\,M_\odot$). The Gaussian profile in the lower right indicates the beam size along the offset axis.}
\label{fig:SO_map_PV}
\end{figure*}

We also found that the compact SO emission shows a velocity gradient along a slightly different direction from the disk traced by \tco and \ceo. The left panel of Figure \ref{fig:SO_map_PV} shows the velocity-integrated intensity map of the high-velocity component ($\pm$8--12\,km s$^{-1}$ with respect to $v_\mathrm{sys}$) of the redshifted and blueshifted SO emission. The position angle of the velocity gradient ($\approx82\arcdeg$) is slightly ($\approx15\arcdeg$) tilted with respect to the position angle ($67\arcdeg$) of the disk traced by \tco, \ceo, and dust continuum. The PV diagram of the SO emission along the position angle of $82\arcdeg$ (the right panel of Figure \ref{fig:SO_map_PV}) indicates that the highest-velocity component of the SO emission is fairly consistent with the Keplerian velocity derived from the \ceo analysis (see Section \ref{subsec:SLAMfit}), suggesting that the compact SO emission originates from the innermost rotating disk. 

The slight tilt of the velocity gradient direction is highly intriguing. While the non-axisymmetric emission distributions in the upper layers of the disk may result in such an appearance, the most straightforward interpretation is that the disk is warped; the rotation axis of the innermost SO disk is slightly ($\approx15\arcdeg$) misaligned with respect to that of the outer disk, while the misalignment may be underestimated due to the projection effect. Observations of protostellar/protoplanetary disks have shown that many of those have misaligned or warped structures: e.g., the L1527~IRS disk \citep{Sakai2018} and the GW~Ori disk \citep{Bi2020, Kraus2020}. In the L1489~IRS disk, a warped structure at the outer radius $\sim 200$\,au is also found by \citet{Sai2020}. 

\begin{figure*}
\plotone{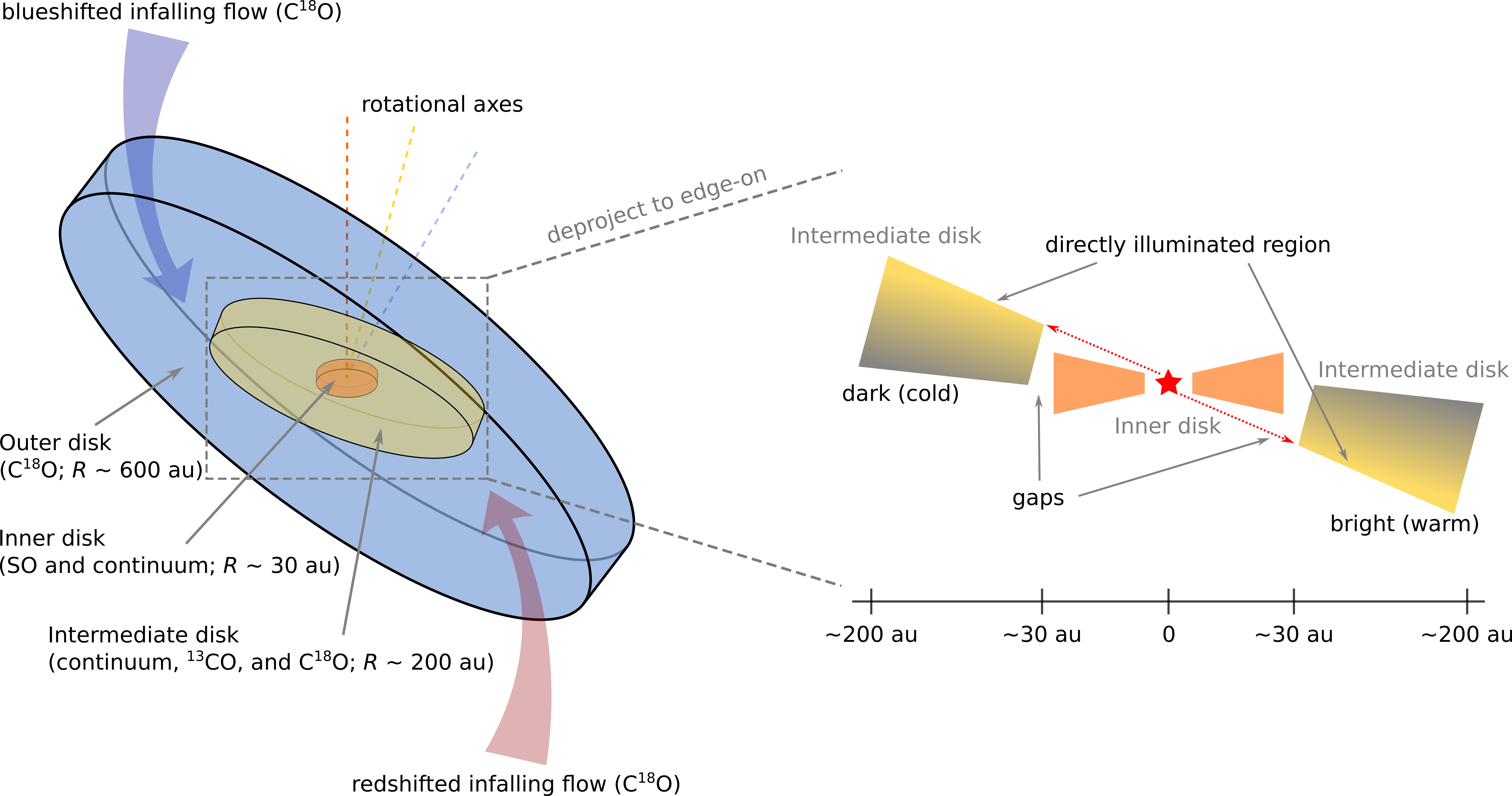}
\caption{Schematic illustration of the L1489~IRS system. The observing configuration of the warped disk is shown in the left figure, while the right figure shows the edge-on view of the inner and intermediate disks. In the right figure, the observer is in front of the screen, but slightly on the southern side (downside), and thus the brightness asymmetry (i.e., the western side is brighter than the eastern side) is observed.}
\label{fig:schematic_figure}
\end{figure*}

With this innermost tilted disk, the observed features in both dust and molecular line emission may be explained consistently. The dust gap and the potential gas gap suggested by the \ceo emission at $\approx30$\,au (see Section \ref{subsec:discussion_dust_ring}) can be the boundary between the outer disk and the innermost disk \citep[e.g.,][]{Nealon2018}. Furthermore, a tilted disk can cast a shadow on the outer disk and make an azimuthal variation in the temperature of the outer disk via irradiation (Figure \ref{fig:schematic_figure}). As a consequence, azimuthal brightness asymmetries in dust continuum and molecular line emission are expected \citep[e.g.,][]{Facchini2018, Young2021}. Indeed, our observations show a slightly weaker dust continuum emission on the northeastern side of the disk (Figure \ref{fig:continuum}), which is consistent with the expectation from the misalignment direction of the innermost SO disk and the configuration of the outer disk; i.e., the northeastern side is shadowed considering that the southern side is the far side of the disk (see Figure \ref{fig:schematic_figure}). The \tco and \ceo emissions also show lower brightness temperatures on the northeastern side (Figure \ref{fig:zoom_maps}\textrm{; see also Section \ref{subsubsec:discussion_azimuthal_aymmetry}}). 

We discuss the possible origins of the disk warp at $\sim$30\,au based on the present observations and theoretical studies. Numerical simulations suggest that warped disks can be formed by the interaction with an embedded planet \citep[e.g,][]{Nealon2018, Zhu2019}. If the orbital plane of the planet is inclined with respect to the outer disk, the inner disk can warp from the outer disk. If we assume that the observed gap is opened by a planet with an inclined orbit, the expected planet mass from the gap depth (Section \ref{subsec:discussion_dust_ring}) is slightly modified. \citet{Zhu2019} revised Equation \ref{eq:planet_mass} considering the inclination of the planet orbit (see also \citealt{Kanagawa2015}). Based on Equation (12) and (13) in \citet{Zhu2019}, \textrm{the expected planet mass would be a factor of $\sim1.4$--$2$ larger than the estimate in Section \ref{subsec:discussion_dust_ring}, i.e., $\sim0.042$--$4.7\,M_\mathrm{J}$, assuming orbital inclinations of $15$--$90\arcdeg$ and a typical disk aspect ratio of $h/r\sim0.05$--$0.1$.}

Theoretical works have shown that the warped disk structure can also be formed by an inclined binary \citep[e.g.,][]{Nixon2013, Facchini2018}. \citet{Nixon2013} suggests that the binary torque can tear the disk and break it into two, inner and outer, disks with a misalignment. Numerical simulations of a circumbinary disk also showed that the circumbinary disk is broken into misaligned outer and inner disks \citep{Facchini2018}. The binarity of L1489~IRS has been discussed in previous studies, although there is still no direct evidence that L1489~IRS is a binary. \citet{Sai2020} derived an upper limit on the separation of the binary ($a\lesssim30$\,au) based on a scattered-light image obtained with the Hubble Space Telescope \citep{Padgett1999}. Although our higher resolution observations of the dust emission in mm wavelength do not spatially resolve the central compact emission either, we can provide a tighter constraint on the upper limit of the binary separation. Considering that the spatial resolution of the \texttt{robust} $=-2$ image is $\approx0\farcs04$ or $\approx7$\,au and that the emission is spatially unresolved, i.e., a point source (see Appendix \ref{appendix:continuum_maps_different_robust}), the separation would be $\lesssim7$\,au.

Can such a close, inclined binary form the gap at $\sim30$\,au observed in the dust (and potentially \ceo) emission? \citet{Nixon2013} simulated the binary system with a misaligned orbit with respect to its circumbinary disk and derived a formula to estimate the disk-breaking radius. Assuming the bending-wave regime ($h/r \gg \alpha$ where $\alpha$ is the viscosity parameter), the possible maximum disk-breaking (or gap) radius would be $\lesssim2a \approx 14$\,au. Here we assume an inclination angle of the binary orbit with respect to the disk of $45\arcdeg$, an equal-mass binary, and a typical disk aspect ratio of $h/r \sim 0.1$, which maximizes the disk-breaking radius. While this expected maximum gap radius of $\approx14$\,au seemingly does not explain the observed gap radius of $\sim30$\,au, the latter may be overestimated due to the beam smearing effect: the dust ring width is spatially resolved while the central compact component is unresolved (see Table \ref{tab:visibility_fit_results}), and the actual gap radius (i.e., the local minima of the radial profile) could thus be smaller. Therefore, it may be possible to form the observed gap by the inner inclined binary.

Warped structures are also formed by the change of the angular momentum in the envelope and the disk during the accretion process. Hydrodynamical simulations show that the temporal variation of the angular momentum axis direction during the gas infall indeed results in a warped disk system \citep{Sai2020}. The outer warped structure at $\sim200$\,au observed in \ceo is likely to be formed by this mechanism rather than a binary \citep{Sai2020}. Similarly, a misalignment of the magnetic field and the angular momentum axis of the core results in a warped disk \citep{Matsumoto2017, Hirano2019, Hirano2020}. The warped structure of the L1489~IRS disk with tilts at $\sim30$ au and $\sim200$ au suggests the gradual change of the angular momentum axis of the infalling material.

\section{Summary} \label{sec:summary}
We have observed the disk around the Class~I protostar L1489~IRS in the 1.3~mm continuum and \tco $J=$ 2--1, \ceo $J=$ 2--1, and SO $J_N=$ 6$_5$--5$_4$ lines at spatial resolutions of $\sim$0\farcs1, as part of the eDisk ALMA Large Program. Our findings are summarized as follows:

\begin{enumerate}
    \item In the 1.3\,mm continuum emission, we detected a disk-like elongated structure with a non-axisymmetric, faint tails in the outermost region ($r\gtrsim2\arcsec$ or $\sim300$\,au) and a ring-like structure at $r \sim$0\farcs4 (or $\sim56$\,au). The dust continuum ring also shows azimuthal asymmetry, where the southwestern side is brighter than the northeastern side. We also detected a centrally peaked compact emission which is not spatially resolved. Furthermore, two tentative shoulder-like substructures at $\sim$1\farcs5 (or $\sim$220\,au) and $\sim$2\farcs3 (or $\sim$340\,au) are also identified in the radial intensity profile of the tapered continuum image.

    \item The high-velocity components of the \tco and \ceo emissions trace the protostellar disk. While \tco originates from the disk surface as suggested by the shift of the emission peak along the disk minor axis, \ceo traces the region near the midplane. The \tco and \ceo emissions also show (slightly) higher brightness temperatures on the southwestern side of the disk. In the outer region ($\gtrsim1\farcs5$), \tco shows an extended structure, suggesting that it traces the envelope. \ceo traces the warped disk structure as well as the streamer-like accretion flow in the outer region, which is consistent with the previous studies \citep{Yen2014, Sai2020}.

    \item The SO emission distribution shows a highly complex structure. While a compact, prominent emission appears in the innermost region ($\lesssim0\farcs2$), emission in the outer region is diffuse and non-uniform. The central compact component exhibits a velocity gradient over the extremely high velocities ($\pm10$--$12$\,km\,s$^{-1}$ with respect to the systemic velocity). In the outer region, multiple localized enhancements of the peak brightness temperature are identified.
    
    % \item The molecular lines exhibit different emission distributions. While the \tco and \ceo emission mainly traces the protostellar disk, the SO emission shows the combination of the centrally peaked emission ($\sim$0\farcs1) and outer diffuse emission ($\sim$2\arcsec). The radial intensity profiles of molecular line emission are presented to clearly exhibit the substructures in molecular lines. The molecular line emission also shows the rich radial substructures, including the different radial emission peaks. The radial profiles for each side (NW and SE) of the disk also exhibit the azimuthal asymmetries. 
    
    \item We performed an analytical fit to the dust continuum emission in the visibility domain, from which we confirm that the observed visibility can be explained by a simple model that consists of two Gaussian components (central compact one and outer extended one) and one Gaussian ring at $r=0\farcs39$ (or 56\,au). We also precisely characterized the properties of each component and dust disk geometry through the fit. While the central component is quite compact ($\sim1$\,au), the size of the outer component is $\sim$$540$\,au. The disk inclination and position angles are also constrained to be $72\arcdeg$ and $67\arcdeg$, respectively.
    
    \item We estimated the central stellar mass by the analysis of the \ceo PV diagram. The estimated central stellar mass of 1.5--1.9\,$M_\odot$ is consistent with the previous studies \citep{Yen2014, Sai2020}.

    \item We discuss the possible origin of the observed substructures. The coincidence between the dust ring and the peak of the \ceo radial intensity profile suggests that the gas disk also has a ring/gap structure. If we assume that the gas gap is carved by a formed planet, the planet mass would be $\sim0.6\,M_\mathrm{Jup}$. Alternatively, the grain growth may be responsible for the observed dust ring, as the ring radius is consistent with the prediction of the dust coagulation model.

    \item The non-uniform distributions and localized brightness temperature increases of the SO emission suggest that the origin of the outer diffuse SO emission is likely to be an accretion shock caused by the infalling material. On the other hand, we suggest that the central compact component originates from the thermal sublimation of SO and/or its precursor molecules at the warm inner disk. Indeed, the disk model of L1489~IRS shows a midplane dust temperature of $\gtrsim40$\,K at the SO emitting radii, which is comparable to the sublimation temperature of SO ($\sim$50\,K).  

    \item We also find that the velocity structure of the compact SO emission is consistent with the Keplerian rotation and that its velocity gradient direction is slightly ($\approx15\arcdeg$) misaligned with respect to that of the disk traced by \tco and \ceo. We propose that the inner disk is tilted, i.e., the rotation axis of the inner disk traced by SO is misaligned with respect to the outer disk traced by \tco and \ceo. The disk misalignment can make shadowed region in the outer disk, which explains the observed azimuthal asymmetry in the brightness of the dust and molecular line emission. We discuss the possible origins of the misalignment, planet or binary with an inclined orbit, or the temporal variation of the angular momentum axis direction during the gas infall.

    \item Combined with the previous studies, the L1489 disk has a warped structure with tilts at $\sim$30 au and $\sim$200 au (Figure \ref{fig:schematic_figure}). While the tilt at $\sim$30 au can be explained by a planet or binary with an inclined orbit, the temporal variation of the angular momentum axis direction during the gas infall could explain the whole warped structure.
    
    % \item One of the most plausible origin of the innermost dust ring at 56\,au could be the dust growth front mechanism \citep{Ohashi2021}. We found that the observed ring location is similar to the expected location by that mechanism. Alternatively, the ring-gap pair may be explained by the presence/absence of the material including the gas, judged from the coincidence of the radial peak of the dust and \ceo ring.
    % \item Although the origin of the outer dust shoulders are unclear, we found that the inner shoulder at $\sim$220\,au coincide with the CO snowline location inferred from the radiative transfer modeling by the previous works. Also, the outer shoulder at $\sim$340\,au is located at the similar location of the enhancement of the \ceo emission. These radial coincidence may imply that the those dust substructures have the snowline origin.
    % \item The origin of the centrally peaked SO emission could be the compact inner disk, suggested by its velocity gradient, where the temperature at the disk surface is high enough ($\sim$60\,K) for SO molecules to sublimate from the dust grains. Furthermore, we found that the direction of the velocity gradient is slightly tilted ($\sim$15\arcdeg) from the outer disk, implying a warped structure. The outer diffuse SO emission would be explained by the accretion shock origin, as observed in the other sources as well.
\end{enumerate}

\section*{Acknowledgments}
This paper makes use of the following ALMA data: ADS/JAO.ALMA\#2019.1.00261.L, \\
ADS/JAO.ALMA\#2019.A.00034.S. \\
ALMA is a partnership of ESO (representing its member states), NSF (USA) and NINS (Japan), together with NRC (Canada), MOST and ASIAA (Taiwan), and KASI (Republic of Korea), in cooperation with the Republic of Chile. The Joint ALMA Observatory is operated by ESO, AUI/NRAO and NAOJ. The National Radio Astronomy Observatory is a facility of the National Science Foundation operated
under cooperative agreement by Associated Universities, Inc. Y.Y. is supported by the International Graduate Program for Excellence in Earth-Space Science (IGPEES), World-leading Innovative Graduate Study (WINGS) Program of the University of Tokyo. Y.A. acknowledges support by NAOJ ALMA Scientific Research Grant code 2019-13B, Grant-in-Aid for Scientific Research (S) 18H05222, and Grant-in-Aid for Transformative Research Areas (A) 20H05844 and 20H05847. N.O. acknowledges support from National Science and Technology Council (NSTC) in Taiwan through the grants NSTC 109-2112-M-001-051 and 110-2112-M-001-031. J.J.T. acknowledges support from NASA XRP 80NSSC22K1159. J.K.J. and R.S. acknowledge support from the Independent Research Fund Denmark (grant No. 0135-00123B). S.T. is supported by JSPS KAKENHI Grant Numbers 21H00048 and 21H04495. This work was supported by NAOJ ALMA Scientific Research Grant Code 2022-20A. I.d.G. acknowledges support from grant PID2020-114461GB-I00, funded by
MCIN/AEI/10.13039/501100011033. P.M.K. acknowledges support from NSTC 108-2112- M-001-012, NSTC 109-2112-M-001-022 and NSTC 110-2112-M-001-057. W.K. was supported by the National Research Foundation of Korea (NRF) grant funded by the Korea government (MSIT) (NRF-2021R1F1A1061794). S.P.L. and T.J.T. acknowledge grants from the National Science and Technology Council of Taiwan
106-2119-M-007-021-MY3 and 109-2112-M-007-010-MY3. C.W.L. is supported by the Basic Science Research Program through the National Research Foundation of Korea (NRF) funded by the Ministry of Education, Science and Technology (NRF- 2019R1A2C1010851), and by the Korea Astronomy and Space Science Institute grant funded by the Korea government (MSIT; Project No. 2022-1-840-05). J.E.L. is supported by the National Research Foundation of Korea (NRF) grant funded by the Korean government (MSIT) (grant number 2021R1A2C1011718). Z.Y.L. is supported in part by NASA NSSC20K0533 and NSF AST-1910106. Z.Y.D.L. acknowledges support from NASA 80NSSC18K1095, the Jefferson Scholars Foundation, the NRAO ALMA Student Observing Support (SOS) SOSPA8-003, the Achievements Rewards for College Scientists (ARCS) Foundation Washington Chapter, the Virginia Space Grant Consortium (VSGC), and UVA research computing (RIVANNA). L.W.L. acknowledges support from NSF AST-2108794. S.M. is supported by JSPS KAKENHI Grant Numbers JP21J00086 and 22K14081. S.N. acknowledges support from the National Science Foundation through the Graduate Research Fellowship Program under Grant No. 2236415. Any opinions, findings, and conclusions or recommendations expressed in this material are those of the authors and do not necessarily reflect the views of the National Science Foundation. H.-W.Y. acknowledges support from the National Science and Technology Council (NSTC) in Taiwan through the grant NSTC 110-2628-M-001-003-MY3 and from the Academia Sinica Career Development Award (AS-CDA-111-M03). M.L.R.H. acknowledges support from the Michigan Society of Fellows. 
% SPL and TJT acknowledge grants from the National Science and Technology Council of Taiwan 106-2119-M-007-021-MY3 and 109-2112-M-007-010-MY3.


\vspace{5mm}
\facilities{ALMA}


\software{\texttt{astropy} \citep{astropy}, \texttt{bettermoments} \citep{bettermoments}, \texttt{GALARIO} \citep{GALARIO}, \texttt{emcee} \citep{emcee}, \texttt{SLAM} \citep{SLAM}, \texttt{disksurf} \citep{disksurf}}



% \restartappendixnumbering
\appendix

\section{Continuum Maps with different robust parameters}\label{appendix:continuum_maps_different_robust}
\begin{figure*}
\plotone{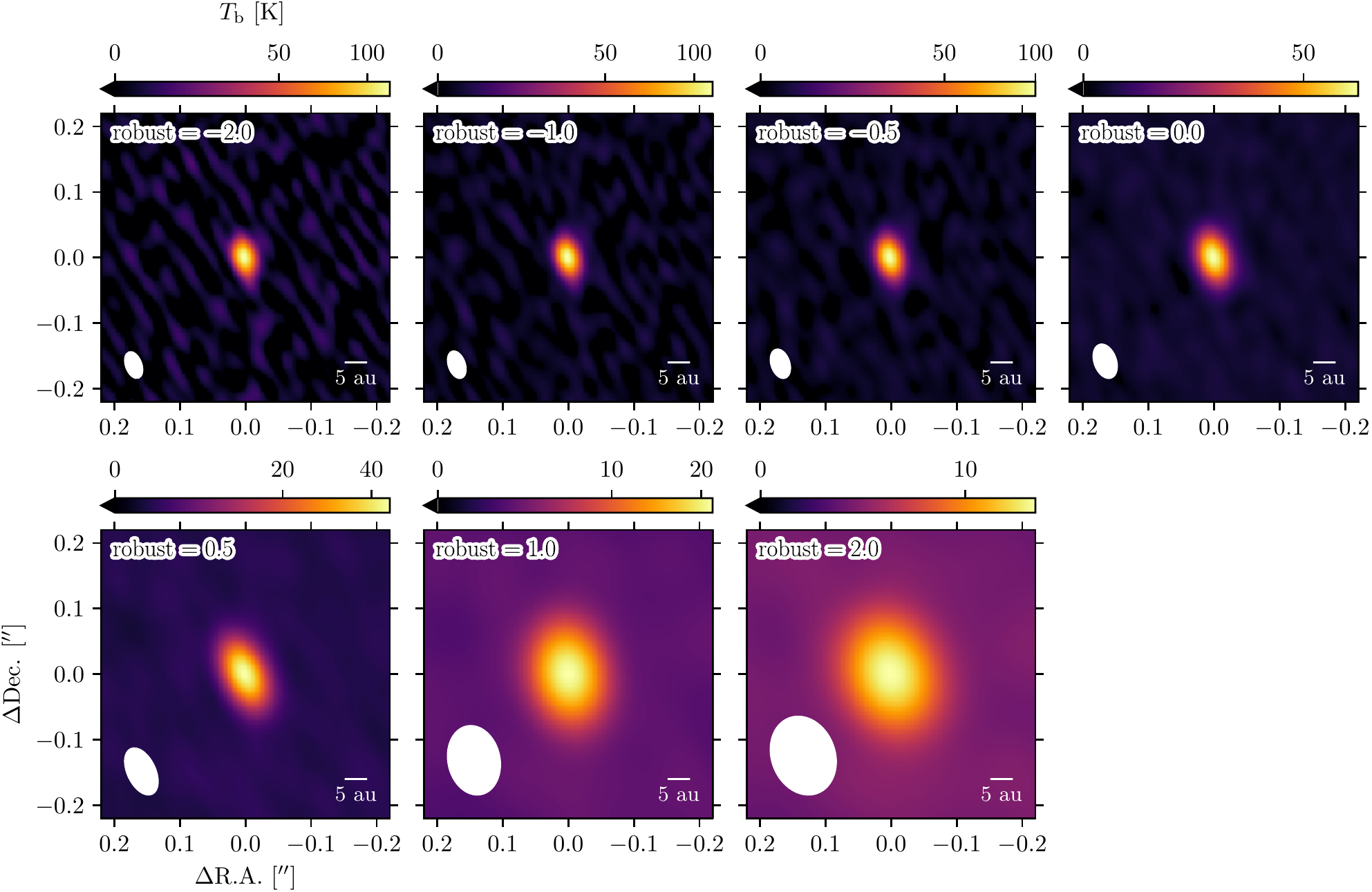}
\caption{Continuum emission at the central region imaged with different robust parameters ($-2.0$, $-1.0$, $-0.5$, $0.0$, $0.5$, $1.0$, and $2.0$). In each panel, the beam size and 5 au scale are indicated in the lower left and right, respectively.}
\label{fig:continum_different_robust}
\end{figure*}

Figure \ref{fig:continum_different_robust} shows the continuum emission maps at the central region imaged with different robust parameters, ranging from $-2.0$ (similar to uniform weighting) to $2.0$ (similar to natural weighting). The central compact component is not spatially resolved even with robust $=-2.0$ where the beam size is $0\farcs041\times0\farcs023$ (or 6.9\,au $\times$ 3.4\,au). 

% \section{Ancillary molecular line maps}\label{appendix:other_lines}

\section{Generation and selection of radial intensity profiles}\label{appendix:generation_of_profiles}
\begin{figure}
\plotone{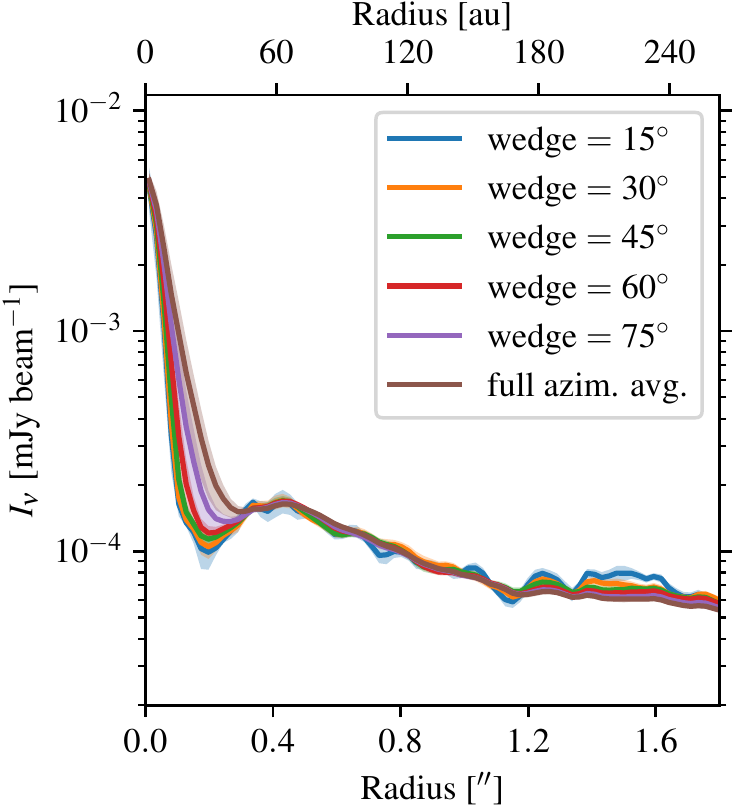}
\caption{Comparison of the radial continuum intensity profiles generated with different azimuthal wedges (averaged over both sides of the disk). Each color represents the radial profile extracted from $\pm15\arcdeg, \pm30\arcdeg, \pm45\arcdeg$, $\pm60\arcdeg$, $\pm75\arcdeg$, and $\pm90\arcdeg$ (i.e., full azimuthal average) wedges with respect to the major axis of the disk. The color-shaded regions represent the uncertainty of the profiles.}
\label{fig:cont_profile_wedge_comparison}
\end{figure}

In this section, we describe the detailed method to generate and select the radial intensity profiles of the continuum and line emission. We follow the methodology described in \citet{Law2021_MAPSIII}. We generate the profiles by deprojecting the emission maps (continuum) or velocity-integrated emission maps (line). To deproject the disk coordinates, we assume the disk geometry ($i=72\arcdeg$ and PA $=67\arcdeg$) based on the results of the visibility analysis (Section \ref{subsec:visfit}) for both continuum and line profiles. The radial bin size is one-quarter of the major axis of the synthesized beam. We averaged the emission in each radial bin. The uncertainty is calculated as the standard error on the region over which the emission is averaged. We generated the continuum intensity profiles averaged over several certain azimuthal wedges ($\pm15\arcdeg, \pm30\arcdeg, \pm45\arcdeg$ and $\pm90\arcdeg$ (i.e., full azimuthal average)) with respect to the disk major axis, and compare them in Figure \ref{fig:cont_profile_wedge_comparison}. While the wider wedges result in low-contrast substructures due to the coarser effective resolutions in the inclined disk, narrower wedges show wavy features that are not considered actual substructures in the outer radii. We selected the profile with $\pm45\arcdeg$ wedge as the representative profiles to compromise those two effects. For the line emission, we choose the same wedge ($\pm45\arcdeg$) as the continuum emission to generate the radial intensity profiles at similar effective resolutions. We note that the deprojection here assumes a geometrically thin disk and does not consider the 3D structures of the disk. 
% Although it is difficult to derive the precise radial intensity profile since the emission height is unknown, we confirmed that the radial intensity profiles do not change much with typical emission heights of $h/r=0.1, 0.2,$ and $0.3$.

\section{Radiative transfer model of the L1489~IRS disk}\label{appendix:disk_model}
\begin{figure}
\plotone{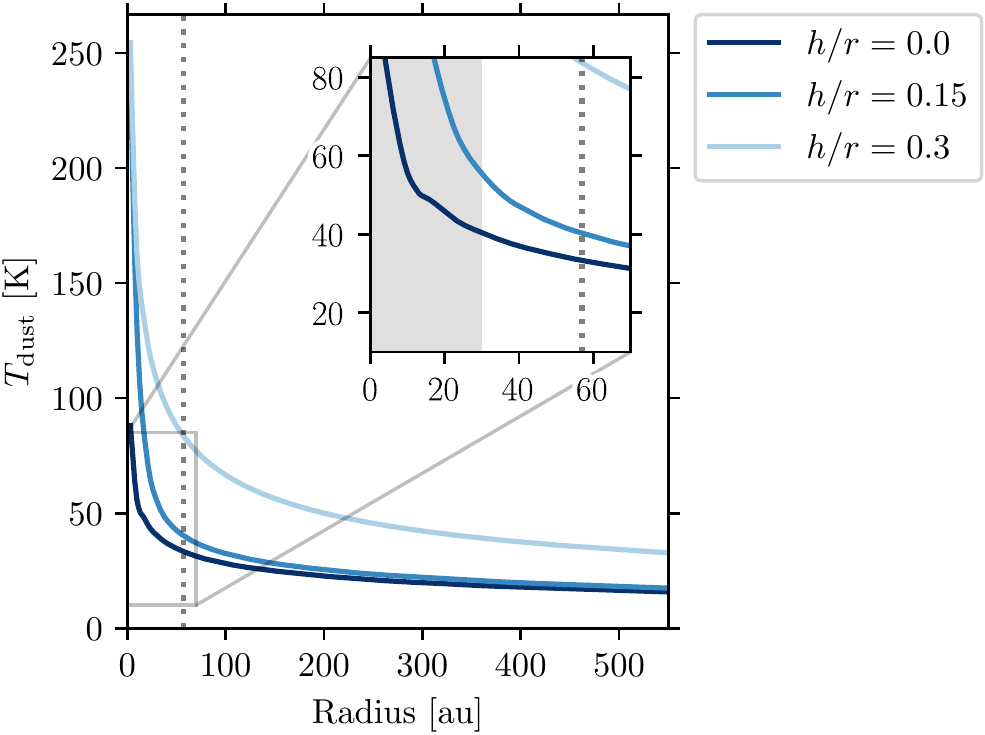}
\caption{Dust temperature profiles of the L1489~IRS disk at selected scale heights ($h/r=0.0$ (midplane)$, 0.15,$ and $0.3$), based on the radiative transfer calculation by \citet{Sai2020}. The inset axis shows the inner region for visual clarity. The vertical dotted lines in the main and inset axes indicate the location of the dust ring. The grey-shaded region in the inset axis is an approximate SO emitting region.}
\label{fig:model_temperature}
\end{figure}

Figure \ref{fig:model_temperature} shows the dust temperature profiles of the L1489~IRS disk at different disk scale heights ($h/r=0.0$ (midplane)$, 0.15,$ and $0.3$) calculated by the radiative transfer modeling by \citet{Sai2020}. The readers are referred to the original paper for the details of modeling.  In the emitting region of SO (the inset of Figure \ref{fig:model_temperature}), the dust temperature is higher than $\sim$40\,K, comparable to the sublimation temperature of SO molecules ($\sim$50\,K; see Section \ref{subsec:discussion_SO_emission}). 

% \section{JvM effect}


\bibliography{sample631}{}

\begin{thebibliography}{}
\expandafter\ifx\csname natexlab\endcsname\relax\def\natexlab#1{#1}\fi
\providecommand{\url}[1]{\href{#1}{#1}}
\providecommand{\dodoi}[1]{doi:~\href{http://doi.org/#1}{\nolinkurl{#1}}}
\providecommand{\doeprint}[1]{\href{http://ascl.net/#1}{\nolinkurl{http://ascl.net/#1}}}
\providecommand{\doarXiv}[1]{\href{https://arxiv.org/abs/#1}{\nolinkurl{https://arxiv.org/abs/#1}}}

\bibitem[{{ALMA Partnership} {et~al.}(2015){ALMA Partnership}, {Brogan},
  {P{\'e}rez}, {Hunter}, {Dent}, {Hales}, {Hills}, {Corder}, {Fomalont},
  {Vlahakis}, {Asaki}, {Barkats}, {Hirota}, {Hodge}, {Impellizzeri}, {Kneissl},
  {Liuzzo}, {Lucas}, {Marcelino}, {Matsushita}, {Nakanishi}, {Phillips},
  {Richards}, {Toledo}, {Aladro}, {Broguiere}, {Cortes}, {Cortes}, {Espada},
  {Galarza}, {Garcia-Appadoo}, {Guzman-Ramirez}, {Humphreys}, {Jung}, {Kameno},
  {Laing}, {Leon}, {Marconi}, {Mignano}, {Nikolic}, {Nyman}, {Radiszcz},
  {Remijan}, {Rod{\'o}n}, {Sawada}, {Takahashi}, {Tilanus}, {Vila Vilaro},
  {Watson}, {Wiklind}, {Akiyama}, {Chapillon}, {de Gregorio-Monsalvo}, {Di
  Francesco}, {Gueth}, {Kawamura}, {Lee}, {Nguyen Luong}, {Mangum}, {Pietu},
  {Sanhueza}, {Saigo}, {Takakuwa}, {Ubach}, {van Kempen}, {Wootten},
  {Castro-Carrizo}, {Francke}, {Gallardo}, {Garcia}, {Gonzalez}, {Hill},
  {Kaminski}, {Kurono}, {Liu}, {Lopez}, {Morales}, {Plarre}, {Schieven},
  {Testi}, {Videla}, {Villard}, {Andreani}, {Hibbard}, \&
  {Tatematsu}}]{ALMAPartnership2015}
{ALMA Partnership}, {Brogan}, C.~L., {P{\'e}rez}, L.~M., {et~al.} 2015, \apjl,
  808, L3, \dodoi{10.1088/2041-8205/808/1/L3}

\bibitem[{{Alves} {et~al.}(2017){Alves}, {Girart}, {Caselli}, {Franco}, {Zhao},
  {Vlemmings}, {Evans}, \& {Ricci}}]{Alves2017}
{Alves}, F.~O., {Girart}, J.~M., {Caselli}, P., {et~al.} 2017, \aap, 603, L3,
  \dodoi{10.1051/0004-6361/201731077}

\bibitem[{{Andrews} \& {Williams}(2005)}]{Andrews2005}
{Andrews}, S.~M., \& {Williams}, J.~P. 2005, \apj, 631, 1134,
  \dodoi{10.1086/432712}

\bibitem[{{Andrews} {et~al.}(2012){Andrews}, {Wilner}, {Hughes}, {Qi},
  {Rosenfeld}, {{\"O}berg}, {Birnstiel}, {Espaillat}, {Cieza}, {Williams},
  {Lin}, \& {Ho}}]{Andrews2012}
{Andrews}, S.~M., {Wilner}, D.~J., {Hughes}, A.~M., {et~al.} 2012, \apj, 744,
  162, \dodoi{10.1088/0004-637X/744/2/162}

\bibitem[{{Andrews} {et~al.}(2018){Andrews}, {Huang}, {P{\'e}rez}, {Isella},
  {Dullemond}, {Kurtovic}, {Guzm{\'a}n}, {Carpenter}, {Wilner}, {Zhang}, {Zhu},
  {Birnstiel}, {Bai}, {Benisty}, {Hughes}, {{\"O}berg}, \&
  {Ricci}}]{Andrews2018}
{Andrews}, S.~M., {Huang}, J., {P{\'e}rez}, L.~M., {et~al.} 2018, \apjl, 869,
  L41, \dodoi{10.3847/2041-8213/aaf741}

\bibitem[{{Ansdell} {et~al.}(2016){Ansdell}, {Williams}, {van der Marel},
  {Carpenter}, {Guidi}, {Hogerheijde}, {Mathews}, {Manara}, {Miotello},
  {Natta}, {Oliveira}, {Tazzari}, {Testi}, {van Dishoeck}, \& {van
  Terwisga}}]{Ansdell2016}
{Ansdell}, M., {Williams}, J.~P., {van der Marel}, N., {et~al.} 2016, \apj,
  828, 46, \dodoi{10.3847/0004-637X/828/1/46}

\bibitem[{{Ansdell} {et~al.}(2018){Ansdell}, {Williams}, {Trapman}, {van
  Terwisga}, {Facchini}, {Manara}, {van der Marel}, {Miotello}, {Tazzari},
  {Hogerheijde}, {Guidi}, {Testi}, \& {van Dishoeck}}]{ansdell2018}
{Ansdell}, M., {Williams}, J.~P., {Trapman}, L., {et~al.} 2018, \apj, 859, 21,
  \dodoi{10.3847/1538-4357/aab890}

\bibitem[{{Aota} {et~al.}(2015){Aota}, {Inoue}, \& {Aikawa}}]{Aota2015}
{Aota}, T., {Inoue}, T., \& {Aikawa}, Y. 2015, \apj, 799, 141,
  \dodoi{10.1088/0004-637X/799/2/141}

\bibitem[{{Aso} \& {Machida}(2020)}]{Aso2020}
{Aso}, Y., \& {Machida}, M.~N. 2020, \apj, 905, 174,
  \dodoi{10.3847/1538-4357/abc6fc}

\bibitem[{Aso \& Sai(2023)}]{SLAM}
Aso, Y., \& Sai, J. 2023, jinshisai/SLAM: First Release of SLAM, v1.0.0,
  Zenodo, \dodoi{10.5281/zenodo.7783868}

\bibitem[{{Aso} {et~al.}(2015){Aso}, {Ohashi}, {Saigo}, {Koyamatsu}, {Aikawa},
  {Hayashi}, {Machida}, {Saito}, {Takakuwa}, {Tomida}, {Tomisaka}, \&
  {Yen}}]{Aso2015}
{Aso}, Y., {Ohashi}, N., {Saigo}, K., {et~al.} 2015, \apj, 812, 27,
  \dodoi{10.1088/0004-637X/812/1/27}

\bibitem[{{Astropy Collaboration} {et~al.}(2013){Astropy Collaboration},
  {Robitaille}, {Tollerud}, {Greenfield}, {Droettboom}, {Bray}, {Aldcroft},
  {Davis}, {Ginsburg}, {Price-Whelan}, {Kerzendorf}, {Conley}, {Crighton},
  {Barbary}, {Muna}, {Ferguson}, {Grollier}, {Parikh}, {Nair}, {Unther},
  {Deil}, {Woillez}, {Conseil}, {Kramer}, {Turner}, {Singer}, {Fox}, {Weaver},
  {Zabalza}, {Edwards}, {Azalee Bostroem}, {Burke}, {Casey}, {Crawford},
  {Dencheva}, {Ely}, {Jenness}, {Labrie}, {Lim}, {Pierfederici}, {Pontzen},
  {Ptak}, {Refsdal}, {Servillat}, \& {Streicher}}]{astropy}
{Astropy Collaboration}, {Robitaille}, T.~P., {Tollerud}, E.~J., {et~al.} 2013,
  \aap, 558, A33, \dodoi{10.1051/0004-6361/201322068}

\bibitem[{{Bae} {et~al.}(2022){Bae}, {Teague}, {Andrews}, {Benisty},
  {Facchini}, {Galloway-Sprietsma}, {Loomis}, {Aikawa}, {Alarc{\'o}n},
  {Bergin}, {Bergner}, {Booth}, {Cataldi}, {Cleeves}, {Czekala}, {Guzm{\'a}n},
  {Huang}, {Ilee}, {Kurtovic}, {Law}, {Gal}, {Liu}, {Long}, {M{\'e}nard},
  {{\"O}berg}, {P{\'e}rez}, {Qi}, {Schwarz}, {Sierra}, {Walsh}, {Wilner}, \&
  {Zhang}}]{Bae2022}
{Bae}, J., {Teague}, R., {Andrews}, S.~M., {et~al.} 2022, \apjl, 934, L20,
  \dodoi{10.3847/2041-8213/ac7fa3}

\bibitem[{{Beckwith} {et~al.}(1990){Beckwith}, {Sargent}, {Chini}, \&
  {Guesten}}]{Beckwith1990}
{Beckwith}, S. V.~W., {Sargent}, A.~I., {Chini}, R.~S., \& {Guesten}, R. 1990,
  \aj, 99, 924, \dodoi{10.1086/115385}

\bibitem[{{Benisty} {et~al.}(2021){Benisty}, {Bae}, {Facchini}, {Keppler},
  {Teague}, {Isella}, {Kurtovic}, {P{\'e}rez}, {Sierra}, {Andrews},
  {Carpenter}, {Czekala}, {Dominik}, {Henning}, {Menard}, {Pinilla}, \&
  {Zurlo}}]{Benisty2021}
{Benisty}, M., {Bae}, J., {Facchini}, S., {et~al.} 2021, \apjl, 916, L2,
  \dodoi{10.3847/2041-8213/ac0f83}

\bibitem[{{Bi} {et~al.}(2020){Bi}, {van der Marel}, {Dong}, {Muto}, {Martin},
  {Smallwood}, {Hashimoto}, {Liu}, {Nomura}, {Hasegawa}, {Takami}, {Konishi},
  {Momose}, {Kanagawa}, {Kataoka}, {Ono}, {Sitko}, {Takahashi}, {Tomida}, \&
  {Tsukagoshi}}]{Bi2020}
{Bi}, J., {van der Marel}, N., {Dong}, R., {et~al.} 2020, \apjl, 895, L18,
  \dodoi{10.3847/2041-8213/ab8eb4}

\bibitem[{{Booth} {et~al.}(2023){Booth}, {Ilee}, {Walsh}, {Kama}, {Keyte}, {van
  Dishoeck}, \& {Nomura}}]{Booth2023}
{Booth}, A.~S., {Ilee}, J.~D., {Walsh}, C., {et~al.} 2023, \aap, 669, A53,
  \dodoi{10.1051/0004-6361/202244472}

\bibitem[{{Booth} {et~al.}(2021{\natexlab{a}}){Booth}, {van der Marel},
  {Leemker}, {van Dishoeck}, \& {Ohashi}}]{Booth2021_IRS48}
{Booth}, A.~S., {van der Marel}, N., {Leemker}, M., {van Dishoeck}, E.~F., \&
  {Ohashi}, S. 2021{\natexlab{a}}, \aap, 651, L6,
  \dodoi{10.1051/0004-6361/202141057}

\bibitem[{{Booth} {et~al.}(2021{\natexlab{b}}){Booth}, {Walsh}, {Terwisscha van
  Scheltinga}, {van Dishoeck}, {Ilee}, {Hogerheijde}, {Kama}, \&
  {Nomura}}]{Booth2021_HD100546}
{Booth}, A.~S., {Walsh}, C., {Terwisscha van Scheltinga}, J., {et~al.}
  2021{\natexlab{b}}, Nature Astronomy, 5, 684,
  \dodoi{10.1038/s41550-021-01352-w}

\bibitem[{{Brinch} {et~al.}(2007){Brinch}, {Crapsi}, {J{\o}rgensen},
  {Hogerheijde}, \& {Hill}}]{Brinch2007}
{Brinch}, C., {Crapsi}, A., {J{\o}rgensen}, J.~K., {Hogerheijde}, M.~R., \&
  {Hill}, T. 2007, \aap, 475, 915, \dodoi{10.1051/0004-6361:20078249}

\bibitem[{{Brunken} {et~al.}(2022){Brunken}, {Booth}, {Leemker}, {Nazari}, {van
  der Marel}, \& {van Dishoeck}}]{Brunken2022}
{Brunken}, N. G.~C., {Booth}, A.~S., {Leemker}, M., {et~al.} 2022, \aap, 659,
  A29, \dodoi{10.1051/0004-6361/202142981}

\bibitem[{{Casassus} \& {C{\'a}rcamo}(2022)}]{Casassus2022}
{Casassus}, S., \& {C{\'a}rcamo}, M. 2022, \mnras, 513, 5790,
  \dodoi{10.1093/mnras/stac1285}

\bibitem[{{Cataldi} {et~al.}(2021){Cataldi}, {Yamato}, {Aikawa}, {Bergner},
  {Furuya}, {Guzm{\'a}n}, {Huang}, {Loomis}, {Qi}, {Andrews}, {Bergin},
  {Booth}, {Bosman}, {Cleeves}, {Czekala}, {Ilee}, {Law}, {Le Gal}, {Liu},
  {Long}, {M{\'e}nard}, {Nomura}, {{\"O}berg}, {Schwarz}, {Teague},
  {Tsukagoshi}, {Walsh}, {Wilner}, \& {Zhang}}]{Cataldi2021}
{Cataldi}, G., {Yamato}, Y., {Aikawa}, Y., {et~al.} 2021, \apjs, 257, 10,
  \dodoi{10.3847/1538-4365/ac143d}

\bibitem[{{Cieza} {et~al.}(2016){Cieza}, {Casassus}, {Tobin}, {Bos},
  {Williams}, {Perez}, {Zhu}, {Caceres}, {Canovas}, {Dunham}, {Hales},
  {Prieto}, {Principe}, {Schreiber}, {Ruiz-Rodriguez}, \& {Zurlo}}]{Cieza2016}
{Cieza}, L.~A., {Casassus}, S., {Tobin}, J., {et~al.} 2016, \nat, 535, 258,
  \dodoi{10.1038/nature18612}

\bibitem[{{Czekala} {et~al.}(2021){Czekala}, {Loomis}, {Teague}, {Booth},
  {Huang}, {Cataldi}, {Ilee}, {Law}, {Walsh}, {Bosman}, {Guzm{\'a}n}, {Gal},
  {{\"O}berg}, {Yamato}, {Aikawa}, {Andrews}, {Bae}, {Bergin}, {Bergner},
  {Cleeves}, {Kurtovic}, {M{\'e}nard}, {Nomura}, {P{\'e}rez}, {Qi}, {Schwarz},
  {Tsukagoshi}, {Waggoner}, {Wilner}, \& {Zhang}}]{Czekala2021}
{Czekala}, I., {Loomis}, R.~A., {Teague}, R., {et~al.} 2021, \apjs, 257, 2,
  \dodoi{10.3847/1538-4365/ac1430}

\bibitem[{{de Gregorio-Monsalvo} {et~al.}(2013){de Gregorio-Monsalvo},
  {M{\'e}nard}, {Dent}, {Pinte}, {L{\'o}pez}, {Klaassen}, {Hales},
  {Cort{\'e}s}, {Rawlings}, {Tachihara}, {Testi}, {Takahashi}, {Chapillon},
  {Mathews}, {Juhasz}, {Akiyama}, {Higuchi}, {Saito}, {Nyman}, {Phillips},
  {Rod{\'o}n}, {Corder}, \& {Van Kempen}}]{deGregorioMonsalvo2013}
{de Gregorio-Monsalvo}, I., {M{\'e}nard}, F., {Dent}, W., {et~al.} 2013, \aap,
  557, A133, \dodoi{10.1051/0004-6361/201321603}

\bibitem[{{Dutrey} {et~al.}(2017){Dutrey}, {Guilloteau}, {Pi{\'e}tu},
  {Chapillon}, {Wakelam}, {Di Folco}, {Stoecklin}, {Denis-Alpizar}, {Gorti},
  {Teague}, {Henning}, {Semenov}, \& {Grosso}}]{Dutrey2017}
{Dutrey}, A., {Guilloteau}, S., {Pi{\'e}tu}, V., {et~al.} 2017, \aap, 607,
  A130, \dodoi{10.1051/0004-6361/201730645}

\bibitem[{{Endres} {et~al.}(2016){Endres}, {Schlemmer}, {Schilke}, {Stutzki},
  \& {M{\"u}ller}}]{CDMS3}
{Endres}, C.~P., {Schlemmer}, S., {Schilke}, P., {Stutzki}, J., \&
  {M{\"u}ller}, H. S.~P. 2016, Journal of Molecular Spectroscopy, 327, 95,
  \dodoi{10.1016/j.jms.2016.03.005}

\bibitem[{{Facchini} {et~al.}(2017){Facchini}, {Birnstiel}, {Bruderer}, \& {van
  Dishoeck}}]{Facchini2017}
{Facchini}, S., {Birnstiel}, T., {Bruderer}, S., \& {van Dishoeck}, E.~F. 2017,
  \aap, 605, A16, \dodoi{10.1051/0004-6361/201630329}

\bibitem[{{Facchini} {et~al.}(2018){Facchini}, {Juh{\'a}sz}, \&
  {Lodato}}]{Facchini2018}
{Facchini}, S., {Juh{\'a}sz}, A., \& {Lodato}, G. 2018, \mnras, 473, 4459,
  \dodoi{10.1093/mnras/stx2523}

\bibitem[{{Flores} {et~al.}(2021){Flores}, {Duch{\^e}ne}, {Wolff}, {Villenave},
  {Stapelfeldt}, {Williams}, {Pinte}, {Padgett}, {Connelley}, {van der Plas},
  {M{\'e}nard}, \& {Perrin}}]{Flores2021}
{Flores}, C., {Duch{\^e}ne}, G., {Wolff}, S., {et~al.} 2021, \aj, 161, 239,
  \dodoi{10.3847/1538-3881/abeb1e}

\bibitem[{{Foreman-Mackey} {et~al.}(2013){Foreman-Mackey}, {Hogg}, {Lang}, \&
  {Goodman}}]{emcee}
{Foreman-Mackey}, D., {Hogg}, D.~W., {Lang}, D., \& {Goodman}, J. 2013, \pasp,
  125, 306, \dodoi{10.1086/670067}

\bibitem[{{Garufi} {et~al.}(2022){Garufi}, {Podio}, {Codella}, {Segura-Cox},
  {Vander Donckt}, {Mercimek}, {Bacciotti}, {Fedele}, {Kasper}, {Pineda},
  {Humphreys}, \& {Testi}}]{Garufi2022}
{Garufi}, A., {Podio}, L., {Codella}, C., {et~al.} 2022, \aap, 658, A104,
  \dodoi{10.1051/0004-6361/202141264}

\bibitem[{{Harsono} {et~al.}(2018){Harsono}, {Bjerkeli}, {van der Wiel},
  {Ramsey}, {Maud}, {Kristensen}, \& {J{\o}rgensen}}]{Harsono2018}
{Harsono}, D., {Bjerkeli}, P., {van der Wiel}, M. H.~D., {et~al.} 2018, Nature
  Astronomy, 2, 646, \dodoi{10.1038/s41550-018-0497-x}

\bibitem[{{Hirano} \& {Machida}(2019)}]{Hirano2019}
{Hirano}, S., \& {Machida}, M.~N. 2019, \mnras, 485, 4667,
  \dodoi{10.1093/mnras/stz740}

\bibitem[{{Hirano} {et~al.}(2020){Hirano}, {Tsukamoto}, {Basu}, \&
  {Machida}}]{Hirano2020}
{Hirano}, S., {Tsukamoto}, Y., {Basu}, S., \& {Machida}, M.~N. 2020, \apj, 898,
  118, \dodoi{10.3847/1538-4357/ab9f9d}

\bibitem[{{Hogerheijde} \& {Sandell}(2000)}]{Hogerheijde2000}
{Hogerheijde}, M.~R., \& {Sandell}, G. 2000, \apj, 534, 880,
  \dodoi{10.1086/308795}

\bibitem[{{Hogerheijde} {et~al.}(1998){Hogerheijde}, {van Dishoeck}, {Blake},
  \& {van Langevelde}}]{Hogerheijde1998}
{Hogerheijde}, M.~R., {van Dishoeck}, E.~F., {Blake}, G.~A., \& {van
  Langevelde}, H.~J. 1998, \apj, 502, 315, \dodoi{10.1086/305885}

\bibitem[{{Jorsater} \& {van Moorsel}(1995)}]{Jorsater1995}
{Jorsater}, S., \& {van Moorsel}, G.~A. 1995, \aj, 110, 2037,
  \dodoi{10.1086/117668}

\bibitem[{{Kanagawa} {et~al.}(2015){Kanagawa}, {Muto}, {Tanaka}, {Tanigawa},
  {Takeuchi}, {Tsukagoshi}, \& {Momose}}]{Kanagawa2015}
{Kanagawa}, K.~D., {Muto}, T., {Tanaka}, H., {et~al.} 2015, \apjl, 806, L15,
  \dodoi{10.1088/2041-8205/806/1/L15}

\bibitem[{{Keppler} {et~al.}(2018){Keppler}, {Benisty}, {M{\"u}ller},
  {Henning}, {van Boekel}, {Cantalloube}, {Ginski}, {van Holstein}, {Maire},
  {Pohl}, {Samland}, {Avenhaus}, {Baudino}, {Boccaletti}, {de Boer},
  {Bonnefoy}, {Chauvin}, {Desidera}, {Langlois}, {Lazzoni}, {Marleau},
  {Mordasini}, {Pawellek}, {Stolker}, {Vigan}, {Zurlo}, {Birnstiel},
  {Brandner}, {Feldt}, {Flock}, {Girard}, {Gratton}, {Hagelberg}, {Isella},
  {Janson}, {Juhasz}, {Kemmer}, {Kral}, {Lagrange}, {Launhardt}, {Matter},
  {M{\'e}nard}, {Milli}, {Molli{\`e}re}, {Olofsson}, {P{\'e}rez}, {Pinilla},
  {Pinte}, {Quanz}, {Schmidt}, {Udry}, {Wahhaj}, {Williams}, {Buenzli},
  {Cudel}, {Dominik}, {Galicher}, {Kasper}, {Lannier}, {Mesa}, {Mouillet},
  {Peretti}, {Perrot}, {Salter}, {Sissa}, {Wildi}, {Abe}, {Antichi},
  {Augereau}, {Baruffolo}, {Baudoz}, {Bazzon}, {Beuzit}, {Blanchard}, {Brems},
  {Buey}, {De Caprio}, {Carbillet}, {Carle}, {Cascone}, {Cheetham}, {Claudi},
  {Costille}, {Delboulb{\'e}}, {Dohlen}, {Fantinel}, {Feautrier}, {Fusco},
  {Giro}, {Gluck}, {Gry}, {Hubin}, {Hugot}, {Jaquet}, {Le Mignant}, {Llored},
  {Madec}, {Magnard}, {Martinez}, {Maurel}, {Meyer}, {M{\"o}ller-Nilsson},
  {Moulin}, {Mugnier}, {Orign{\'e}}, {Pavlov}, {Perret}, {Petit}, {Pragt},
  {Puget}, {Rabou}, {Ramos}, {Rigal}, {Rochat}, {Roelfsema}, {Rousset}, {Roux},
  {Salasnich}, {Sauvage}, {Sevin}, {Soenke}, {Stadler}, {Suarez}, {Turatto}, \&
  {Weber}}]{Keppler2018}
{Keppler}, M., {Benisty}, M., {M{\"u}ller}, A., {et~al.} 2018, \aap, 617, A44,
  \dodoi{10.1051/0004-6361/201832957}

\bibitem[{{Kraus} {et~al.}(2020){Kraus}, {Kreplin}, {Young}, {Bate}, {Monnier},
  {Harries}, {Avenhaus}, {Kluska}, {Laws}, {Rich}, {Willson}, {Aarnio},
  {Adams}, {Andrews}, {Anugu}, {Bae}, {ten Brummelaar}, {Calvet}, {Cur{\'e}},
  {Davies}, {Ennis}, {Espaillat}, {Gardner}, {Hartmann}, {Hinkley}, {Labdon},
  {Lanthermann}, {LeBouquin}, {Schaefer}, {Setterholm}, {Wilner}, \&
  {Zhu}}]{Kraus2020}
{Kraus}, S., {Kreplin}, A., {Young}, A.~K., {et~al.} 2020, Science, 369, 1233,
  \dodoi{10.1126/science.aba4633}

\bibitem[{{Kwon} {et~al.}(2011){Kwon}, {Looney}, \& {Mundy}}]{Kwon2011}
{Kwon}, W., {Looney}, L.~W., \& {Mundy}, L.~G. 2011, \apj, 741, 3,
  \dodoi{10.1088/0004-637X/741/1/3}

\bibitem[{{Lambrechts} \& {Johansen}(2014)}]{Lambrechts2014}
{Lambrechts}, M., \& {Johansen}, A. 2014, \aap, 572, A107,
  \dodoi{10.1051/0004-6361/201424343}

\bibitem[{{Law} {et~al.}(2021{\natexlab{a}}){Law}, {Teague}, {Loomis}, {Bae},
  {{\"O}berg}, {Czekala}, {Andrews}, {Aikawa}, {Alarc{\'o}n}, {Bergin},
  {Bergner}, {Booth}, {Bosman}, {Calahan}, {Cataldi}, {Cleeves}, {Furuya},
  {Guzm{\'a}n}, {Huang}, {Ilee}, {Le Gal}, {Liu}, {Long}, {M{\'e}nard},
  {Nomura}, {P{\'e}rez}, {Qi}, {Schwarz}, {Soto}, {Tsukagoshi}, {Yamato},
  {van't Hoff}, {Walsh}, {Wilner}, \& {Zhang}}]{Law2021_MAPSIV}
{Law}, C.~J., {Teague}, R., {Loomis}, R.~A., {et~al.} 2021{\natexlab{a}},
  \apjs, 257, 4, \dodoi{10.3847/1538-4365/ac1439}

\bibitem[{{Law} {et~al.}(2021{\natexlab{b}}){Law}, {Loomis}, {Teague},
  {{\"O}berg}, {Czekala}, {Andrews}, {Huang}, {Aikawa}, {Alarc{\'o}n}, {Bae},
  {Bergin}, {Bergner}, {Boehler}, {Booth}, {Bosman}, {Calahan}, {Cataldi},
  {Cleeves}, {Furuya}, {Guzm{\'a}n}, {Ilee}, {Le Gal}, {Liu}, {Long},
  {M{\'e}nard}, {Nomura}, {Qi}, {Schwarz}, {Sierra}, {Tsukagoshi}, {Yamato},
  {van't Hoff}, {Walsh}, {Wilner}, \& {Zhang}}]{Law2021_MAPSIII}
{Law}, C.~J., {Loomis}, R.~A., {Teague}, R., {et~al.} 2021{\natexlab{b}},
  \apjs, 257, 3, \dodoi{10.3847/1538-4365/ac1434}

\bibitem[{{Law} {et~al.}(2022){Law}, {Crystian}, {Teague}, {{\"O}berg}, {Rich},
  {Andrews}, {Bae}, {Flaherty}, {Guzm{\'a}n}, {Huang}, {Ilee}, {Kastner},
  {Loomis}, {Long}, {P{\'e}rez}, {P{\'e}rez}, {Qi}, {Rosotti},
  {Ru{\'\i}z-Rodr{\'\i}guez}, {Tsukagoshi}, \& {Wilner}}]{Law2022}
{Law}, C.~J., {Crystian}, S., {Teague}, R., {et~al.} 2022, \apj, 932, 114,
  \dodoi{10.3847/1538-4357/ac6c02}

\bibitem[{{Lee} {et~al.}(2017){Lee}, {Li}, {Ho}, {Hirano}, {Zhang}, \&
  {Shang}}]{Lee2017}
{Lee}, C.-F., {Li}, Z.-Y., {Ho}, P. T.~P., {et~al.} 2017, Science Advances, 3,
  e1602935, \dodoi{10.1126/sciadv.1602935}

\bibitem[{{Lee} {et~al.}(2018){Lee}, {Kim}, {Myers}, {Saito}, {Kim}, {Kwon},
  {Lyo}, {Soam}, \& {Kim}}]{Lee2018}
{Lee}, C.~W., {Kim}, G., {Myers}, P.~C., {et~al.} 2018, \apj, 865, 131,
  \dodoi{10.3847/1538-4357/aadcf6}

\bibitem[{{Lee} {et~al.}(2019){Lee}, {Lee}, {Baek}, {Aikawa}, {Cieza}, {Yoon},
  {Herczeg}, {Johnstone}, \& {Casassus}}]{Lee2019}
{Lee}, J.-E., {Lee}, S., {Baek}, G., {et~al.} 2019, Nature Astronomy, 3, 314,
  \dodoi{10.1038/s41550-018-0680-0}

\bibitem[{{Lin} {et~al.}(2023){Lin}, {Li}, {Tobin}, {Ohashi}, {J{\o}rgensen},
  \& W.}]{eDisk_IRAS04302}
{Lin}, Z.-Y.~D., {Li}, Z.-Y., {Tobin}, J.~J., {et~al.} 2023, \apj, 0, 0,
  \dodoi{0}

\bibitem[{{Long} {et~al.}(2022){Long}, {Andrews}, {Rosotti}, {Harsono},
  {Pinilla}, {Wilner}, {{\"O}berg}, {Teague}, {Trapman}, \&
  {Tabone}}]{Long2022}
{Long}, F., {Andrews}, S.~M., {Rosotti}, G., {et~al.} 2022, \apj, 931, 6,
  \dodoi{10.3847/1538-4357/ac634e}

\bibitem[{{Maret} {et~al.}(2020){Maret}, {Maury}, {Belloche}, {Gaudel},
  {Andr{\'e}}, {Cabrit}, {Codella}, {Lef{\'e}vre}, {Podio}, {Anderl}, {Gueth},
  \& {Hennebelle}}]{Maret2020}
{Maret}, S., {Maury}, A.~J., {Belloche}, A., {et~al.} 2020, \aap, 635, A15,
  \dodoi{10.1051/0004-6361/201936798}

\bibitem[{{Matsumoto} {et~al.}(2017){Matsumoto}, {Machida}, \&
  {Inutsuka}}]{Matsumoto2017}
{Matsumoto}, T., {Machida}, M.~N., \& {Inutsuka}, S.-i. 2017, \apj, 839, 69,
  \dodoi{10.3847/1538-4357/aa6a1c}

\bibitem[{{McMullin} {et~al.}(2007){McMullin}, {Waters}, {Schiebel}, {Young},
  \& {Golap}}]{CASA}
{McMullin}, J.~P., {Waters}, B., {Schiebel}, D., {Young}, W., \& {Golap}, K.
  2007, in Astronomical Society of the Pacific Conference Series, Vol. 376,
  Astronomical Data Analysis Software and Systems XVI, ed. R.~A. {Shaw},
  F.~{Hill}, \& D.~J. {Bell}, 127

\bibitem[{{Mercimek} {et~al.}(2022){Mercimek}, {Codella}, {Podio}, {Bianchi},
  {Chahine}, {Bouvier}, {L{\'o}pez-Sepulcre}, {Neri}, \&
  {Ceccarelli}}]{Mercimek2022}
{Mercimek}, S., {Codella}, C., {Podio}, L., {et~al.} 2022, \aap, 659, A67,
  \dodoi{10.1051/0004-6361/202141790}

\bibitem[{{Miura} {et~al.}(2017){Miura}, {Yamamoto}, {Nomura}, {Nakamoto},
  {Tanaka}, {Tanaka}, \& {Nagasawa}}]{Miura2017}
{Miura}, H., {Yamamoto}, T., {Nomura}, H., {et~al.} 2017, \apj, 839, 47,
  \dodoi{10.3847/1538-4357/aa67df}

\bibitem[{{Motte} \& {Andr{\'e}}(2001)}]{Motte2001}
{Motte}, F., \& {Andr{\'e}}, P. 2001, \aap, 365, 440,
  \dodoi{10.1051/0004-6361:20000072}

\bibitem[{{M{\"u}ller} {et~al.}(2005){M{\"u}ller}, {Schl{\"o}der}, {Stutzki},
  \& {Winnewisser}}]{CDMS2}
{M{\"u}ller}, H. S.~P., {Schl{\"o}der}, F., {Stutzki}, J., \& {Winnewisser}, G.
  2005, Journal of Molecular Structure, 742, 215,
  \dodoi{10.1016/j.molstruc.2005.01.027}

\bibitem[{{M{\"u}ller} {et~al.}(2001){M{\"u}ller}, {Thorwirth}, {Roth}, \&
  {Winnewisser}}]{CDMS1}
{M{\"u}ller}, H.~S.~P., {Thorwirth}, S., {Roth}, D.~A., \& {Winnewisser}, G.
  2001, \aap, 370, L49, \dodoi{10.1051/0004-6361:20010367}

\bibitem[{{Nealon} {et~al.}(2018){Nealon}, {Dipierro}, {Alexander}, {Martin},
  \& {Nixon}}]{Nealon2018}
{Nealon}, R., {Dipierro}, G., {Alexander}, R., {Martin}, R.~G., \& {Nixon}, C.
  2018, \mnras, 481, 20, \dodoi{10.1093/mnras/sty2267}

\bibitem[{{Nixon} {et~al.}(2013){Nixon}, {King}, \& {Price}}]{Nixon2013}
{Nixon}, C., {King}, A., \& {Price}, D. 2013, \mnras, 434, 1946,
  \dodoi{10.1093/mnras/stt1136}

\bibitem[{{{\"O}berg} {et~al.}(2015){{\"O}berg}, {Furuya}, {Loomis}, {Aikawa},
  {Andrews}, {Qi}, {van Dishoeck}, \& {Wilner}}]{Oberg2015}
{{\"O}berg}, K.~I., {Furuya}, K., {Loomis}, R., {et~al.} 2015, \apj, 810, 112,
  \dodoi{10.1088/0004-637X/810/2/112}

\bibitem[{{{\"O}berg} {et~al.}(2021){{\"O}berg}, {Guzm{\'a}n}, {Walsh},
  {Aikawa}, {Bergin}, {Law}, {Loomis}, {Alarc{\'o}n}, {Andrews}, {Bae},
  {Bergner}, {Boehler}, {Booth}, {Bosman}, {Calahan}, {Cataldi}, {Cleeves},
  {Czekala}, {Furuya}, {Huang}, {Ilee}, {Kurtovic}, {Le Gal}, {Liu}, {Long},
  {M{\'e}nard}, {Nomura}, {P{\'e}rez}, {Qi}, {Schwarz}, {Sierra}, {Teague},
  {Tsukagoshi}, {Yamato}, {van't Hoff}, {Waggoner}, {Wilner}, \&
  {Zhang}}]{Oberg2021}
{{\"O}berg}, K.~I., {Guzm{\'a}n}, V.~V., {Walsh}, C., {et~al.} 2021, \apjs,
  257, 1, \dodoi{10.3847/1538-4365/ac1432}

\bibitem[{{Ohashi} {et~al.}(2023){Ohashi}, {Tobin}, {J{\o}rgensen}, {Takakuwa},
  {Sheehan}, \& {Aikawa}}]{eDisk_overview}
{Ohashi}, N., {Tobin}, J.~J., {J{\o}rgensen}, J.~K., {et~al.} 2023, \apj, 0, 0,
  \dodoi{0}

\bibitem[{{Ohashi} {et~al.}(2014){Ohashi}, {Saigo}, {Aso}, {Aikawa},
  {Koyamatsu}, {Machida}, {Saito}, {Takahashi}, {Takakuwa}, {Tomida},
  {Tomisaka}, \& {Yen}}]{Ohashi2014}
{Ohashi}, N., {Saigo}, K., {Aso}, Y., {et~al.} 2014, \apj, 796, 131,
  \dodoi{10.1088/0004-637X/796/2/131}

\bibitem[{{Ohashi} {et~al.}(2022{\natexlab{a}}){Ohashi}, {Kobayashi}, {Sai}, \&
  {Sakai}}]{Ohashi2022_L1489IRS}
{Ohashi}, S., {Kobayashi}, H., {Sai}, J., \& {Sakai}, N. 2022{\natexlab{a}},
  arXiv e-prints, arXiv:2206.07800.
\newblock \doarXiv{2206.07800}

\bibitem[{{Ohashi} {et~al.}(2022{\natexlab{b}}){Ohashi}, {Nakatani}, {Liu},
  {Kobayashi}, {Zhang}, {Hanawa}, \& {Sakai}}]{Ohashi2022_L1527IRS}
{Ohashi}, S., {Nakatani}, R., {Liu}, H.~B., {et~al.} 2022{\natexlab{b}}, \apj,
  934, 163, \dodoi{10.3847/1538-4357/ac794e}

\bibitem[{{Ohashi} {et~al.}(2021){Ohashi}, {Kobayashi}, {Nakatani}, {Okuzumi},
  {Tanaka}, {Murakawa}, {Zhang}, {Liu}, \& {Sakai}}]{Ohashi2021}
{Ohashi}, S., {Kobayashi}, H., {Nakatani}, R., {et~al.} 2021, \apj, 907, 80,
  \dodoi{10.3847/1538-4357/abd0fa}

\bibitem[{{Okuzumi} {et~al.}(2016){Okuzumi}, {Momose}, {Sirono}, {Kobayashi},
  \& {Tanaka}}]{Okuzumi2016}
{Okuzumi}, S., {Momose}, M., {Sirono}, S.-i., {Kobayashi}, H., \& {Tanaka}, H.
  2016, \apj, 821, 82, \dodoi{10.3847/0004-637X/821/2/82}

\bibitem[{{Padgett} {et~al.}(1999){Padgett}, {Brandner}, {Stapelfeldt},
  {Strom}, {Terebey}, \& {Koerner}}]{Padgett1999}
{Padgett}, D.~L., {Brandner}, W., {Stapelfeldt}, K.~R., {et~al.} 1999, \aj,
  117, 1490, \dodoi{10.1086/300781}

\bibitem[{{Pani{\'c}} {et~al.}(2009){Pani{\'c}}, {Hogerheijde}, {Wilner}, \&
  {Qi}}]{Panic2009}
{Pani{\'c}}, O., {Hogerheijde}, M.~R., {Wilner}, D., \& {Qi}, C. 2009, \aap,
  501, 269, \dodoi{10.1051/0004-6361/200911883}

\bibitem[{{Pinte} {et~al.}(2016){Pinte}, {Dent}, {M{\'e}nard}, {Hales}, {Hill},
  {Cortes}, \& {de Gregorio-Monsalvo}}]{Pinte2016}
{Pinte}, C., {Dent}, W.~R.~F., {M{\'e}nard}, F., {et~al.} 2016, \apj, 816, 25,
  \dodoi{10.3847/0004-637X/816/1/25}

\bibitem[{{Pinte} {et~al.}(2018{\natexlab{a}}){Pinte}, {Price}, {M{\'e}nard},
  {Duch{\^e}ne}, {Dent}, {Hill}, {de Gregorio-Monsalvo}, {Hales}, \&
  {Mentiplay}}]{Pinte2018_HD163296}
{Pinte}, C., {Price}, D.~J., {M{\'e}nard}, F., {et~al.} 2018{\natexlab{a}},
  \apjl, 860, L13, \dodoi{10.3847/2041-8213/aac6dc}

\bibitem[{{Pinte} {et~al.}(2018{\natexlab{b}}){Pinte}, {M{\'e}nard},
  {Duch{\^e}ne}, {Hill}, {Dent}, {Woitke}, {Maret}, {van der Plas}, {Hales},
  {Kamp}, {Thi}, {de Gregorio-Monsalvo}, {Rab}, {Quanz}, {Avenhaus}, {Carmona},
  \& {Casassus}}]{Pinte2018}
{Pinte}, C., {M{\'e}nard}, F., {Duch{\^e}ne}, G., {et~al.} 2018{\natexlab{b}},
  \aap, 609, A47, \dodoi{10.1051/0004-6361/201731377}

\bibitem[{{Roccatagliata} {et~al.}(2020){Roccatagliata}, {Franciosini},
  {Sacco}, {Randich}, \& {Sicilia-Aguilar}}]{Roccatagliata2020}
{Roccatagliata}, V., {Franciosini}, E., {Sacco}, G.~G., {Randich}, S., \&
  {Sicilia-Aguilar}, A. 2020, \aap, 638, A85,
  \dodoi{10.1051/0004-6361/201936401}

\bibitem[{{Ru{\'\i}z-Rodr{\'\i}guez} {et~al.}(2017){Ru{\'\i}z-Rodr{\'\i}guez},
  {Cieza}, {Williams}, {Principe}, {Tobin}, {Zhu}, \&
  {Zurlo}}]{Ruiz-Rodriguez2017}
{Ru{\'\i}z-Rodr{\'\i}guez}, D., {Cieza}, L.~A., {Williams}, J.~P., {et~al.}
  2017, \mnras, 468, 3266, \dodoi{10.1093/mnras/stx703}

\bibitem[{{Sai} {et~al.}(2022){Sai}, {Ohashi}, {Maury}, {Maret}, {Yen}, {Aso},
  \& {Gaudel}}]{Sai2022}
{Sai}, J., {Ohashi}, N., {Maury}, A.~J., {et~al.} 2022, \apj, 925, 12,
  \dodoi{10.3847/1538-4357/ac341d}

\bibitem[{{Sai} {et~al.}(2020){Sai}, {Ohashi}, {Saigo}, {Matsumoto}, {Aso},
  {Takakuwa}, {Aikawa}, {Kurose}, {Yen}, {Tomisaka}, {Tomida}, \&
  {Machida}}]{Sai2020}
{Sai}, J., {Ohashi}, N., {Saigo}, K., {et~al.} 2020, \apj, 893, 51,
  \dodoi{10.3847/1538-4357/ab8065}

\bibitem[{{Sakai} {et~al.}(2018){Sakai}, {Hanawa}, {Zhang}, {Higuchi},
  {Ohashi}, {Oya}, \& {Yamamoto}}]{Sakai2018}
{Sakai}, N., {Hanawa}, T., {Zhang}, Y., {et~al.} 2018, \nat, 565, 206,
  \dodoi{10.1038/s41586-018-0819-2}

\bibitem[{{Sakai} {et~al.}(2014){Sakai}, {Sakai}, {Hirota}, {Watanabe},
  {Ceccarelli}, {Kahane}, {Bottinelli}, {Caux}, {Demyk}, {Vastel}, {Coutens},
  {Taquet}, {Ohashi}, {Takakuwa}, {Yen}, {Aikawa}, \& {Yamamoto}}]{Sakai2014}
{Sakai}, N., {Sakai}, T., {Hirota}, T., {et~al.} 2014, \nat, 507, 78,
  \dodoi{10.1038/nature13000}

\bibitem[{{Sakai} {et~al.}(2017){Sakai}, {Oya}, {Higuchi}, {Aikawa}, {Hanawa},
  {Ceccarelli}, {Lefloch}, {L{\'o}pez-Sepulcre}, {Watanabe}, {Sakai}, {Hirota},
  {Caux}, {Vastel}, {Kahane}, \& {Yamamoto}}]{Sakai2017}
{Sakai}, N., {Oya}, Y., {Higuchi}, A.~E., {et~al.} 2017, \mnras, 467, L76,
  \dodoi{10.1093/mnrasl/slx002}

\bibitem[{{Segura-Cox} {et~al.}(2020){Segura-Cox}, {Schmiedeke}, {Pineda},
  {Stephens}, {Fern{\'a}ndez-L{\'o}pez}, {Looney}, {Caselli}, {Li}, {Mundy},
  {Kwon}, \& {Harris}}]{Segura-cox2020}
{Segura-Cox}, D.~M., {Schmiedeke}, A., {Pineda}, J.~E., {et~al.} 2020, \nat,
  586, 228, \dodoi{10.1038/s41586-020-2779-6}

\bibitem[{{Seifried} {et~al.}(2016){Seifried}, {S{\'a}nchez-Monge}, {Walch}, \&
  {Banerjee}}]{Seifried2016}
{Seifried}, D., {S{\'a}nchez-Monge}, {\'A}., {Walch}, S., \& {Banerjee}, R.
  2016, \mnras, 459, 1892, \dodoi{10.1093/mnras/stw785}

\bibitem[{{Sheehan} \& {Eisner}(2017)}]{Sheehan2017}
{Sheehan}, P.~D., \& {Eisner}, J.~A. 2017, \apj, 851, 45,
  \dodoi{10.3847/1538-4357/aa9990}

\bibitem[{{Sheehan} \& {Eisner}(2018)}]{Sheehan2018}
---. 2018, \apj, 857, 18, \dodoi{10.3847/1538-4357/aaae65}

\bibitem[{{Sheehan} {et~al.}(2020){Sheehan}, {Tobin}, {Federman}, {Megeath}, \&
  {Looney}}]{Sheehan2020}
{Sheehan}, P.~D., {Tobin}, J.~J., {Federman}, S., {Megeath}, S.~T., \&
  {Looney}, L.~W. 2020, \apj, 902, 141, \dodoi{10.3847/1538-4357/abbad5}

\bibitem[{{Takahashi} \& {Muto}(2018)}]{Takahashi2018}
{Takahashi}, S.~Z., \& {Muto}, T. 2018, \apj, 865, 102,
  \dodoi{10.3847/1538-4357/aadda0}

\bibitem[{{Takakuwa} \& {eDisk team}(2022)}]{eDisk_IRS7B_model}
{Takakuwa}, S., \& {eDisk team}. 2022, \apjs

\bibitem[{{Tamura} {et~al.}(1991){Tamura}, {Gatley}, {Waller}, \&
  {Werner}}]{Tamura1991}
{Tamura}, M., {Gatley}, I., {Waller}, W., \& {Werner}, M.~W. 1991, \apjl, 374,
  L25, \dodoi{10.1086/186063}

\bibitem[{{Tazzari} {et~al.}(2018){Tazzari}, {Beaujean}, \& {Testi}}]{GALARIO}
{Tazzari}, M., {Beaujean}, F., \& {Testi}, L. 2018, \mnras, 476, 4527,
  \dodoi{10.1093/mnras/sty409}

\bibitem[{{Teague} {et~al.}(2019){Teague}, {Bae}, \& {Bergin}}]{Teague2019}
{Teague}, R., {Bae}, J., \& {Bergin}, E.~A. 2019, \nat, 574, 378,
  \dodoi{10.1038/s41586-019-1642-0}

\bibitem[{{Teague} \& {Foreman-Mackey}(2018)}]{bettermoments}
{Teague}, R., \& {Foreman-Mackey}, D. 2018, Research Notes of the American
  Astronomical Society, 2, 173, \dodoi{10.3847/2515-5172/aae265}

\bibitem[{Teague {et~al.}(2021)Teague, Law, Huang, \& Meng}]{disksurf}
Teague, R., Law, C.~J., Huang, J., \& Meng, F. 2021, Journal of Open Source
  Software, 6, 3827, \dodoi{10.21105/joss.03827}

\bibitem[{{Tobin} {et~al.}(2020){Tobin}, {Sheehan}, {Megeath},
  {D{\'\i}az-Rodr{\'\i}guez}, {Offner}, {Murillo}, {van 't Hoff}, {van
  Dishoeck}, {Osorio}, {Anglada}, {Furlan}, {Stutz}, {Reynolds}, {Karnath},
  {Fischer}, {Persson}, {Looney}, {Li}, {Stephens}, {Chandler}, {Cox},
  {Dunham}, {Tychoniec}, {Kama}, {Kratter}, {Kounkel}, {Mazur}, {Maud},
  {Patel}, {Perez}, {Sadavoy}, {Segura-Cox}, {Sharma}, {Stephenson}, {Watson},
  \& {Wyrowski}}]{Tobin2020}
{Tobin}, J.~J., {Sheehan}, P.~D., {Megeath}, S.~T., {et~al.} 2020, \apj, 890,
  130, \dodoi{10.3847/1538-4357/ab6f64}

\bibitem[{{Trapman} {et~al.}(2019){Trapman}, {Facchini}, {Hogerheijde}, {van
  Dishoeck}, \& {Bruderer}}]{Trapman2019}
{Trapman}, L., {Facchini}, S., {Hogerheijde}, M.~R., {van Dishoeck}, E.~F., \&
  {Bruderer}, S. 2019, \aap, 629, A79, \dodoi{10.1051/0004-6361/201834723}

\bibitem[{{Tychoniec} {et~al.}(2021){Tychoniec}, {van Dishoeck}, {van't Hoff},
  {van Gelder}, {Tabone}, {Chen}, {Harsono}, {Hull}, {Hogerheijde}, {Murillo},
  \& {Tobin}}]{Tychoniec2021}
{Tychoniec}, {\L}., {van Dishoeck}, E.~F., {van't Hoff}, M. L.~R., {et~al.}
  2021, \aap, 655, A65, \dodoi{10.1051/0004-6361/202140692}

\bibitem[{{van der Marel} {et~al.}(2013){van der Marel}, {van Dishoeck},
  {Bruderer}, {Birnstiel}, {Pinilla}, {Dullemond}, {van Kempen}, {Schmalzl},
  {Brown}, {Herczeg}, {Mathews}, \& {Geers}}]{vanderMarel2013}
{van der Marel}, N., {van Dishoeck}, E.~F., {Bruderer}, S., {et~al.} 2013,
  Science, 340, 1199, \dodoi{10.1126/science.1236770}

\bibitem[{{van Gelder} {et~al.}(2021){van Gelder}, {Tabone}, {van Dishoeck}, \&
  {Godard}}]{vanGelder2021}
{van Gelder}, M.~L., {Tabone}, B., {van Dishoeck}, E.~F., \& {Godard}, B. 2021,
  \aap, 653, A159, \dodoi{10.1051/0004-6361/202141591}

\bibitem[{{van 't Hoff} {et~al.}(2023){van 't Hoff}, {Li}, {Tobin}, {Ohashi},
  {J{\o}rgensen}, \& {Lin}}]{eDisk_L1527}
{van 't Hoff}, M. L.~R., {Li}, Z.-Y., {Tobin}, J.~J., {et~al.} 2023, \apj, 0,
  0, \dodoi{0}

\bibitem[{{van 't Hoff} {et~al.}(2020){van 't Hoff}, {Harsono}, {Tobin},
  {Bosman}, {van Dishoeck}, {J{\o}rgensen}, {Miotello}, {Murillo}, \&
  {Walsh}}]{vantHoff2020}
{van 't Hoff}, M. L.~R., {Harsono}, D., {Tobin}, J.~J., {et~al.} 2020, \apj,
  901, 166, \dodoi{10.3847/1538-4357/abb1a2}

\bibitem[{{Villenave} {et~al.}(2020){Villenave}, {M{\'e}nard}, {Dent},
  {Duch{\^e}ne}, {Stapelfeldt}, {Benisty}, {Boehler}, {van der Plas}, {Pinte},
  {Telkamp}, {Wolff}, {Flores}, {Lesur}, {Louvet}, {Riols}, {Dougados},
  {Williams}, \& {Padgett}}]{Villenave2020}
{Villenave}, M., {M{\'e}nard}, F., {Dent}, W.~R.~F., {et~al.} 2020, \aap, 642,
  A164, \dodoi{10.1051/0004-6361/202038087}

\bibitem[{{Villenave} {et~al.}(2022){Villenave}, {Stapelfeldt}, {Duch{\^e}ne},
  {M{\'e}nard}, {Lambrechts}, {Sierra}, {Flores}, {Dent}, {Wolff}, {Ribas},
  {Benisty}, {Cuello}, \& {Pinte}}]{Villenave2022}
{Villenave}, M., {Stapelfeldt}, K.~R., {Duch{\^e}ne}, G., {et~al.} 2022, \apj,
  930, 11, \dodoi{10.3847/1538-4357/ac5fae}

\bibitem[{{Weaver} {et~al.}(2018){Weaver}, {Isella}, \& {Boehler}}]{Weaver2018}
{Weaver}, E., {Isella}, A., \& {Boehler}, Y. 2018, \apj, 853, 113,
  \dodoi{10.3847/1538-4357/aaa481}

\bibitem[{{Yen} {et~al.}(2013){Yen}, {Takakuwa}, {Ohashi}, \& {Ho}}]{Yen2013}
{Yen}, H.-W., {Takakuwa}, S., {Ohashi}, N., \& {Ho}, P. T.~P. 2013, \apj, 772,
  22, \dodoi{10.1088/0004-637X/772/1/22}

\bibitem[{{Yen} {et~al.}(2014){Yen}, {Takakuwa}, {Ohashi}, {Aikawa}, {Aso},
  {Koyamatsu}, {Machida}, {Saigo}, {Saito}, {Tomida}, \& {Tomisaka}}]{Yen2014}
{Yen}, H.-W., {Takakuwa}, S., {Ohashi}, N., {et~al.} 2014, \apj, 793, 1,
  \dodoi{10.1088/0004-637X/793/1/1}

\bibitem[{{Young} {et~al.}(2021){Young}, {Alexander}, {Walsh}, {Nealon},
  {Booth}, \& {Pinte}}]{Young2021}
{Young}, A.~K., {Alexander}, R., {Walsh}, C., {et~al.} 2021, \mnras, 505, 4821,
  \dodoi{10.1093/mnras/stab1675}

\bibitem[{{Zhang} {et~al.}(2015){Zhang}, {Blake}, \& {Bergin}}]{Zhang2015}
{Zhang}, K., {Blake}, G.~A., \& {Bergin}, E.~A. 2015, \apjl, 806, L7,
  \dodoi{10.1088/2041-8205/806/1/L7}

\bibitem[{{Zhang} {et~al.}(2021){Zhang}, {Booth}, {Law}, {Bosman}, {Schwarz},
  {Bergin}, {{\"O}berg}, {Andrews}, {Guzm{\'a}n}, {Walsh}, {Qi}, {van't Hoff},
  {Long}, {Wilner}, {Huang}, {Czekala}, {Ilee}, {Cataldi}, {Bergner}, {Aikawa},
  {Teague}, {Bae}, {Loomis}, {Calahan}, {Alarc{\'o}n}, {M{\'e}nard}, {Le Gal},
  {Sierra}, {Yamato}, {Nomura}, {Tsukagoshi}, {P{\'e}rez}, {Trapman}, {Liu}, \&
  {Furuya}}]{Zhang2021_MAPSV}
{Zhang}, K., {Booth}, A.~S., {Law}, C.~J., {et~al.} 2021, \apjs, 257, 5,
  \dodoi{10.3847/1538-4365/ac1580}

\bibitem[{{Zhang} {et~al.}(2018){Zhang}, {Zhu}, {Huang}, {Guzm{\'a}n},
  {Andrews}, {Birnstiel}, {Dullemond}, {Carpenter}, {Isella}, {P{\'e}rez},
  {Benisty}, {Wilner}, {Baruteau}, {Bai}, \& {Ricci}}]{Zhang2018}
{Zhang}, S., {Zhu}, Z., {Huang}, J., {et~al.} 2018, \apjl, 869, L47,
  \dodoi{10.3847/2041-8213/aaf744}

\bibitem[{{Zhu}(2019)}]{Zhu2019}
{Zhu}, Z. 2019, \mnras, 483, 4221, \dodoi{10.1093/mnras/sty3358}

\bibitem[{{Zucker} {et~al.}(2020){Zucker}, {Speagle}, {Schlafly}, {Green},
  {Finkbeiner}, {Goodman}, \& {Alves}}]{Zucker2020}
{Zucker}, C., {Speagle}, J.~S., {Schlafly}, E.~F., {et~al.} 2020, \aap, 633,
  A51, \dodoi{10.1051/0004-6361/201936145}

\bibitem[{{Zucker} {et~al.}(2019){Zucker}, {Speagle}, {Schlafly}, {Green},
  {Finkbeiner}, {Goodman}, \& {Alves}}]{Zucker2019}
---. 2019, \apj, 879, 125, \dodoi{10.3847/1538-4357/ab2388}

\end{thebibliography}
\bibliographystyle{aasjournal}



\end{document}